\documentclass[prd,aps,a4paper,twocolumn,eqsecnum,floatfix]{revtex4}  %,nofootinbib

\newif\ifusesec
\usesectrue  
   
\usepackage{graphicx} 
\usepackage{amsmath,amsfonts,amssymb}
\usepackage{mathtools}
\usepackage{color}

\makeatletter
\newsavebox{\@brx}
\newcommand{\llangle}[1][]{\savebox{\@brx}{\(\m@th{#1\langle}\)}%
  \mathopen{\copy\@brx\kern-0.5\wd\@brx\usebox{\@brx}}}
\newcommand{\rrangle}[1][]{\savebox{\@brx}{\(\m@th{#1\rangle}\)}%
  \mathclose{\copy\@brx\kern-0.5\wd\@brx\usebox{\@brx}}}
\makeatother

\newcommand{\beq}{\begin{equation}}
\newcommand{\eeq}{\end{equation}}
\newcommand{\bea}{\begin{eqnarray}}
\newcommand{\eea}{\end{eqnarray}}
\newcommand{\be}{\begin{equation}}
\newcommand{\ee}{\end{equation}}
\newcommand{\FP}{\mathop{\mathrm{FP}}_{B=0}}

\newcommand{\pinf}{p_{\infty}}
\begin{document}

\title{Quadrupolar bremsstrahlung waveform \\ at the third-and-a-half post-Newtonian accuracy}

\author{Donato Bini$^{1}$, Thibault Damour$^{2}$, Andrea Geralico$^{1}$}  
  \affiliation{
$^1$Istituto per le Applicazioni del Calcolo ``M. Picone'' CNR, I-00185 Rome, Italy\\
$^2$Institut des Hautes Etudes Scientifiques, 91440 Bures-sur-Yvette, France\\
}

\date{\today}

\begin{abstract}
We study the quadrupolar part of the gravitational waveform $h_{ij}$ (encoded in the helicity-($-2)$ radiative quadrupole moment 
$U_2 = \frac{1}{2!} \bar m^{i} \bar m^{j } U_{i j} \in\frac{R}{4G} \bar m^{i} \bar m^{j }  h_{i j}\equiv W $)
emitted during the scattering of two masses. Working within the Multipolar Post-Minkowskian (MPM) formalism, we compute the time-domain value of $U_2$  at the third-and-a-half post-Newtonian (3.5PN) accuracy by using the 3.5PN radiation-reacted quasi-Keplerian representation of the hyperbolic motion.  We then explicitly evaluate the {\it frequency-domain} value of  $U_2$ up to the 2-loop
level, i.e. $ O(G^4)$ contributions to $h_{ij}(\omega, \theta,\phi)$, corresponding to $O(G^3)$ contributions to  $\hat U_2(\omega, \theta,\phi)$.
The nonlinear memory contribution to the waveform in the center-of-mass  frame is computed too, and checked against the soft-limit of the waveform.
The 1-loop truncation of our 3.5PN frequency-domain MPM waveform is found to agree with corresponding existing Effective Field Theory (EFT) results when  subtracting the dipolar part of the Veneziano-Vilkovisky supertranslation connecting the MPM and EFT Bondi-Metzner-Sachs (BMS) frames. 
\end{abstract}

\maketitle

\section{Introduction}

The study of the gravitational two-body problem (both of its dynamics, and its radiation) is currently the focus of intense investigations using 
different theoretical approaches: post-Newtonian
(PN),  post-Minkowskian (PM), Multipolar Post-Minkowskian (MPM), self-force (SF), Effective-Field-Theory (EFT), Effective One-Body (EOB), Tutti Frutti (TF).
For entries into these various approaches  see e.g.  \cite{Blanchet:2013haa,Porto:2016pyg,Barack:2018yvs,Travaglini:2022uwo,Buonanno:2022pgc,Brunello:2025eso,Buonanno:1998gg,Buonanno:2000ef,Bini:2020nsb}.
Theoretical bottlenecks exist in all of the above mentioned approaches, especially when 
dealing with radiation-reaction effects.

Recently, the  (PN-matched) MPM  formalism has been pushed to high perturbative order so as  to reach the 4.5PN accuracy in the phasing of quasi-circular inspiralling binaries
\cite{Blanchet:2022vsm,Blanchet:2023bwj,Blanchet:2023sbv}.
Here we use state-of-the-art MPM results to  
compute the waveform $ W =\frac{R}{4G} \bar m^{i} \bar m^{j }  h_{i j}$ emitted by scattering orbits at the 3.5PN and the $O(G^3)$ (2-loop) accuracy. We present the explicit value of the frequency-domain waveform, obtained by Fourier-transforming the time-domain MPM waveform.

Our results go beyond previous similar computations.
Ref. \cite{Bini:2024rsy} went to the 3PN level but only at 1-loop and  Ref. \cite{Bini:2024ijq} went to 2-loop but was limited at the 2PN accuracy.
In other words, our current results extend  the previous knowledge in two directions: PN and loop (PM) orders. In the present paper we only consider the quadrupolar part of the waveform, leaving to future work the computation of other multipolar parts.

We use the mostly positive metric convention ($-+++$), and denote the PN expansion parameter as $\eta=\frac{1}{c}$.
We work with the MPM formalism and express all source and gauge multipole moments (see below) in terms of the dynamical variables of the binary system in  \lq\lq modified  harmonic coordinates." See  Ref. \cite{Blanchet:2013haa}.

\section{Radiative, canonical and source-type multipole moments}
 
The  transverse-traceless (TT) asymptotic waveform  $\lim_{R\to \infty}( R \, h_{ij}^{\rm TT})$ is conveniently
encoded in the complex waveform
\bea
h_c(t_r,\theta,\phi) &=&\lim_{R\to \infty}(R( h_+ -  i h_\times))\nonumber\\
&=&\lim_{R\to \infty} \bar m^{\mu } \bar m^{\nu }R\, h_{\mu \nu}\,.
\eea
For convenience, we work with the following rescaled waveform
\beq
W(t_r,\theta,\phi)\equiv\frac{c^4}{4G}h_c(t_r,\theta,\phi)\,,
\eeq
and focus on the computation of its frequency-domain version:
\beq
\label{hat_W_FT}
\hat W(\omega,\theta,\phi)\equiv \int_{-\infty}^\infty dt_r e^{i\omega t_r}W(t_r,\theta,\phi)\,.
\eeq
In the above equations 
\be \label{tret}
t_r = t-\frac{r}{c}-\frac{2 G{\cal M} }{c^3}\ln \frac{r}{b_0},
\ee
 is the (Bondi-type) retarded time (involving the total  center-of-mass (cm) mass-energy ${\mathcal M}=E/c^2$ and the
arbitrary length scale $b_0$), and $\bar m^{\mu }$ is a null polarization vector (chosen, like in our previous works \cite{Bini:2023fiz,Bini:2024rsy,Bini:2024ijq},
to have only spatial components $\bar m^j =\frac{1}{\sqrt{2}}(e^j_{\theta} - i e^j_{\phi})$ in the  cm  frame).
Here $\theta,\phi$ are polar coordinates in the cm frame when using a $z$-axis orthogonal to the orbital plane. We use the same frame as in our previous works \cite{Bini:2023fiz,Bini:2024rsy}.   
See Eqs. \eqref{frame1} and \eqref{frame1inv} below for the precise definitions of the cm frame spatial triad, $e_x,e_y,e_z$. 

The MPM formalism computes  the {\it multipolar decomposition} of $h_c(t_r,\theta,\phi)$.
More precisely,  the rescaled complex waveform $W$  
admits the multipolar decomposition (using $\eta \equiv \frac1c$ as a formal PN counting parameter)
\bea
\label{W_deco}
W(t_r,\theta,\phi)&=& U_2+ \eta (V_2 +U_3) + \eta^2 (V_3+U_4)\nonumber\\ 
&+& \eta^3 (V_4+U_5)+ \cdots \,.
\eea
Here, each $U_\ell$ (respectively $V_\ell$) denotes an even-parity (respectively odd-parity) $2^\ell$ radiative multipole contribution.
In the MPM formalism they are expressed  in terms 
of symmetric-trace-free (STF) Cartesian tensors of order $\ell$ (the radiative multipole moments  $U_{ i_1 i_2 \cdots i_{\ell}}(t_r)$
and  $V_{ i_1 i_2 \cdots i_{\ell}}(t_r)$)
according to
\bea
U_\ell(t_r,\theta,\phi) &=& \frac{1}{\ell!} \bar m^{i} \bar m^{j } n^{i_1} n^{i_2} \cdots n^{i_{\ell-2}} U_{i j i_1 i_2 \cdots i_{\ell-2}}(t_r)\,, \nonumber\\
V_\ell(t_r,\theta,\phi) &=& - \frac{1}{\ell!}\frac{2\ell}{\ell+1} \bar m^{i} \bar m^{j } n^c  n^{i_1} n^{i_2} \cdots n^{i_{\ell-2}}\times \nonumber\\
&&  \epsilon_{cd i}V_{j d i_1 i_2 \cdots i_{\ell-2}}(t_r)\,.
\eea

The PN-matched MPM formalism \cite{Blanchet:1985sp,Blanchet:1989ki,Damour:1990ji,Blanchet:1998in,Poujade:2001ie} computes the  radiative moments (observed at future null infinity) as nonlinear retarded functionals
(obtained by iteratively solving  Einstein's vacuum equations in the exterior zone) of two other types of multipole moments: (i) two sequences of source moments, 
\beq
 I_{i_1 i_2 \cdots i_{\ell}}(t_r)\,,\qquad J_{i_1 i_2 \cdots i_{\ell}}(t_r)\,,
\eeq 
together with four sequences of gauge multipole moments:  
\bea
&& W_{i_1 i_2 \cdots i_{\ell}}(t_r)\,,\quad X_{i_1 i_2 \cdots i_{\ell}}(t_r)\,,\nonumber\\
&& Y_{i_1 i_2 \cdots i_{\ell}}(t_r)\,,\quad  Z_{i_1 i_2 \cdots i_{\ell}}(t_r)\,.
\eea
In the MPM formalism the (time-domain) source and gauge moments are computed as explicit integrals of the PN-expanded effective stress-energy tensor of the source:
$\tau^{\mu\nu}=|g|(T^{\mu\nu}+T_{\rm grav}^{\mu\nu}) $.
In the case of a binary system, this means explicit expressions in terms of the positions and velocities. 
As already said we shall compute all the needed multipole moments in the
incoming cm frame of the system, and in modified harmonic coordinates \cite{Blanchet:2013haa}.

The frequency-domain waveform $W(\omega,  \theta,\phi) $ is then obtained by Fourier-transforming (over the retarded time variable) 
the radiative moments,
\bea
\hat U_{i_1 i_2 \cdots i_{\ell}}(\omega)= \int_{- \infty}^{+ \infty} dt_r e^{i \omega t_r} U_{i_1 i_2 \cdots i_{\ell}}(t_r)\,, \nonumber \\
\hat V_{i_1 i_2 \cdots i_{\ell}}(\omega)= \int_{- \infty}^{+ \infty} dt_r e^{i \omega t_r} V_{i_1 i_2 \cdots i_{\ell}}(t_r)\,.
\eea
This  leads to the following multipole expansion of the frequency-domain rescaled complex waveform 
$\hat h_c(\omega,\theta,\phi) \equiv 4 G \eta^4 \hat W(\omega,  \theta,\phi)$:
\bea
\label{Wom_th_phi}
\hat W(\omega,  \theta,\phi) &\equiv& \hat U_2(\omega,\theta,\phi)\nonumber\\
&+& \eta (\hat V_2(\omega,\theta,\phi) +\hat U_3(\omega,\theta,\phi))\nonumber\\ 
&+& \eta^2 (\hat V_3(\omega,\theta,\phi)+\hat U_4(\omega,\theta,\phi)) + \cdots\,,\nonumber\\ 
\eea
where, for instance, the quadrupole contribution reads
\beq
\hat U_2(\omega,\theta,\phi)=  \frac{1}{2!} \bar m^{i} \bar m^{j } \hat U_{i j}(\omega)\,.
\eeq

In the present work, we reach the  fractional  3.5PN accuracy level, i.e., $O(\eta^7)$, in $\hat U_2$, 
i.e. the
fractional 3.5PN accuracy level in the radiative quadrupole moment $U_{ij}$ (which starts at the ``Newtonian" level by the famous
Einstein quadrupole formula  $U^{\rm Newt}_{ij}= {\rm STF}\frac{d^2}{dt^2} (m_1 x_1^i x_2^j + m_2 x_2^i x_2^j)  $.
In the time domain  reaching the 3.5PN accuracy  means including terms up to $G^4$ in $U_{ij}(x(t), v(t))$.

To reach this accuracy on the complete $W=\frac{h_c}{4 G \eta^4}$, one would need, in view of the PN factors associated with higher multipoles,
\bea
W(t_r,\theta,\phi)&=& U_2+ \eta (V_2 +U_3) + \eta^2 (V_3+U_4)\nonumber\\ 
&+& \eta^3 (V_4+U_5)+ \eta^4 (V_5+U_6)+ \eta^5 (V_6+U_7)\nonumber\\
&+&  \eta^6 (V_7+U_8)+ \eta^7 (V_8+U_9)\nonumber\\
&+& O(\eta^8)\,,
\eea
the following fractional PN accuracies on the various multipoles:
$U_2 \sim  {\rm 3.5PN}\sim \eta^7$, 
$V_2, U_3  \sim  {\rm 3PN}\sim \eta^6$,  
$V_3,U_4 \sim  {\rm 2.5PN}\sim \eta^5$, 
$V_4,U_5  \sim  {\rm 2PN}\sim \eta^4$\, etc.
Here, we focus on the more demanding quadrupole term $U_2$, leaving to future work the other multipoles.
  Let us also note that while we will reach the full 3.5PN accuracy (including $G^4$ 
contributions) on the time-domain quadrupole $U_2(t_r)$, we will limit the accuracy  of the corresponding frequency-domain
quadrupole bremsstrahlung $\hat U_2(\omega)$ to the $G^3$ level, keeping, however, the full $\eta^7$ accuracy. In other words, we shall compute
\bea
\hat U_2(\omega) &\sim& G(1+\cdots+\eta^7) + G^2(1+\cdots+\eta^7)\nonumber\\
& +& G^3(1+\cdots+\eta^7) + O(G^4).
\eea
In the latter expansion, the $G^1$ term corresponds to the tree-level bremsstrahlung waveform (first computed in the
time-domain by Kovacs and Thorne  \cite{Kovacs:1977uw,Kovacs:1978eu}, see also \cite{Jakobsen:2021smu}; see e.g. Refs. \cite{Mougiakakos:2021ckm,Bini:2024ijq}  for its frequency-domain value). The $G^2$ term corresponds 
to the one-loop level \cite{Brandhuber:2023hhy,Herderschee:2023fxh,Georgoudis:2023lgf,Georgoudis:2023eke,Bini:2024rsy}, and  the $G^3$ term
to the two-loop level (which has not yet been computed by amplitude methods).

Let us recall that the knowledge of the radiative moments allows one to compute  the radiative losses of energy, angular momentum, linear momentum (and center-of-mass position). E.g., we have the following angle-integrated time-domain fluxes (with simple, corresponding 
frequency-domain integrated losses)
\bea
{\mathcal F}_E(t_r) &=& \frac{G}{c^5}\left\{
\frac15 U^{(1)}_{ij}U^{(1)}_{ij}\right.\nonumber\\
&+& \eta^2 \left[\frac{1}{189}U^{(1)}_{ijk}U^{(1)}_{ijk}+\frac{16}{45}V^{(1)}_{ij}V^{(1)}_{ij}\right]\nonumber\\
&+& \eta^4 \left[ \frac{1}{9072}U^{(1)}_{ijkm}U^{(1)}_{ijkm}+\frac{1}{84}V^{(1)}_{ijk}V^{(1)}_{ijk} \right]\nonumber\\
&+&\eta^6 \left[\frac{1}{594000}U^{(1)}_{ijkml}U^{(1)}_{ijkml}+\frac{4}{14175}  V^{(1)}_{ijkl}V^{(1)}_{ijkl}  \right]\nonumber\\
&+& \left. O(\eta^8)\right\}\,,
\eea
%%%
\bea
{\mathcal F}_{J^i}(t_r) &=& \frac{G}{c^5}\epsilon_{iab}\left\{
\frac25 U_{aj}U^{(1)}_{bj}\right.\nonumber\\
&+& \eta^2 \left[\frac{1}{63}U_{ajk}U^{(1)}_{bjk}+\frac{32}{45}V_{aj}V^{(1)}_{bj}\right]\nonumber\\
&+& \eta^4 \left[ \frac{1}{2268}U_{ajkl}U^{(1)}_{bjkl}+\frac{1}{28}V_{ajk}V^{(1)}_{bjk} \right]\nonumber\\
&+&\eta^6 \left[\frac{1}{118800}U_{ajklm}U^{(1)}_{bjklm}+\frac{16}{14175}  V_{ajkl}V^{(1)}_{bjkl}  \right]\nonumber\\
&+& \left. O(\eta^8)\right\}\,,
\eea
%and
\bea
&&{\mathcal F}_{P^i}(t_r) = \frac{G}{c^7}\left\{
\frac2{63} U^{(1)}_{ijk}U^{(1)}_{jk}\right.\nonumber\\
&&+ \eta^2 \left[\frac{1}{1134}U^{(1)}_{ijkl}U^{(1)}_{jkl}+\frac{1}{126}\epsilon_{ijk}U^{(1)}_{jab}V^{(1)}_{kab}+\frac{4}{63}V^{(1)}_{ijk}V^{(1)}_{jk}\right]\nonumber\\
&&+ \eta^4 \left[ \frac{1}{59400}U^{(1)}_{ijklm}U^{(1)}_{jklm}+\frac{2}{14175}\epsilon_{ijk}U^{(1)}_{jabc}V^{(1)}_{kabc}\right.\nonumber\\
&&+\left.\left. \frac{2}{945}V^{(1)}_{ijkl}V^{(1)}_{jkl} \right]+ O(\eta^6)\right\}\,.
\eea
Concerning the flux associated to the cm position see  Sec. 2.4.1, Eq. 181b, of Ref. \cite{Blanchet:2013haa} 
and related discussion (and references therein).
%%%%%%%%%%%%%
\section{Functional structure of the quadrupole radiative moment in the MPM formalism}

The quadrupole radiative moment $U_2(t_r)$ is first expressed in the MPM formalism as a {\it retarded nonlinear functional} of the {\it canonical moments} $M_L$ (mass moments) and $S_L$ (current moments),
\beq
U_L=U_L[M_L,S_L]\,,\qquad V_L=V_L[M_L,S_L]\,,
\eeq
where $L=i_1i_2\dots i_l$ is a multi-index.
The canonical moments $M_L$ and $S_L$, in turn, are expressed as {\it local functionals} of  six other sequences of STF moments:
the two sequences of {\it source moments}, $I_L$ and $J_L$ , and four other sequences of {\it gauge moments}, usually denoted as $W_L, X_L, Y_L, Z_L$:
\bea
M_L&=&M_L[I_L,J_L, W_L, X_L,Y_L, Z_L]\,,\nonumber\\ 
S_L&=& S_L[I_L,J_L,, W_L X_L,Y_L, Z_L]\,.
\eea
In particular, the canonical quadrupole moment $M_{ij}$ is related to the source and gauge moments by
\bea
\label{blocks_0}
M_{ij} &=& 
	I_{ij}
	+ 4G \eta^5 M^{W_0I_2}_{ij}\nonumber\\
&+&
4G \eta^7[M^{W_2I_2}_{ij} 
+M^{Y_2I_2}_{ij}+M^{X_0I_2}_{ij}+M^{W_1I_3}_{ij}\nonumber\\
&+& M^{Y_1I_3}_{ij}+ M^{W_0W_2}_{ij}+ M^{W_1W_1}_{ij}+M^{W_0Y_2}_{ij}+M^{W_1Y_1}_{ij}\nonumber\\
&+& M^{Z_1I_2}_{ij}+M^{W_1J_2}_{ij}+M^{Y_1J_2}_{ij} ]\,,
\eea
where (using the notation $T_{\langle i j\rangle}\equiv {\rm STF}_{ij} T_{ij}\equiv T_{(ij)}-\frac13 \delta_{ij} T^{ss} $)
\bea
\label{blocks}
M^{W_0I_2}_{ij}&=&  W^{(2)} I^{}_{ij}- W^{(1)} I^{(1)}_{ij}
\,, \nonumber\\
M^{W_2I_2}_{ij}&=& \frac{4}{7} W^{(1)}_{a\langle i} I^{(3)}_{j\rangle a}+ \frac{6}{7} W^{}_{a\langle i} I^{(4)}_{j\rangle a}
\,, \nonumber\\
M^{Y_2I_2}_{ij}&=& - \frac{1}{7} Y^{(3)}_{a\langle i} I^{}_{j\rangle a}-  Y^{}_{a\langle i} I^{(3)}_{j\rangle a}
\,, \nonumber\\
M^{X_0I_2}_{ij}&=& - 2 X\, I^{(3)}_{ij}
\,, \nonumber\\
M^{W_1I_3}_{ij}&=& - \frac{5}{21}W^{(4)}_{a} I^{}_{ija}+ \frac{1}{63} W^{(3)}_{a} I^{(1)}_{ija}
\,, \nonumber\\
M^{Y_1I_3}_{ij}&=& - \frac{25}{21} Y^{(3)}_{a} I^{}_{ija}- \frac{22}{63} Y^{(2)}_{a} I^{(1)}_{ija}+ \frac{5}{63} Y^{(1)}_{a} I^{(2)}_{ija}
\,, \nonumber\\
M^{W_0W_2}_{ij}&=& + 2 W^{(3)} W^{}_{ij}+ 2 W^{(2)} W^{(1)}_{ij}
\,, \nonumber\\
M^{W_1W_1}_{ij}&=&  - \frac{4}{3} W_{\langle i} W^{(3)}_{j\rangle}
\,, \nonumber\\
M^{W_0Y_2}_{ij}&=& + 2 W^{(2)} Y^{}_{ij}
\,, \nonumber\\
M^{W_1Y_1}_{ij}&=& - 4 W_{\langle i} Y^{(2)}_{j\rangle}
\,, \nonumber\\
M^{Z_1I_2}_{ij}&=& \epsilon_{ab\langle i}\bigg[\frac{1}{3} I_{j\rangle a} Z^{(3)}_b- I_{j\rangle a}^{(3)}Z^{}_b\bigg]
\,, \nonumber\\
M^{W_1J_2}_{ij}&=&  \epsilon_{ab\langle i}\frac{4}{9} J^{}_{j\rangle a} W^{(3)}_b
\,, \nonumber\\
M^{Y_1J_2}_{ij}&=& \epsilon_{ab\langle i}\bigg[- \frac{4}{9} J^{}_{j\rangle a} Y^{(2)}_b+ \frac{8}{9} J^{(1)}_{j\rangle a} Y^{(1)}_b\bigg]\,.
\eea
Here, the superscript within parentheses denotes a repeated time derivative: $F^{(n)}\equiv \frac{d^n}{dt^n} F(t)$.
Beware that the quantity $W$ entering the first Eq. \eqref{blocks} denotes the $l=0$ (monopolar) component of the sequence of gauge multipoles of $W_L$.
In the latter expressions, only the term $M^{W_0I_2}_{ij}$ needs to be computed at the fractional 1PN level of accuracy, while all the others can be taken at the leading Newtonian accuracy. 
In view of Eq. \eqref{blocks_0} one needs to know the source quadrupole $I_{ij}$ at accuracy $O(G^3\eta^7)$ and the various gauge multipoles at the following accuracies:  the gauge monopole $W$ needs to be computed at the 1PN fractional accuracy, while all the others  can be taken at the Newtonian accuracy.
Completing the Newtonian level values of $W$ 
\cite{Arun:2007sg},  $W_i$ and $Y_i$ \cite{Mishra:2015bqa}, we computed the 1PN contribution to $W$ and the Newtonian values of $W_{ij}$, $Z_i$, $Y_{ij}$ and $X$.
Table \ref{gauge_pot} summarizes our results which have been checked to be in agreement with very recent independent results \cite{Blanchet:2026suq}. 
We give in appendix \ref{gauge_moments} some details of their derivation, as well as references to  the original literature. 

The expressions, Eq. \eqref{blocks_0}, of $M_{ij}$ in terms of source and gauge moments must then be inserted in the functional relation between $U_{ij}$ and the canonical moments given at the  $O(\eta^7)$ fractional accuracy by the results derived in Refs. \cite{Faye:2012we,Faye:2014fra} (see also Ref. \cite{Blanchet:2013haa}), namely:
\bea
\label{U_def}
U_{ij} &=& M_{ij}^{(2)} + \eta^3 U_{ij}^{\overline{\rm 1.5PN}}  +\eta^5 U_{ij}^{\overline{\rm 2.5PN}}\nonumber\\ 
&+&  \eta^6 U_{ij}^{\overline{\rm 3PN}}+\eta^7 U_{ij}^{\overline{\rm 3.5PN}}+ O\left(\eta^8\right)\,,
\eea
where we have put a line over the 1.5PN, 2.5PN and 3.5PN superscripts as a warning that these contributions start at the shown PN order and need to be further expanded  up to the requested accuracy of our work. 
In addition,
the quantities entering \eqref{U_def} are defined, at the fractional 3.5PN accuracy as follows.

The other terms in Eq. \eqref{U_def} read at our accuracy (with ${\cal M} \equiv \frac{E}{c^2}$ denoting the total ADM mass in the cm system)
\begin{widetext}
\bea
\label{lista_up_3pn}
U_{ij}^{\overline{\rm 1.5PN}}(t_r) &=&  2 G {\cal M}\eta^3  \int_0^{+\infty} d\tau\,  M_{ij}^{(4)}(t_r-\tau)\bigg[\ln \left(\frac{c \tau}{2 b_0}\right)+ \frac{11}{12}\biggl]\,,\nonumber\\
U_{ij}^{\overline{{\rm 2.5PN}}}(t_r) &=&  G\eta^5  \left\lbrace 
U_{ij}^{\overline{{\rm 2.5PN}} \rm (mem)}
+U_{ij}^{ \overline{{\rm 2.5PN}} I_2I_2}+\text{U}_{ij}^{\text{2.5PN}I_2J_1}\right\rbrace\,,\nonumber\\
U_{ij}^{\overline{{\rm 2.5PN}} \rm  (mem)}(t_r)&=&- \frac{2}{7} \int_0^{+\infty} \! d\tau\!\left[ M^{(3)}_{a\langle i} M^{(3)}_{j\rangle a}\right]\!(t_r-\tau)\,,\nonumber\\
U_{ij}^{\overline{\rm 2.5PN}I_2I_2}(t_r)&=&  \frac{1}{7}\, M^{(5)}_{a\langle i} M^{}_{j\rangle a} - \frac{5}{7} \, M^{(4)}_{a\langle i} M^{(1)}_{j\rangle a} -\frac{2}{7}\, M^{(3)}_{a\langle i} M^{(2)}_{j\rangle a}  
\,,\nonumber\\
U_{ij}^{\overline{\rm 2.5PN}I_2J_1}(t_r)&=&  \frac{1}{3}\epsilon^{}_{ab\langle i} M^{(4)}_{j\rangle a}   S^{}_{b}  \,,\nonumber\\
U_{ij}^{\overline{\rm 3PN}}(t_r) &=&  2 G^2 {\mathcal M}^2\eta^6 \!\!\int_{0}^{+\infty} \!\!\!d\tau\, M_{ij}^{(5)}(t_r-\tau) \! \left[\ln^2\left(\frac{c \tau}{2b_0}\right) + \frac{11}{6} \ln\left(\frac{c \tau}{2b_0}\right)- \frac{107}{105} \ln\left(\frac{c \tau}{2r_0}\right) + \frac{124627}{44100}\right]\,.
\eea
\end{widetext}
The contributions $U_{ij}^{\overline{\rm 1.5PN}}$  and $U_{ij}^{\overline {\rm 3PN}}$ are  known as \lq\lq tail" and \lq\lq tail-of-tail," respectively.
Note  that the tail-of-tail term involves two different length scales entering the logarithms: the $b_0$ scale entering
the definition of the retarded time, \eqref{tret}, and the $r_0$ scale entering the Partie Finie  prescription of the MPM formalism.
The $r_0$ scale entering the last tail-of-tail logarithm (with coefficient $- \frac{107}{105}$) cancels with a corresponding term~\footnote{
While  $r_0$ enters the $U_{ij}[M_{ij}]$ link as an ultraviolet cutoff, it enters the PN-expanded integral expression of $I_{ij}$ as an infrared cutoff.}
entering the scale dependence of the source quadrupole moment \cite{Blanchet:1997jj,Goldberger:2009qd}. We will discuss below the infrared
meaning of the $b_0$ scale.

As already explained above, we have put a line over the 2.5PN and 3.5PN superscripts as a warning that these contributions contain an overall prefactor
$\eta^5$, or $\eta^7$ but are made themselves of building blocks having their own $\eta$ expansion. In addition, the first term
$M_{ij}^{(2)} $ also admits its own $\eta$ expansion (both because of the PN expansion of the building block $M_{ij}$,
and because of the double time derivative which involves the PN-expanded equations of motion of the particles).
Finally, 
\bea
U_{ij}^{\overline {\rm 3.5PN}}&=& G\eta^7 \left[\text{U}_{ij}^{\overline{{\rm 3.5PN}}\rm (mem)} +\text{U}_{ij}^{\overline{{\rm 3.5PN}} I_2I_4}\right.\nonumber\\
&+&\text{U}_{ij}^{\overline{{\rm 3.5PN}} I_3I_3}+\text{U}_{ij}^{\overline{{\rm 3.5PN}} J_1 J_3}\nonumber\\
&+&\left.  \text{U}_{ij}^{\overline{{\rm 3.5PN}} J_2 J_2}+\text{U}_{ij}^{\overline{{\rm 3.5PN}} J_3 I_2}+\text{U}_{ij}^{\overline{{\rm 3.5PN}} I_3 J_2}\right]\,,\qquad
\eea
where all the various contributions can be evaluated at the Newtonian accuracy level and where
\begin{widetext}
\bea
\label{lista_35pn}
U_{ij}^{\overline{{\rm 3.5PN}}\rm (mem)}&=&   \int_0^{+\infty} \! d\tau\!  \left[ - \frac{5}{756}  M_{ab}^{(4)}  M_{ijab}^{(4)} - \frac{32}{63}  S_{a \langle i}^{(3)}  S_{j\rangle a}^{(3)}\right]\!(t_r-\tau)\nonumber\\
&+&
\epsilon_{ac \langle i}   \int_0^{+\infty}\!d\tau \left[ \frac{5}{42}  S_{j\rangle cb}^{(4)}  M_{ab}^{(3)} - \frac{20}{189}  M_{j \rangle cb}^{(4)}  S_{ab}^{(3)} \right]\!(t_r-\tau)\,,
\nonumber\\
U_{ij}^{\overline{{\rm 3.5PN}} I_2I_4} &=& - \frac{1}{432}	 M_{ab}  M_{ijab}^{(7)} + \frac{1}{432} M_{ab}^{(1)}  M_{ijab}^{(6)} - \frac{5}{756}  M_{ab}^{(2)}  M_{ijab}^{(5)} + \frac{19}{648} M_{ab}^{(3)}  M_{ijab}^{(4)}+ \frac{1957}{3024}  M_{ab}^{(4)}  M_{ijab}^{(3)}  \nonumber \\
&+& \frac{1685}{1008} M_{ab}^{(5)}  M_{ijab}^{(2)} + \frac{41}{28} M_{ab}^{(6)} M_{ijab}^{(1)} + \frac{91}{216}  M_{ab}^{(7)}  M_{ijab}\,,\nonumber\\
U_{ij}^{\overline{{\rm 3.5PN}} I_3I_3} &=& - \frac{5}{252} M_{ab \langle i}  M_{j \rangle ab}^{(7)}+ \frac{5}{189}  M_{ab \langle i}^{(1)}  M_{j \rangle ab}^{(6)} +\frac{5}{126}  M_{ab \langle i}^{(2)}  M_{j \rangle ab}^{(5)} + \frac{5}{2268}  M_{ab \langle i}^{(3)}  M_{j \rangle ab}^{(4)}\,, \nonumber\\
U_{ij}^{\overline{{\rm 3.5PN}} J_1 J_3} &=&  \frac{5}{42}  S_a  S_{ija}^{(5)}\,,\nonumber\\
U_{ij}^{\overline{{\rm 3.5PN}} J_2 J_2} &=& \frac{80}{63}  S_{a \langle i}  S_{j \rangle a}^{(5)} + \frac{16}{63}  S_{a \langle i}^{(1)}  S_{j \rangle a}^{(4)} - \frac{64}{63}  
S_{a\langle i}^{(2)}  S_{j \rangle a}^{(3)}\,,\nonumber\\
U_{ij}^{\overline{{\rm 3.5PN}} J_3 I_2}&=&\epsilon_{ac \langle i} \left( \frac{1}{168}  S_{j \rangle	bc}^{(6)}  M_{ab} + \frac{1}{24}  S_{j\rangle bc}^{(5)} M_{ab}^{(1)}	+ \frac{1}{28}  S_{j \rangle bc}^{(4)}  M_{ab}^{(2)}  - \frac{1}{6} S_{j \rangle bc}^{(3)}  M_{ab}^{(3)} + \frac{3}{56}  S_{j \rangle bc}^{(2)}  M_{ab}^{(4)}\right.\nonumber\\
&+&\left. \frac{187}{168}  S_{j \rangle bc}^{(1)}  M_{ab}^{(5)} + \frac{65}{84}  S_{j \rangle bc}  M_{ab}^{(6)}\right)\,,\nonumber\\
U_{ij}^{\overline{{\rm 3.5PN}} I_3 J_2} &=&\epsilon_{ac \langle i} \left(  \frac{1}{189}M^{(6)}_{j\rangle bc}S_{ab}-\frac{1}{189} M^{(5)}_{j\rangle bc}S^{(1)}_{ab} +\frac{10}{189} M^{(4)}_{j\rangle bc}S^{(2)}_{ab}
+\frac{32}{189} M^{(3)}_{j\rangle bc}S^{(3)}_{ab}+\frac{65}{189} M^{(2)}_{j\rangle bc}S^{(4)}_{ab}  \right.\nonumber\\
&-&\left.  \frac{5}{189}  M_{j \rangle bc}^{(1)}  S_{ab}^{(5)} - \frac{10}{63}  M_{j \rangle bc}	 S_{ab}^{(6)}\right)\,.  
\eea
Using the relations
$U_L=I_L^{(l)}+O(\eta^3)$ and $V_L=J_L^{(l)}+O(\eta^3)$, 
(see Ref. \cite{Blanchet:2023sbv},  Eqs. (2.4) and beyond), in all the terms $U_{ij}^\text{1.5PN}, U_{ij}^{\overline {\rm 2.5PN}}, U_{ij}^{\overline {\rm 3PN}}, U_{ij}^{\overline  {\rm 3.5PN}}$ one can insert the approximations $M_{ij}=I_{ij}$ and  $S_{ij}=J_{ij}$. 
 
\end{widetext}

We find it convenient to compute $U_2=\frac12 \sum_X \text{U}_{ij}^\text{ X} \bar m^i \bar m^j$ which automatically takes care of the  STF projection, having denoted generically by $X$ all the various components listed above.

%
% table 1
%
\begin{table*}
\caption{\label{gauge_pot} $U_2^{G^3}$: Gauge momenta at the accuracy level needed for the present work. Notation here is $X_{12}=X_1-X_2=\frac{m_1-m_2}{M}$ while $\hat T^{ij}=T^{\langle i j \rangle}$ stands for the symmetric and tracefree part of the tensor $T^{ij}$.
In the circular case they reduce to the values listed in Eqs. (5.7a)--(5.7g) of Ref. \cite{Faye:2012we}. 
}
\begin{ruledtabular}
\begin{tabular}{ll}
$X$ &$ \frac{M\nu r^2}{20}\left[ (1-3\nu)\dot r^2 -\frac{1-3\nu}{3}v^2 -\frac{2(2-\nu)}{3}\frac{GM}{r}\right]$\\
$Y_{ij}$  &$ \frac{M\nu r^2}{7} \left[-2(1-3\nu)\hat v^{ij}    +6(1-3\nu)\dot r v^{\langle i}n^{j\rangle}+\left(\frac{1+4\nu}{2}\frac{GM}{r}-2v^2(1-3\nu) \right)\hat n^{ij}\right]$\\
$Y_i$  &$ \frac{1}{10}M\nu X_{12} r \left[ \left( v^2+ \frac{GM}{r} \right)n^i-3\dot r v^i\right]$\\
$Z_i$ &$ -\frac15 M\nu (1-3\nu)r^2 \dot r \epsilon_{abi}v^an^b$ \\
$W_{ij}$ &$ \frac{5}{21}M\nu (1-3\nu) r^3 [\dot r \hat n^{ij}-\frac25 n^{\langle i}v^{j\rangle}]$\\
$W_i$ &$  -\frac{3}{10}M\nu X_{12}r^2 [\dot r n^i-\frac13 v^i]$\\
$W$&$ \frac13 M\nu r \dot r +\frac15 \eta^2 M\nu r \dot r \left( \frac{21+17\nu}{3} \frac{GM}{r}+\frac{3}{2}(1-3\nu)v^2\right)$\\
\end{tabular}
\end{ruledtabular}
\end{table*}

\section{The 3.5PN hyperbolic motion}

Starting from the 2.5PN accuracy, the motion of a gravitational two-body system undergoes radiation-reaction effects. Therefore, in general,  the (relative) motion, $(x^i(t), v^i(t))$, can be split into two parts:  a (3PN level) conservative part, which we write by using the Quasi-Keplerian (QK) parametrization, and a radiation reacted one, which in the present work includes  contributions both at the 2.5PN and the 3.5PN level.
We recall that we use modified harmonic coordinates.

The 3PN-accurate QK parametrization (in terms of the auxiliary temporal variable $v$) of the {\it conservative} hyperboliclike motion \cite{Cho:2018upo,Bini:2022enm} is defined 
as follows. First, one defines a function $t \rightarrow v^{\rm cons}(t)$ as the solution in $v$ of the following hyperbolic Kepler-type equation 
\bea
\bar n^{\rm 3PN, cons}\,  (t-t_0)&=&e_t \sinh v-v + f_t V+g_t \sin V\nonumber\\
&&
+h_t \sin 2V+i_t \sin 3V\,,
\eea  
where 
\beq
\label{Vdef}
V(v)\equiv 2\, {\rm arctan}\left[\sqrt{\frac{e_\phi+1}{e_\phi-1}}\tanh \frac{v}{2}  \right]\,.
\eeq
Second, one defines the conservative 3PN radial and angular motions  (such that the relative motion reads $x(t)=r \cos\phi$, $y(t)=r \sin\phi$ and $z(t)=0$) 
\begin{eqnarray}
\label{QK_orb} 
&&r^{\rm 3PN, cons}(t)= \bar a_r (e_r \cosh v^{\rm cons}(t)-1)\,,\nonumber\\
&&\phi^{\rm 3PN, cons}(t)-\phi_0=K[V(v^{\rm cons}(t))\nonumber\\
&&+f_\phi \sin 2V(v^{\rm cons}(t))+g_\phi \sin 3V(v^{\rm cons}(t))\nonumber\\
&&+
h_\phi \sin 4V(v^{\rm cons}(t))+i_\phi \sin 5V(v^{\rm cons}(t))]\,.
\end{eqnarray}

The expressions for the orbital parameters in modified harmonic coordinates can be found in Appendix D of Ref. \cite{Bini:2022enm} as functions of the specific binding energy $\bar E \equiv (E-Mc^2)/(\mu c^2)$ and the dimensionless angular momentum $j=c J/(GM\mu)$ of the system, or equivalently in terms of $\bar a_r$ and $e_r$. One also passes from $j$ to the impact parameter $b$ via the relation
\beq
GE j=b p_\infty\,.
\eeq
Here $p_\infty$ is linked to the total (cm) incoming energy of the system via
\beq
E=M\sqrt{1+2\nu (\gamma-1)}\,,\qquad \gamma=\sqrt{1+p_\infty^2}\,.
\eeq
All the conserved quantities entering the above 3PN motion are taken to be functions of the {\it incoming} values of $E$ and $j$. Hereafter, we will    set to zero the two other integration constants entering Eq. \eqref{QK_orb}, i.e., $t_0=0$ and $\phi_0=0$.

Following previous works our cm spatial frame is anchored to the averaged (conservative) momenta $\bar p_a$,
\beq
\label{bar_pa}
\bar p_a=\frac12 (p_a+p_a')\,,\qquad a=1,2\,,
\eeq
rather than to the incoming momenta $p_a$.
In particular, the vector $e_y$ is aligned with 
the spatial direction of $\bar p_1$  (i.e., the bisector between the incoming and the outgoing spatial momentum
of the first particle in the cm  frame).
The two spatial frames: $e_x, e_y,e_z$ (defined via $\bar p_a$) and the corresponding one $e_X,e_Y,e_Z$ (anchored on the incoming momenta  $p_a$) differ by a $O(G^1)$ rotation  involving  half  the (relative) conservative scattering angle,  $\chi_{\rm cons}/2$, around the $z$-axis, common to both frames: $e_Z=e_z$.
Explicitly, we have 
\bea
\label{frame1}
e_x&=& \cos \frac{\chi^{\rm cons}}{2} e_X +\sin \frac{\chi^{\rm cons}}{2} e_Y\,,\nonumber\\
e_y&=& -\sin \frac{\chi^{\rm cons}}{2} e_X+\cos \frac{\chi^{\rm cons}}{2} e_Y\,,
\eea
with inverse relations
\bea
\label{frame1inv}
e_X&=& \cos \frac{\chi^{\rm cons}}{2} e_x -\sin \frac{\chi^{\rm cons}}{2} e_y\,,\nonumber\\
e_Y&=& \sin \frac{\chi^{\rm cons}}{2} e_x+\cos \frac{\chi^{\rm cons}}{2} e_y\,.
\eea

Note that in this frame the 3PN conservative motion is time-symmetric with $x(-t)=x(t)$ and $y(-t)=-y(t)$.

When taking into account radiation-reaction effects at the 3.5PN accuracy by the Lagrange method of variation of  constants the expressions for $r$ and $\phi$ as functions of time can be decomposed as follows: 
\bea
r(t)&=& r^{\rm 3PN, cons}(t)+\eta^5 \delta^{\rm rr\,2.5} r(t) +\eta^7 \delta^{\rm rr\,3.5 } r (t)\,,\nonumber\\
\phi(t)&=& \phi^{\rm 3PN, cons}(t)+\eta^5 \delta^{\rm rr\,2.5 } \phi(t) +\eta^7 \delta^{\rm rr\,3.5 } \phi (t)\,.\nonumber\\
\eea
The leading order radiation-reaction corrections, $\delta^{\rm rr \,2.5} X(t)$ ($X=r,\phi$) were computed in Ref. \cite{Bini:2022enm}  while the next-to-leading order radiation-reaction contributions $\delta^{\rm rr \,3.5} X(t)$ have been recently obtained in  Ref. \cite{Bini:2025rng}. 
[More completely, the radiation-reacted motion involves perturbations in $r$, $\phi$ and the mean anomaly variable $l(t)$, and in the four constants of the motion, $c_1=E$, $c_2=j$, $c_3=c_\phi=\phi_0$, $c_4= nt_0$.]
While the QK 3PN conservative motion was defined so as to be symmetric under  $t\to -t$, the additional radiation-reaction contributions are defined so that  they vanish at $t\to -\infty$. More precisely, we have $\delta^{\rm rr} x(t)=O(\frac{1}{t^2})$ and $\delta^{\rm rr} y(t)=O(\frac{1}{t})$ as $t\to -\infty$.
Their explicit expressions are conveniently written as functions of $v^{\rm cons}(t)$.
We have checked that the direct integration of the radiation reacted equations of motion (in modified harmonic coordinates) and  the results obtained by the method of variation of the constants coincide, modulo an adequate choice of the integration constants $t_0$ and $\phi_0$.

Inserting the results of Refs. \cite{Bini:2022enm,Bini:2025rng} in the motion as a function  of time yields a radiation-reacted dynamics which is {\it non-perturbative}  in $G$ but limited to the $O(\eta^7)$ PN accuracy. Then, inserting the latter $G$-exact $O(\eta^7)$ dynamics in the explicit time-domain expression of $U_{ij}(t_r)$ derived 
below would define a  $G$-exact $(O(\eta^7)$ time-domain waveform (containing however the tail and tail-of-tail non-local-in-time integrals).
In view of the complexity of the $G$-exact $(O(\eta^7)$ time-domain waveform, henceforth we will systematically limit our accuracy to the third order in $G$, 
that is to the 2-loop level. This truncation will also allow us to  compute with the same 2-loop accuracy the frequency-domain waveform, in which we will now be able to explicitly compute the tail and tail-of-tail contributions.

As an example  of explicit $O(G^2+G^3)$ radiation reacted dynamics, let us exhibit the motion along the $x$  and $y$ axes in terms of the rescaled time variable
\be
  T\equiv \frac{ p_\infty t}{b }.
  \ee
We find
\begin{widetext}
\bea
\label{deltarrG2x}
\delta^{\rm rr,G^2} x &=& -\frac85\left(1+\frac{T}{\sqrt{1+T^2}}\right)\frac{G^2M^2\nu p_\infty}{b} \eta^5+\left[-\frac45 \nu \frac{T}{(1+T^2)^{3/2}}\right.\nonumber\\
&-&  \frac{88}{35}-\frac{88}{35} \frac{T}{\sqrt{1+T^2}} -\frac{34}{15} \frac{T}{(1+T^2)^{3/2}} +\left.  \frac{16}{5}\frac{T}{(1+T^2)^{5/2}}  \right]\frac{\nu G^2M^2 p_\infty^3}{b}\eta^7\,,\\
\label{deltarrG3x}
\delta^{\rm rr,G^3} x &=&\left(-\frac85\frac{1}{(1+T^2)^{3/2}}{\rm arcsinh}(T)-\frac{47}{15}{\rm arctan}(T)-\frac{47}{30} \pi\right.\nonumber\\
&-& \frac{8}{5} (1+T^2)^{1/2}+\frac{8}{5(1+T^2)^{1/2}}-\frac{8}{5} T-\left.  \frac{59}{15 (1+T^2)} T\right)\frac{G^3M^3\nu}{p_\infty b^2}\eta^5\nonumber\\
&+&\left[ 
\left(
\left(\frac{464}{105} +\frac45 \nu \right)\frac{1}{(1+T^2)^{3/2}}
+\left(-\frac{12}{5}\nu -\frac{98}{5} \right)\frac{1}{(1+T^2)^{5/2}} 
+\frac{16}{(1+T^2)^{7/2}}\right) {\rm arcsinh}(T)
\right.\nonumber\\
&+&\left(-\frac{172}{35} -\frac{12}{5}\nu \right) (1+T^2)^{1/2}-\frac{172}{35} T-\frac{4997}{1680}\pi-\frac{4997}{840} {\rm arctan}(T)\nonumber\\
&-&\frac{12}{5}\nu T+\frac{40}{ 7 (1+T^2)^{1/2}}+\left(-\frac{2357}{840} -2\nu \right)\frac{T}{(1+T^2)} +\left(\frac{12}{5}\nu -\frac{4}{5} \right)\frac{1}{(1+T^2)^{3/2}}\nonumber\\
&+& \left(-\frac{47}{5}\nu -\frac{97}{12} \right)\frac{T}{(1+T^2)^2}+\left.\frac{16T}{(1+T^2)^3}
 \right] \frac{\nu G^3M^3 p_\infty}{b^2}\eta^7\,,
\eea
and
\bea
\label{deltarrG2y}
\delta^{\rm rr,G^2} y &=&  \frac{8}{5}\frac{G^2 M^2 \nu p_\infty }{b}\frac{1}{\sqrt{1+T^2}}\eta^5 \nonumber\\
&+& \frac{G^2 M^2 \nu p_\infty^3 }{b}\left[\frac{4\nu}{5 (1+T^2)^{3/2}}-\left(\frac{108}{35\sqrt{1+T^2}}+\frac{2}{5 (1+T^2)^{3/2}}+\frac{16}{5 (1+T^2)^{5/2}} \right)  \right]\eta^7\,,\\
\label{deltarrG3y}
\delta^{\rm rr,G^3} y &=& \frac{G^2 M^3 \nu}{b^2 p_\infty}\left[-\frac{37}{15}T \left({\rm arctan }T+\frac{\pi}{2} \right) -\frac{13}{15}+\frac{8}{5\sqrt{1+T^2}}+\frac{59}{15 (1+T^2)}-\frac{8}{5}\frac{T{\rm arcsinh}T}{(1+T^2)^{3/2}} \right]\eta^5 \nonumber\\
&+&  \frac{G^2 M^3 \nu p_\infty}{b^2 }\left[\nu \left(-\frac{37}{5}T \left({\rm arctan }T+\frac{\pi}{2} \right) +\frac{12}{5}\frac{T}{\sqrt{1+T^2}}+\frac{1}{5(1+T^2)}
-\frac{4T}{5(1+T^2)^{3/2}}({\rm arcsinh}T-3)\right.\right.\nonumber\\
&+&\left. \frac{47}{5(1+T^2)^2}-\frac{12}{5}\frac{T{\rm arcsinh T}}{(1+T^2)^{5/2}}\right)\nonumber\\
&+&  \frac{143}{168}\left({\rm arctan }T+\frac{\pi}{2} \right)
+\frac{139}{840}-\frac{24T}{35 \sqrt{1+T^2}}-\frac{3811}{280 (1+T^2)}
-\frac{4 T}{5(1+T^2)^{3/2}}\left(1-\frac{48}{7} {\rm arcsinh T}\right)\nonumber\\
&+&\left. \frac{197}{60 (1+T^2)^2}+\frac{6T {\rm arcsinh T}}{5(1+T^2)^{5/2}}
-\frac{16}{(1+T^2)^3}+\frac{16 T {\rm arcsinh} T}{(1+T^2)^{7/2}}
 \right]\,.
\eea

\end{widetext}
The asymptotic behaviours of $\delta^{\rm rr} x$ and $\delta^{\rm rr} y$ when $T\to \mp \infty$ are such that  
\bea
\delta^{{\rm rr} G^2} x &\underset{_{-\infty}}{\approx} & (\eta^5+\eta^7)\left(\frac{1}{T^2}+\frac{1}{T^4}\right)\,,\nonumber\\
\delta^{{\rm rr} G^3} x & \underset{_{-\infty}}{\approx}& (\eta^5+\eta^7)\left[\frac{1}{T}+\frac{1+\ln (-T)}{T^3}\right]\,,\nonumber\\
%%%%
\delta^{{\rm rr} G^2} x &\underset{_{+\infty}}{\approx} & (\eta^5+\eta^7)\left(1+\frac{1}{T^2}\right)\,,\nonumber\\
\delta^{{\rm rr} G^3} x & \underset{_{+\infty}}{\approx}& \eta^5\left( T+1+\frac{1+\ln T}{T^3}\right)\nonumber\\
&+& \eta^7 \left(  T+1+\frac{1}{T}+\frac{1+\ln T}{T^3} \right)\,,\nonumber\\
\eea
\bea
\delta^{{\rm rr} G^2} y &\underset{_{-\infty}}{\approx} &  (\eta^5+\eta^7)\left(\frac{1}{T}+\frac{1}{T^3}+\frac{1}{T^5}\right)\,,\nonumber\\ 
\delta^{{\rm rr} G^3} y & \underset{_{-\infty}}{\approx} &  (\eta^5+\eta^7)\frac{1+\ln(-T)}{T^2} \,,\nonumber\\
%%%%%
\delta^{{\rm rr} G^2} y &\underset{_{+\infty}}{\approx} & (\eta^5+\eta^7)\left(\frac{1}{T}+\frac{1}{T^3}+\frac{1}{T^5}\right)\,,\nonumber\\
\delta^{{\rm rr} G^3} y &\underset{_{+\infty}}{\approx} &  (\eta^5+\eta^7)\left(T+1+\frac{1+\ln T}{T^2}\right) \,.
\eea

\section{Computation of the frequency domain waveform at the 2-loop 3.5PN accuracy, i.e., including  $O(G^3\eta^7)$ terms}

The time-domain waveform is conveniently expressed in terms of the dimensionless retarded time variable
\beq
T_r \equiv\frac{p_\infty t_r}{b}\,,
\eeq
where we recall that $p_\infty=\sqrt{\gamma^2-1}$.
It is then  natural  to express the Fourier-domain waveform \eqref{hat_W_FT}
 in terms of the corresponding rescaled frequency variable $u$ (which should not be confused with either often used notations for both the eccentric anomaly and the retarded time)
\beq
u\equiv \frac{\omega b}{p_\infty}\,.
\eeq 
Indeed, we have
\beq
\hat W(\omega)|_{\omega= u p_\infty/b}=\frac{p_\infty}{b}\int_{- \infty}^{+\infty} dT_r e^{iuT_r} W(T_r) \,.
\eeq
In the following $\hat W(\omega)|_{\omega= u p_\infty/b}$ will be often  denoted as $\hat W(u)$, or even simply as $W(u)$.

Evidently, the multipolar decomposition \eqref{W_deco} commutes with the Fourier transform. Therefore, we can write 
\beq
\hat U_2(u)\equiv \hat U_2(\omega)|_{\omega= u p_\infty/b}=\frac{p_\infty}{b}\int_{- \infty}^{+\infty} dT_r e^{iuT_r} U_2(T_r) \,.
\eeq

The structure of the frequency domain quadrupolar waveform is
\bea
\hat  U_2(u, {\bf n}, \bar {\bf m})&=&G U_2^{G^1}(u, {\bf n}, \bar {\bf m})+G^2 U_2^{G^2}(u, {\bf n}, \bar {\bf m})\nonumber\\
&+& G^3 U_2^{G^3}(u, {\bf n}, \bar {\bf m})+O(G^4)\,,
\eea
where  ${\bf n}$  denotes the direction of emission of the gravitational   waves and where, for brevity, we henceforth suppress on the rhs the hat on the various MPM contributions to $\hat U_2(u)$.
 We recall that we work in the (incoming) cm frame of the system with spatial frame $e_x,e_y,e_z$, such that $x=r\cos\phi$ and $y=r\sin\phi$, $z=0$. We use canonical polar coordinates associated with $x,y,z$ coordinates also denoted $x^1,x^2,x^3$, i.e., 
\beq
{\bf n}=(n_1,n_2,n_3) =(\sin \theta \cos\phi, \sin \theta \sin\phi,\cos \theta)\,,
\eeq
using the same choice of null polarization vector as in our previous works
\beq
{\bar {\bf m}}=\frac{1}{\sqrt{2}}(e_\theta -ie_\phi)\,.
\eeq
The explicit components of  ${\bar {\bf m}}$  read
\bea
\label{choicebarms}
\bar m_1&=& \frac{1}{\sqrt{2}} (\cos \theta \cos\phi +i \sin \phi)), \nonumber\\
\bar m_2&=&  \frac{1}{\sqrt{2}}  (\cos \theta \sin \phi  -i\cos\phi), \nonumber\\
\bar m_3&=&  - \frac{1}{\sqrt{2}} \sin \theta\,.
\eea
They satisfy the following relations
\bea
(\bar m^1)^2+(\bar m^2)^2&=&-\frac12 \sin^2 \theta \,,  \nonumber\\
(\bar m^1)^2-(\bar m^2)^2&=& \frac14\cos(2\phi)   (3 + \cos(2\theta)) \nonumber\\
&+& i\cos\theta \sin(2\phi)\,,\nonumber\\
 \bar m^1 \bar m^2 &=& \frac18 (-4 i \cos(2\phi)\cos\theta \nonumber\\
&+& (3 + \cos(2\theta)) \sin(2\phi))\,.
\eea
Each $U_2^{G^n}(u, {\bf n}, \bar {\bf m})$  is bilinear in $\bar m^i \bar m^j$.
The radiative quadrupole moment of the binary system has only three
independent physical components, namely $U_{xx}$,  $U_{xy}$, and $U_{yy}$ (with $U_{zz}= -U_{xx} -U_{yy}$ and vanishing other components). The latter three independent components are {\it real }
in the time domain. The corresponding time-domain (complex) quadrupolar waveform $U_2(t_r)=\frac12 {U}_{ij}(t_r) \bar m^i \bar m^j$
is then equal to the following sum of three terms
\bea
U_2(t_r)&=& U_2^{\bar m_1^2 }(t_r)\bar m_1^2 +U_2^{\bar m_1 \bar m_2 }(t_r)\bar m_1 \bar m_2\nonumber\\
& +& U_2^{\bar m_2^2 }(t_r)\bar m_2^2 \,,
\eea
where
\bea
U_2^{\bar m_1^2 }(t_r)&=& \frac12 (2 U_{xx}(t_r) + U_{yy}(t_r), \nonumber\\
U_2^{\bar m_2^2 }(t_r)&=& \frac12 (2 U_{yy}(t_r) + U_{xx}(t_r), \nonumber\\
U_2^{\bar m_1 \bar m_2 }(t_r)&=& U_{xy}(t_r)\,.
\eea
Note that the three coefficients $U_2^{\bar m_1^2 }(t_r)$,
 $U_2^{\bar m_2^2 }(t_r)$, $U_2^{\bar m_1 \bar m_2 }(t_r)$ entering the
complex quadrupolar waveform $U_2(t_r)$ are {\it real} in the time-domain.
The corresponding three frequency-domain coefficients
$\hat U_2^{\bar m_1^2 }(u),
\hat U_2^{\bar m_2^2 }(u), \hat U_2^{\bar m_1 \bar m_2 }(u)$ (where
$u= \omega b/\pinf$) will be complex but will satisfy the \lq\lq reality conditions"
\bea
\label{real}
(\hat U_2^{\bar m_1^2 }(u))^*= \hat U_2^{\bar m_1^2 }(-u), \nonumber\\
(\hat U_2^{\bar m_2^2 }(u))^*= \hat U_2^{\bar m_2^2 }(-u), \nonumber\\
(\hat U_2^{\bar m_1 \bar m_2 }(u))^*= \hat U_2^{\bar m_1 \bar m_2 }(-u)\,.
\eea
Because of these reality conditions, it is enough to consider only the positive
frequency axis: $\omega >0$ and therefore $u >0$.

The  $(l,m)=(2,m)$ decomposition of $\hat  U_2(u, {\bf n}, \bar {\bf m})$
along the spin-weighted spherical harmonics ${}_{s}Y_{lm}(\theta,\phi)$ of spin $s=-2 \equiv \bar 2$ [The $m=\pm1$ coefficients vanish
because of the equatorial nature of the orbital motion.]
includes only three (complex) coefficients, namely:
$U_{22}$, $U_{20}$ and $U_{2\bar 2}$. The latter three coefficients 
are in one-to-one correspondence with $\hat U_2^{\bar m_1^2 },
\hat U_2^{\bar m_2^2 }, \hat U_2^{\bar m_1 \bar m_2 }$.
The relation (valid both in the time domain and in the frequency domain) between these two triplets of coefficients is 
\bea
\label{U22022m2}
U_{22}&=&\sqrt{\frac{\pi}{5}}[U_2^{\bar m_1^2 }-U_2^{\bar m_2^2 }-iU_2^{\bar m_1 \bar m_2 }]\,, \nonumber\\
U_{20}&=& -\sqrt{\frac{2\pi}{15}}[U_2^{\bar m_1^2 }+U_2^{\bar m_2^2 }]\,,\nonumber\\
U_{2\bar 2}&=& \sqrt{\frac{\pi}{5}}[U_2^{\bar m_1^2 }-U_2^{\bar m_2^2 }+iU_2^{\bar m_1 \bar m_2 }]\,, 
\eea
with inverse   
\bea
\label{U22022m2inverse}
U_2^{\bar m_1^2} &=&   \frac14 \sqrt{\frac{5}{\pi}} (U_{22} + U_{2\bar 2} - \sqrt{6}  U_{20})\,,\nonumber\\
U_2^{\bar m_1 \bar m_2} &=&  \frac12 i \sqrt{\frac{5}{\pi}} (U_{22} - U_{2\bar 2})\,, \nonumber\\
U_2^{\bar m_2^2} &=&  - \frac14 \sqrt{\frac{5}{\pi}} (U_{22} + U_{2\bar 2} + \sqrt{6} U_{20})\,.
\eea
Note that while 
$U_2^{\bar m_1^2 },
U_2^{\bar m_2^2 }, U_2^{\bar m_1 \bar m_2 }$ are real in the time-domain,
$U_{22}$, $U_{20}$ and $U_{2\bar 2}$ are complex even in the time-domain. The reality conditions \eqref{real} then say for instance that
$[U_{22}(u)]^*=U_{2\bar 2}(-u)$.

All the coefficients entering the relations above only depend, at each
order in $G$, on $u$,  $p_\infty$, the total mass $M=m_1+m_2$, the symmetric mass ratio $\nu\equiv \frac{m_1m_2}{(m_1+m_2)^2}$ and the incoming impact parameter $b$. 

The three (complex) coefficients $U_2^{\bar m_1^2}(u)$, $U_2^{\bar m_1 \bar m_2}(u)$ and $U_2^{\bar m_2^2}(u)$ are in 1-1 correspondence with the three expansion coefficients, $U_{22}$, $U_{20}$ and $U_{2\bar 2}$, of
$U_2$ along the  spin-weighted spherical harmonics ${}_{s}Y_{lm}(\theta,\phi)$ of spin $s=-2 \equiv \bar 2$. 
Explicitly we have
\bea
U_2&=&\sum_{m=-2}^2 U_{2m} \; {}_{\bar 2}Y_{2m}(\theta,\phi)\nonumber\\
&=&  U_{22}\; {}_{\bar 2} Y_{22}(\theta,\phi)+ U_{20}\;  {}_{\bar 2} Y_{20}(\theta,\phi)+ U_{2\bar 2}\;  {}_{\bar 2} Y_{2 \bar 2}(\theta,\phi)\,,\nonumber\\
\eea
where the relations between $U_{2m}$  and $U_2^{\bar m_a \bar m_b}$ are displayed in Eqs. \eqref{U22022m2} and \eqref{U22022m2inverse} above.

In the following we shall work with the PM expansion of the frequency-domain quadrupolar waveform components $U_2^{\bar m_i \bar m_j}(u)$, namely
\beq
U_2^{\bar m_i \bar m_j}(u)=U_2^{\bar m_i \bar m_j G}(u)+U_2^{\bar m_i \bar m_j G^2}(u)+\ldots
\eeq  
The structure of the various $U_2^{\bar m_i \bar m_j G^n}(u)$ is as follows
\bea
G U_2^{\bar m_i \bar m_j G}(u)&\sim &\frac{GM^2 \nu}{p_\infty}[1+\eta^2 p_\infty^2 + \eta^4 p_\infty^4\nonumber\\
&+&  \eta^6 p_\infty^6+O(\eta^8)]\,,\nonumber\\
G^2 U_2^{\bar m_i \bar m_j G^2}(u)&\sim &\frac{GM^2 \nu}{p_\infty} \left(\frac{GM}{b p_\infty^2} \right)[1+\eta^2 p_\infty^2 + \eta^3 p_\infty^3\nonumber\\
&+&\eta^4 p_\infty^4 +\eta^5 p_\infty^5\nonumber\\
&+& \eta^6 p_\infty^6+  \eta^7 p_\infty^7+O(\eta^8)]\,,\nonumber\\
G^3 U_2^{\bar m_i \bar m_j G^3}(u)&\sim &\frac{GM^2 \nu}{p_\infty}\left(\frac{GM}{b p_\infty^2} \right)^2 [1+\eta^2 p_\infty^2 + \eta^3 p_\infty^3 \nonumber\\
&+& \eta^4 p_\infty^4 +\eta^5 p_\infty^5\nonumber\\
&+& \eta^6 p_\infty^6+  \eta^7 p_\infty^7+O(\eta^8)]\,.\nonumber\\
\eea
Here the powers of $\eta$ count the fractional powers of $p_\infty$ in the various terms. The coefficients entering the expansions   $[\sum_{n=0} c_n \eta^n p_\infty^n]$  in the above expressions now depend only on $u$ and  $\nu$. They  are polynomials in $\nu$.
For example, the lowest order contributions $O(G^1 \eta^0)$ involve the Bessel K functions $K_0(u)$ and $K_1(u)$ and read
\bea
U_2^{\bar m_1^2 G^1 \eta^0}&=& -\frac{2 \nu  u }{p_\infty}K_1(u)\,,\nonumber\\
U_2^{\bar m_1\bar m_2 G^1 \eta^0}&=&-\frac{4 i \nu  u }{p_\infty}K_0(u)-\frac{4 i \nu }{p_\infty} K_1(u)\,,\nonumber\\
U_2^{\bar m_2^2 G^1 \eta^0}&=&\frac{2 \nu }{p_\infty} K_0(u)+\frac{2 \nu  u}{p_\infty}  K_1(u)\,.\\
\eea
Starting at order $O(G^2)$ there appears  $K_{1/2}(u)=\sqrt{\frac{\pi}{2}}\frac{e^{-u}}{\sqrt{u}}$. In addition, the following logarithm starts appearing at order $O(G^2\eta^3)$
\beq
{\mathcal L}\equiv \ln \left( \frac{2b_0 e^{\gamma_E}u p_\infty}{b}\right)=\ln \left( 2b_0 e^{\gamma_E}\omega\right)\,.
\eeq
See below for further discussion of the $b_0$ dependence of the Fourier domain waveform. Starting at order $G^3$ new special functions, besides $K_n(u)$ and $K_{n+\frac12}(u)$, appear, as was indicated in Ref. \cite{Bini:2024ijq}.

More precisely, there appear three types of \lq\lq iterated Bessel functions" which can be expressed in terms of the three master integrals
$Q^{\rm as}_1(u)$, $Q^{\rm as2}_{1/2}(u)$, $Q^{\rm at}_{1/2}(u)$ and their first derivatives with respect to $u$. Here, we defined
\bea
Q^{\rm as}_1(u)&\equiv&\int_{-\infty}^{+\infty}dT e^{iuT}\frac{{\rm arcsinh}(T)}{(1+T^2)}\,, \nonumber\\
Q^{\rm as2}_{1/2}(u)&\equiv& \int_{-\infty}^{+\infty}dT e^{iuT}\frac{{\rm arcsinh}^2(T)}{(1+T^2)^{1/2}} \,, \nonumber\\
Q^{\rm at}_{1/2}(u)&\equiv&\int_{-\infty}^{+\infty}dT e^{iuT}\frac{{\rm arctan}(T)}{(1+T^2)^{1/2}} \,.
\eea 
For all of these integrals explicit analytic expressions exist in terms of the (second) derivatives of the Bessel functions $K_\nu(u)$ with respect to the order $\nu$, or,  equivalently in terms of MeijerG functions. The Supplemental Material/ancillary file \cite{ancillary} associated with this work contains their explicit evaluation, as well as the complete results for $U_2$ up to $O(G^3\eta^7)$.

For illustration, we list  the $O(G^3,\eta^0)$ and $O(G^3,\eta^7)$ parts of $U_2$ (the complete expressions at any PN order up to the 3.5PN can be found in the associated ancillary file) in Tables \ref{U2Gcubeeta0} and \ref{U2Gcubeeta7}.
%
% table 2
%
\begin{table*}
\caption{\label{U2Gcubeeta0}
$U_2^{G^3}$: List of all components (third order in $G$, Newtonian order),  projected on the basis $\bar m^i$. 
}
\begin{ruledtabular}
\begin{tabular}{ll}
$U_2^{\bar m_1^2 G^3 \eta^0}$ &$\frac{\nu}{b^2 p_\infty^5}\left[(2 + u^2) K_0(u) + 2 u K_1(u) - \frac12  u^3 
Q^{\rm as2}_{1/2}{}'(u)\right]$\\
$U_2^{\bar m_1 \bar m_2  G^3 \eta^0}$ &$\frac{\nu}{b^2 p_\infty^5}\left[
6 i u K_0(u)+ (4 i + 2 i u^2)K_1(u) + i u^3 Q^{\rm as2}_{1/2}(u)   - i u^2 Q^{\rm as2}_{1/2}{}'(u)\right]$\\
$U_2^{\bar m_2^2 G^3 \eta^0}$ &$\frac{\nu}{b^2 p_\infty^5}\left[
(-2 - u^2) K_0(u) - 3 u K_1(u)  - \frac12 u^2 Q^{\rm as2}_{1/2}(u) + \frac12 u^3 Q^{\rm as2}_{1/2}{}'(u)\right]$\\
\end{tabular}
\end{ruledtabular}
\end{table*}
%
%
% table 3
%
%
\begin{table*}
\caption{\label{U2Gcubeeta7} $U_2^{G^3}$: List of all components (third order in $G$,   PN order  terms at $O(\eta^7)$) projected on the basis $\bar m^i$.   
Here ${\mathcal L}=\ln\left(\frac{2b_0e^{\gamma_E} p_\infty u}{b} \right)$.
}
\begin{ruledtabular}
\begin{tabular}{ll}
$U_2^{\bar m_1^2 G^3 \eta^7}$ &$\frac{\nu p_\infty^2}{b^2}[A_{K_0}^{\bar m_1^2}K_0(u)+A_{K_1}^{\bar m_1^2}K_1(u)+A_{e^{-u}}^{\bar m_1^2}e^{-u}]$\\
$A_{K_0}^{\bar m_1^2}$ & $\left(\frac{32 \nu ^2}{7}-\frac{1472 \nu }{105}\right) u^2+\left(\left(-\frac{\pi ^2}{7}-\frac{3853 i \pi }{630}\right) \nu ^2+\left(-\frac{59
   \pi ^2}{42}+\frac{1117 i \pi }{180}\right) \nu +\frac{47 \pi ^2}{84}-\frac{517 i \pi }{504}\right) u^4+{\mathcal L} \left(-\frac{2}{7} i \pi  \nu ^2-\frac{59 i \pi  \nu }{21}+\frac{47 i \pi }{42}\right) u^4$\\
$A_{K_1}^{\bar m_1^2}$ & $\left(\left(-\frac{41 \pi ^2}{189}+\frac{1592 i \pi }{567}\right) \nu ^2+\left(\frac{247 \pi ^2}{378}-\frac{16523 i \pi }{5670}\right) \nu
   -\frac{71 \pi ^2}{378}+\frac{781 i \pi }{2268}\right) u^5$\\
&$\left(\left(-\frac{96}{35}+\frac{386 i \pi }{945}-\frac{8 \pi ^2}{63}\right) \nu
   ^2+\left(\frac{152}{105}-\frac{541 i \pi }{2520}-\frac{31 \pi ^2}{252}\right) \nu +\frac{2071 \pi ^2}{504}-\frac{22781 i \pi }{3024}\right)
   u^3$\\
&$+{\mathcal L} \left(\left(-\frac{82}{189} i \pi  \nu ^2+\frac{247 i \pi  \nu }{189}-\frac{71 i \pi }{189}\right) u^5+\left(-\frac{16}{63} i \pi  \nu
   ^2-\frac{31 i \pi  \nu }{126}+\frac{2071 i \pi }{252}\right) u^3\right)$\\
$A_{e^{-u}}^{\bar m_1^2}$ & $-\frac{409 i \pi  \nu ^2}{1344}-\frac{3397 i \pi  \nu }{2688}+\left(-\frac{16073 i \pi  \nu ^2}{12096}+\left(-\frac{39 \pi ^2}{14}+\frac{114349
   i \pi }{17280}\right) \nu +\frac{61 \pi ^2}{28}-\frac{671 i \pi }{168}\right) u^3$\\
&$+\left(\frac{14221 i \pi  \nu ^2}{10080}+\left(-\frac{33
   \pi ^2}{14}-\frac{84703 i \pi }{60480}\right) \nu -\frac{97 \pi ^2}{28}+\frac{1067 i \pi }{168}\right) u^2+\left(-\frac{477}{320} i \pi  \nu
   ^2+\left(-\frac{33 \pi ^2}{14}+\frac{13939 i \pi }{2688}\right) \nu -\frac{97 \pi ^2}{28}+\frac{1067 i \pi }{168}\right) u$\\
&$+\frac{-\frac{409
   i \pi  \nu ^2}{1344}-\frac{3397 i \pi  \nu }{2688}}{u}+{\mathcal L} \left(\left(\frac{61 i \pi }{14}-\frac{39 i \pi  \nu }{7}\right) u^3+\left(-\frac{33}{7} i \pi  \nu -\frac{97 i \pi }{14}\right)
   u^2+\left(-\frac{33}{7} i \pi  \nu -\frac{97 i \pi }{14}\right) u\right)$\\
\hline 
$U_2^{\bar m_1\bar m_2 G^3 \eta^7}$ &$\frac{\nu p_\infty^2}{b^2}[A_{K_0}^{\bar m_1\bar m_2}K_0(u)+A_{K_1}^{\bar m_1\bar m_2}K_1(u)+A_{e^{-u}}^{\bar m_1\bar m_2}e^{-u}]$\\
$A_{K_0}^{\bar m_1\bar m_2}$ & $\left(\left(-\frac{3184 \pi }{567}-\frac{82 i \pi ^2}{189}\right) \nu ^2+\left(\frac{16523 \pi }{2835}+\frac{247 i \pi ^2}{189}\right) \nu
   -\frac{71 i \pi ^2}{189}-\frac{781 \pi }{1134}\right) u^5$\\
&$+\left(\left(-\frac{192 i}{35}-\frac{388 \pi }{189}-\frac{32 i \pi ^2}{63}\right)
   \nu ^2+\left(\frac{304 i}{105}+\frac{8209 \pi }{1260}+\frac{577 i \pi ^2}{126}\right) \nu +\frac{236 i \pi ^2}{63}+\frac{1298 \pi
   }{189}\right) u^3+\left(-\frac{32 i \nu ^2}{35}-\frac{1088 i \nu }{105}\right) u$\\
&$+{\mathcal L} \left(\left(\frac{164 \pi  \nu ^2}{189}-\frac{494 \pi  \nu }{189}+\frac{142 \pi }{189}\right) u^5+\left(\frac{64 \pi  \nu ^2}{63}-\frac{577
   \pi  \nu }{63}-\frac{472 \pi }{63}\right) u^3\right)$\\
$A_{K_1}^{\bar m_1\bar m_2}$ & $-\frac{128 i \nu ^2}{35}-\frac{144 i \nu }{7}+\left(\left(\frac{107 \pi }{2835}-\frac{14 i \pi ^2}{27}\right) \nu ^2+\left(\frac{9881 \pi
   }{810}+\frac{671 i \pi ^2}{189}\right) \nu -\frac{40 i \pi ^2}{27}-\frac{220 \pi }{81}\right) u^4$\\
&$+\left(\left(\frac{96 i}{35}-\frac{712 \pi
   }{315}-\frac{16 i \pi ^2}{21}\right) \nu ^2+\left(-\frac{1376 i}{105}+\frac{4933 \pi }{3780}+\frac{17 i \pi ^2}{42}\right) \nu +\frac{172 i
   \pi ^2}{21}+\frac{946 \pi }{63}\right) u^2$\\
&$+{\mathcal L} \left(\left(\frac{28 \pi  \nu ^2}{27}-\frac{1342 \pi  \nu }{189}+\frac{80 \pi }{27}\right) u^4+\left(\frac{32 \pi  \nu ^2}{21}-\frac{17 \pi 
   \nu }{21}-\frac{344 \pi }{21}\right) u^2\right)$\\
$A_{e^{-u}}^{\bar m_1\bar m_2}$ & $-\frac{40423 \pi  \nu ^2}{10080}+\left(-\frac{216371 \pi }{20160}-\frac{45 i \pi ^2}{7}\right) \nu +\left(\frac{16073 \pi  \nu
   ^2}{6048}+\left(-\frac{114349 \pi }{8640}-\frac{39 i \pi ^2}{7}\right) \nu +\frac{61 i \pi ^2}{14}+\frac{671 \pi }{84}\right)
   u^3$\\
&$+\left(-\frac{36671 \pi  \nu ^2}{5040}+\left(-\frac{1003 \pi }{6048}-\frac{54 i \pi ^2}{7}\right) \nu +\frac{4 i \pi ^2}{7}+\frac{22 \pi
   }{21}\right) u^2+\left(-\frac{44263 \pi  \nu ^2}{10080}+\left(-\frac{649753 \pi }{60480}-\frac{45 i \pi ^2}{7}\right) \nu -\frac{75 i \pi
   ^2}{14}-\frac{275 \pi }{28}\right) u$\\
&$-\frac{75 i \pi ^2}{14}-\frac{275 \pi }{28}+{\mathcal L} \left(\frac{90 \pi  \nu }{7}+\left(\frac{78 \pi  \nu }{7}-\frac{61 \pi }{7}\right) u^3+\left(\frac{108 \pi  \nu }{7}-\frac{8 \pi }{7}\right)
   u^2+\left(\frac{90 \pi  \nu }{7}+\frac{75 \pi }{7}\right) u+\frac{75 \pi }{7}\right)$\\
\hline 
$U_2^{\bar m_2^2 G^3 \eta^7}$ &$\frac{\nu p_\infty^2}{b^2}[A_{K_0}^{\bar m_2^2}K_0(u)+A_{K_1}^{\bar m_2^2}K_1(u)+A_{e^{-u}}^{\bar m_2^2}e^{-u}]$\\
$A_{K_0}^{\bar m_2^2}$ & $\left(\left(\frac{10 \pi ^2}{63}-\frac{883 i \pi }{270}\right) \nu ^2+\left(-\frac{271 \pi ^2}{63}+\frac{58553 i \pi }{3780}\right) \nu
   +\frac{467 \pi ^2}{252}-\frac{5137 i \pi }{1512}\right) u^4$\\
&$+\left(\left(-\frac{32}{35}-\frac{344 i \pi }{315}+\frac{8 \pi ^2}{21}\right) \nu
   ^2+\left(\frac{8}{15}+\frac{24473 i \pi }{2520}-\frac{479 \pi ^2}{84}\right) \nu +\frac{215 \pi ^2}{168}-\frac{2365 i \pi }{1008}\right) u^2$\\
&$+{\mathcal L} \left(\left(\frac{20}{63} i \pi  \nu ^2-\frac{542 i \pi  \nu }{63}+\frac{467 i \pi }{126}\right) u^4+\left(\frac{16}{21} i \pi  \nu
   ^2-\frac{479 i \pi  \nu }{42}+\frac{215 i \pi }{84}\right) u^2\right)$\\
$A_{K_1}^{\bar m_2^2}$ & $\left(\left(\frac{41 \pi ^2}{189}-\frac{1592 i \pi }{567}\right) \nu ^2+\left(-\frac{247 \pi ^2}{378}+\frac{16523 i \pi }{5670}\right) \nu
   +\frac{71 \pi ^2}{378}-\frac{781 i \pi }{2268}\right) u^5$\\
&$+\left(\left(\frac{96}{35}-\frac{1019 i \pi }{315}+\frac{52 \pi ^2}{63}\right) \nu
   ^2+\left(-\frac{152}{105}+\frac{104897 i \pi }{7560}-\frac{1735 \pi ^2}{252}\right) \nu +\frac{253 \pi ^2}{72}-\frac{2783 i \pi
   }{432}\right) u^3+\left(\frac{64 \nu ^2}{35}-\frac{32 \nu }{35}\right) u$\\
&$+{\mathcal L} \left(\left(\frac{82}{189} i \pi  \nu ^2-\frac{247 i \pi  \nu }{189}+\frac{71 i \pi }{189}\right) u^5+\left(\frac{104}{63} i \pi  \nu
   ^2-\frac{1735 i \pi  \nu }{126}+\frac{253 i \pi }{36}\right) u^3\right)$\\
$A_{e^{-u}}^{\bar m_2^2}$ & $-\frac{24491 i \pi  \nu ^2}{6720}-\frac{21311 i \pi  \nu }{13440}+\left(\frac{16073 i \pi  \nu ^2}{12096}+\left(\frac{39 \pi
   ^2}{14}-\frac{114349 i \pi }{17280}\right) \nu -\frac{61 \pi ^2}{28}+\frac{671 i \pi }{168}\right) u^3$\\
&$+\left(-\frac{6569 i \pi  \nu
   ^2}{1120}+\left(\frac{75 \pi ^2}{14}-\frac{94733 i \pi }{60480}\right) \nu -\frac{113 \pi ^2}{28}+\frac{1243 i \pi }{168}\right)
   u^2+\left(-\frac{16293 i \pi  \nu ^2}{2240}+\left(\frac{75 \pi ^2}{14}-\frac{11253 i \pi }{4480}\right) \nu -\frac{113 \pi
   ^2}{28}+\frac{1243 i \pi }{168}\right) u$\\
&$+\frac{-\frac{24491 i \pi  \nu ^2}{6720}-\frac{21311 i \pi  \nu }{13440}}{u}+{\mathcal L} \left(\left(\frac{39 i \pi  \nu }{7}-\frac{61 i \pi }{14}\right) u^3+\left(\frac{75 i \pi  \nu }{7}-\frac{113 i \pi }{14}\right)
   u^2+\left(\frac{75 i \pi  \nu }{7}-\frac{113 i \pi }{14}\right) u\right)$\\
\end{tabular}
\end{ruledtabular}
\end{table*}

Our results go beyond comparable results in the literature in that they push the PN accuracy to $O(\eta^7)$ for all considered $G$-orders. We have checked that they agree with previously published results. We recall that Refs. \cite{Bini:2023fiz,Bini:2024rsy}
gave the results at accuracy $G^2 \eta^5$ (showing only equatorial plane expressions for simplicity).

In view of the complicated intermediate building blocks that went into the computation of our  quadrupolar frequency-domain waveform,
we felt important to perform several partial (but non trivial) checks of our final results. We indicate in the next four Sections the four
different checks we did using various properties, or limits, of the exact waveform. The first check concerns the $b_0$ dependence, the second
the extreme mass-ratio limit $\nu \ll 1$, the third one concerns the soft limit, $\omega \to 0$ (i.e. $u \to 0$), while the fourth and most stringent test is a comparison with the existing high-PN accuracy expansion of the 1-loop ($h_c=O(G^3)$ i.e.,  $O(G^2)$) $W$ waveform \cite{Heissenberg:2025fcr}. Needless to say:
all our checks have been successful.

\section{Renormalization group associated with the tail terms}

We already mentioned the appearance of  several scale-dependent logarithms at order $O(G^2 \eta^3)$. The origin of these logarithms lies in  
the tail and tail-of-tail integrals, where they enter in the following form (see e.g., Ref. \cite{Bini:2021qvf})
\beq
A_m(\omega, C_X)= \int_0^{+\infty}  d\tau\,  e^{ i\omega \tau }  \ln^m\left(\frac{\tau}{C_{X}} \right)\,,
\eeq
that is, for $m=1,2$, 
\bea
A_1(\omega, C_X)
&=& -\frac{\pi}{2|\omega|}-\frac{i}{\omega}\ln (C_X |\omega|e^\gamma)\,,\nonumber\\
A_2(\omega, C_X)&=& \frac{\pi}{|\omega|}\ln (C_X |\omega|e^\gamma)\nonumber\\
&+& \frac{i}{\omega}\left[ \ln^2 (C_X |\omega|e^\gamma)-\frac{\pi^2}{12}\right]\,,
\eea
with the property
\beq
A_2(\omega, C_X)=(-i\omega)A_1^2(\omega, C_X)+\frac{1}{(-i\omega)}\frac{\pi^2}{6}\,.
\eeq
In the case of the source quadrupole $I_2$
\beq
\label{C_I2_def}
C_{I_2}=\frac{2b_0 e^{-11/12}}{c}\,.  
\eeq

The logarithmic kernel entering the tail-of-tail can be rewritten as
\bea
&&\ln^2\left(\frac{c \tau}{2b_0}\right) + \frac{11}{6} \ln\left(\frac{c \tau}{2b_0}\right)- \frac{107}{105} \ln\left(\frac{c \tau}{2r_0}\right)\nonumber\\
&& + \frac{124627}{44100} =\ln^2\left(\frac{\tau}{C_{I_2}}\right)-\frac{107}{105}\ln\left(\frac{ \tau}{C_{I_2}^*}\right)\,,
\eea
with
\beq
C_{I_2}^*=2r_0 e^{\frac{116761}{59920}}\,.
\eeq

We have already mentioned  that (as we checked) the UV-scale $r_0$, entering with coefficient $-107/105$, cancels against the scale dependence of the source quadrupole moment $I_{ij}$ \cite{Blanchet:1997jj,Goldberger:2009qd}.
The final expression of the full quadrupolar waveform $\hat U_2(\omega)$ only depends on the logarithmic scale $b_0$. 

As is well known, the $b_0$ dependence of $\hat U_2(\omega)$ should only enter through an exponentiated phase factor of the form
\beq
e^{\frac{i2G{\mathcal M}\omega}{c^3}\ln b_0}\,,
\eeq
where ${\mathcal M}=E/c^2$. As a consequence, $\hat U_2(\omega, b_0)$  should
satisfy the renormalization group equation
\bea \label{RGb0}
b_0 \frac{\partial}{\partial b_0}  {\hat  U}_{2}{} (\omega)
&=&  i\frac{2G{\mathcal M}\omega}{c^3}  \hat U_{2} (u)\,.
\eea
We have explicitly checked that our results for $\hat U_2(\omega)$ satisfies (within our accuracy)  the equation \eqref{RGb0}.
In particular, this computation checks that all the logarithms of the form $G{\mathcal L}$+ $G^2({\mathcal L}^2+{\mathcal L})+\ldots $
that appear in our result come from expanding the exponentiated  phase factor $e^{\frac{i2G{\mathcal M}\omega}{c^3}\ln b_0}$.

\section{Confirmation of our results via  black-hole perturbation theory}

In the extreme mass-ratio limit, say $m_2 \ll m_1 $, i.e., to first order in the symmetric mass ratio  $\nu$, the waveform can be computed in the framework of first-order Black-Hole perturbation theory (see, e.g., Ref. \cite{Sasaki:2003xr}).
The smaller mass $m_2$ is assumed to move along a hyperboliclike geodesic orbit on the equatorial plane of a Schwarzschild spacetime
of mass $m_1$, with parametric equations $x^\mu =x_p^{\mu}(\tau)$, $\tau$ denoting the proper time.
 
The Weyl scalar $\psi_4$ is asymptotically related to the two independent polarizations $h_+$ and $h_\times$ of the gravitational waves 
(encoded in the complex waveform $h_{\bar m \bar m} \equiv h_{\mu\nu}{\bar m}^\mu{\bar m}^\nu \equiv h_+-ih_\times$) by
\beq
\label{hdef}
\psi_4(r\to\infty)\approx-\frac12\ddot h_{\bar m \bar m}\,,
\eeq
a dot denoting a retarded time derivative.
$\psi_4$ satisfies the Teukolsky equation with spin-weight $s=-2$, and can be decomposed as 
\beq
\label{sep}
\psi_4= \frac1{r^4}\int\frac{d\omega}{2\pi}e^{-i\omega t}\sum_{lm}\,\,R_{lm\omega}(r)\,\, {}_{-2}Y_{lm}(\theta,\phi)\,,
\eeq
where (as above) ${}_{s}Y_{lm}(\theta,\phi)$ are spin-weighted spherical harmonics.
The radial function $R_{lm\omega}(r)$ satisfies an inhomogeneous Teukolsky equation with source term $T_{lm\omega}(r)$. 
The asymptotic solution representing purely outgoing waves is given by 
\beq
R_{lm\omega }(r\to\infty)\approx  Z^\infty_{lm\omega} r^3e^{i\omega r_*}\,,
\eeq
where $r_*$ is the tortoise coordinate, and where the amplitude $Z^\infty_{lm\omega}$ is given by
\beq
\label{Zinf}
Z^\infty_{lm\omega}=\frac{C^{\rm trans}_{lm\omega}}{W_{lm\omega}}\int_{2 G m_1}^\infty dr\frac{R^{\rm in}_{lm\omega}(r)T_{lm\omega}(r)}{\Delta^2}\,.
\eeq
Here $\Delta=r(r-2G m_1)$, $W_{lm\omega}$ denotes the (constant) Wronskian, and $C^{\rm trans}_{lm\omega}$ is the  transmission coefficient.

The asymptotic form of $\psi_4$ then implies
\beq
\label{h}
 h_{\bar m \bar m}=\frac{4G}{r}\sum_{lm}\int\frac{d\omega}{2\pi}{\mathcal W}_{lm}(\omega)e^{-i\omega(t-r_*)}\,\,{}_{-2}Y_{lm}(\theta,\phi)\,,
\eeq
where
\beq
\label{Wlmdef}
{\mathcal W}_{lm}(\omega)\equiv\frac{Z^\infty_{lm\omega}}{2 G\omega^2}\,,
\eeq
are the waveform modes in the frequency domain.
The latter are integrals of the type
\beq
\label{Wlmdef2}
{\mathcal W}_{lm}(\omega)=\int dt e^{i(\omega t-m\phi_p(t))}{\mathcal F}_{lm\omega}(r_p(t))\,,
\eeq
with the function ${\mathcal F}_{lm\omega}(r_p(t))$ evaluated at the particle position $r=r_p(t)$. The latter functions are computed by using the Mano-Suzuki-Takasugi (MST) solutions satisfying the retarded boundary conditions of ingoing radiation at the horizon and upgoing at infinity \cite{Mano:1996mf,Mano:1996vt}.

In order to compute the integrals \eqref{Wlmdef2} it is convenient to parametrize the geodesics in a quasi-Keplerian form, leading to Bessel as well as iterated-Bessel integrals. 
For details see Ref. \cite{Geralico:2026kbm}.

We have explicitly checked that, within our 3.5PN accuracy, the three independent ${\mathcal W}_{2m}(\omega)$ modes (for $m=2,0, \bar 2$) 
agree with the $O(\nu)$ piece of our ${U}_{2m}(\omega)$ final result (given by Eq. \eqref{U22022m2} with coefficients listed 
in the associated ancillary file.
More precisely, they agree modulo a phase factor
$e^{i\omega\delta t}$ associated with a time shift $\delta t$, or equivalently (in view of Eq. \eqref{RGb0}) modulo a relation between
the arbitrary MPM scale $b_0$ and the scale $  2G m_1$ selected by the MST solution. Explicitly, we find that, with
\beq
\delta t=2 G m_1 \, \ln\left(\frac{2Gm_1}{\sqrt{e}b_0}\right) \,,
\eeq
we have (for $m=2,0, \bar 2$)  the relation
\beq
{\mathcal W}_{2m}=e^{i\omega\delta t}U_{2m}^{\nu^1}
\equiv U_{2m}^{\nu^1}+\delta U_{2m}^{\nu^1}\,,
\eeq
where
\bea
\delta U_{2m}^{\nu^1}&=&i\omega\delta t\left[U_{2m}^{\nu^1\, G^1}\left(1+\frac{i\omega\delta t}{2}\right)G+U_{2m}^{\nu^1\, G^2}G^2\right]\nonumber\\
&&+O(G^4)\,.
\eea
This agrees with Ref. \cite{Fujita:2010xj} (see there the text below Eq. 5.7c).

\section{Soft-limit}

The soft expansion (small-frequency expansion, $\omega\to 0$, at some fixed emission direction ${\mathbf n}(\theta,\phi)$) of the waveform has the structure \cite{Weinberg:1965nx,Saha:2019tub,Sahoo:2021ctw} 
\bea
\label{soft_W}
\hat W(\omega,{\mathbf n})&=& \frac{i{\mathcal A}({\mathbf n})}{\omega}+{\mathcal B}({\mathbf n})\ln \omega +{\mathcal C}({\mathbf n})\omega \ln^2 \omega\nonumber\\
& +& {\mathcal D}({\mathbf n})\omega \ln \omega+\ldots 
\eea 
The leading-order term in this expansion measures the memory, namely
\beq
{\mathcal A}({\mathbf n})= [W(t_r, {\mathbf n})]_{-\infty}^{+\infty}\,.
\eeq

The memory (see Refs. \cite{Christodoulou:1991cr,Wiseman:1991ss,Thorne:1992sdb,Blanchet:1992br,Favata:2008yd}), $[W(t_r,{\mathbf n})]_{-\infty}^{+\infty}$ is made of two types of contributions, namely
\bea
\label{W_def2}
[W(t_r,{\mathbf n})]_{-\infty}^{+\infty}&=&[W]^{\rm tot}({\mathbf n})\nonumber\\
&=& [W]^{\rm lin}({\mathbf n})+[W]^{\rm nonlin}({\mathbf n})\,.
\eea
The first contribution (\lq\lq linear memory") is linked to the momenta of the ingoing and outgoing massive particles
\beq
\label{weinb}
[W]^{\rm lin}({\mathbf n})= \left[\sum_{a=1}^2  \frac{(p_a\cdot \bar m)^2}{E_a -{\mathbf p}_a \cdot  {\mathbf n}}\right]_{-\infty}^{+\infty}\,.
\eeq
The second contribution (\lq\lq nonlinear memory" \cite{Christodoulou:1991cr,Wiseman:1991ss,Blanchet:1992br,Thorne:1992sdb,Favata:2008yd} is linked to the momenta of the gravitons  emitted during the scattering process.
Here, we assume that the initial state does not contain incoming gravitons. It reads
\bea
[W]^{\rm nonlin}({\mathbf n})=\int dt \int d\Omega' \frac{dE_{\rm gw}(t,{\mathbf n}')}{d\Omega' dt}\,\, \frac{({\mathbf n}'\cdot \bar m)^2}{1-{\mathbf n}'\cdot {\mathbf n}}
\,.
\eea
 
We can decompose the total memory $[W]^{\rm tot}({\mathbf n})$ (which is a spin-weight $-2$ field on the sphere) in spin-weighted spherical harmonics (recalling the notation ${}_{-2}Y_{lm}(\theta,\phi)={}_{\bar 2}Y_{lm}(\theta,\phi)$)
\beq
[W]^{\rm tot}({\mathbf n})=\sum_{l\ge 2} [W]^{\rm tot}_l\,,
\eeq
where 
\beq
\label{W_tot_fin}
[W]^{\rm tot}_l=\sum_{m=-l}^l \left(\int d\Omega' W^{\rm tot}({\mathbf n}')\,\,{}_{\bar 2}Y_{lm}^*(\theta',\phi')\right) {}_{\bar 2}Y_{lm}(\theta,\phi)\,.
\eeq
The first four terms of the $G^3$ ($2$-loop) part of the soft expansion ($u\to 0$, $\omega\to 0$) of the quadrupolar MPM waveform (as given in the ancillary file)
read
\bea
{\mathcal A}^{G^3,\le {\rm 3.5PN}}_{l=2}({\mathbf n}) 
&=& \frac{1}{b^3}\left\{\frac{4 }{p_\infty^4}{\bar m}_1 {\bar m}_2 \nu\right.\nonumber\\
&+& \left(-\frac{10}{7}   + \frac{44}{7}   \nu \right)\frac{1}{p_\infty^2}\nu {\bar m}_1 {\bar m}_2\nonumber\\
&+& \left(-\frac{379}{6} + \frac{589}{21}  \nu  + \frac{32}{21}  \nu^2\right) \nu {\bar m}_1 {\bar m}_2 \nonumber\\
&+& \left(-\frac{32}{5} {\bar m}_1 {\bar m}_2  - \frac{23}{28} {\bar m}_1^2   \pi - 
    \frac{117}{20} {\bar m}_2^2  \pi \right)\nu^2 p_\infty \nonumber\\
&+& \left(-\frac{12847}{308}  + \frac{5359}{154}  \nu  \right.\nonumber\\
&+&\left. 
   \frac{1296}{77}  \nu^2 - \frac{32}{77} \nu^3\right)\nu{\bar m}_1 {\bar m}_2  p_\infty^2 \nonumber\\
&+& \left[\left(- \frac{144}{7}  - \frac{128}{35}   \nu\right) {\bar m}_1 {\bar m}_2 \right.\nonumber\\  
&+&\left( -\frac{   3397}{2688}  - \frac{ 409  }{1344}   \nu   \right) {\bar m}_1^2   \pi\nonumber\\
&+&\left. \left(- \frac{ 21311 }{13440} 
- \frac{ 24491}{6720}  \nu   \right){\bar m}_2^2  \pi \right] \nu^2 p_\infty^3\nonumber\\
&+& \left. O(p_\infty^4)\right\}\,,
\eea
\bea
{\mathcal B}^{G^3, \le  {\rm 3.5PN}}_{l=2}({\mathbf n}) &=& \frac{\nu}{b^3}\left[-\left( {\bar m}_1^2 -{\bar m}_2^2  \right)\frac{2}{p_\infty^5} \right.\nonumber\\
&+&  ({\bar m}_1^2-{\bar m}_2^2) (- \frac{30}{7}   - \frac{22}{7} \nu) \frac{1}{p_\infty^3} \nonumber\\
&+& ({\bar m}_1^2-{\bar m}_2^2) \left(\frac{160}{21} - \frac{247}{42} \nu   - \frac{16}{21} \nu^2\right) \frac{1}{p_\infty} \nonumber\\
&+& 12 {\bar m}_1 {\bar m}_2 \pi \nonumber\\
&+& ({\bar m}_1^2-{\bar m}_2^2)\left(\frac{4016}{231}   + \frac{12347}{924} \nu  \right.\nonumber\\
&-& \left. \frac{272}{231} \nu^2  + \frac{16}{77} \nu^3  \right) p_\infty\nonumber\\
&+&\left( \frac{75}{7}    + \frac{90}{7}  \nu  \right)\pi  {\bar m}_1 {\bar m}_2 p_\infty^2\nonumber\\
&+&\left.  O(p_\infty^3)\right]\,,
\eea
\bea
{\mathcal C}^{G^3, \le  {\rm 3.5PN}}_{l=2}({\mathbf n}) &=& \frac{\nu}{b}i {\bar m}_1 {\bar m}_2 \left[
+ \frac{2}{p_\infty^6}  +\left(-\frac{19}{7}    +  \frac{22}{7} \nu\right)\frac{1}{p_\infty^4} \right.\nonumber\\
&+&\left(- \frac{913}{84} - \frac{215}{42}\nu + \frac{16}{21}\nu^2\right)\frac{1}{p_\infty^2} \nonumber\\
&+&\frac{36193}{1848} - \frac{14305}{924}\nu - \frac{344}{231} \nu^2 - \frac{16}{77}   \nu^3 \nonumber\\
&+& \left.  O(p_\infty)\right]\,,
\eea
and
\begin{widetext}
\bea
{\mathcal D}^{G^3,\le  {\rm 3.5PN}}_{l=2}({\mathbf n}) &=& \frac{\nu}{b}\left[
\left( -4  + 4  {\mathcal L}_1\right) i \frac{{\bar m}_1 {\bar m}_2}{p_\infty^6}\right.\nonumber\\
&+& \left( \frac{152  }{21} - \frac{68  \nu}{7} + \left(-\frac{38 }{7} + \frac{44\nu}{7}\right) {\mathcal L}_1\right) i \frac{{\bar m}_1 {\bar m}_2}{p_\infty^4}\nonumber\\
&+& 4 \pi\frac{{\bar m}_1 {\bar m}_2}{p_\infty^3}\nonumber\\
&+& \left(  \frac{2069  }{126} + \frac{73   \nu}{9} - \frac{208   \nu^2}{63}  
+ \left(-\frac{913 }{42} - \frac{215 \nu}{21} + \frac{32 \nu^2}{21}\right)  {\mathcal L}_1\right)i\frac{{\bar m}_1 {\bar m}_2}{p_\infty^2}\nonumber\\
&+& \left(-6 i {\bar m}_1^2\pi + 6 i {\bar m}_2^2\pi + 
 {\bar m}_1 {\bar m}_2 \left(-\frac{16 i \nu}{5} + \frac{4\pi}{7} + \frac{44 \nu\pi}{7}\right) \right)\frac{1}{p_\infty}\nonumber\\
%%%%%
&+& {\bar m}_1 {\bar m}_2 \left(-\frac{20969 i}{462} + \frac{3541 i \nu}{99} + \frac{584 i \nu^2}{231} + \frac{
   752 i \nu^3}{693} + 
   16 i {\mathcal L}_2\right.\nonumber\\
& +&\left.  \left(\frac{21409 }{924} - \frac{14305   \nu}{462} -  
      \frac{688  \nu^2}{231} - \frac{32 \nu^3}{77}\right) i {\mathcal L}_1+8\pi
\right)\nonumber\\
%%%%%
&+& \left( {\bar m}_1^2 i\pi \left(-\frac{97 }{14} - \frac{33   \nu }{7}\right) + 
 {\bar m}_2^2 i\pi\left(-\frac{113 }{14} + \frac{75 \nu }{7}\right) + 
 {\bar m}_1 {\bar m}_2 \left(\frac{8 i \nu}{105} - \frac{32 i \nu^2}{35} - \frac{344\pi}{21} 
- \frac{17 \nu\pi}{21} + \frac{32 \nu^2\pi}{21} \right)\right) p_\infty \nonumber\\
&+& \left. O(p_\infty^2)\right]\,,
\eea
\end{widetext}
where we used the notation
\beq
{\mathcal L}_1=\ln \left(\frac{be^{\gamma_E}}{2p_\infty} \right)\,,\qquad {\mathcal L}_2=\ln (2b_0e^{\gamma_E})\,.
\eeq

As a partial check on our results we have focussed on the $G^3$ contribution  to the leading soft term ${\mathcal A}^{G^3,\le {\rm 3.5PN}}_{l=2}({\mathbf n})$ and verified its agreement with the universal expression of the (linear-plus-nonlinear) memory as given at the beginning of this section.

More precisely,  we decomposed the computation of the right-hand-side of Eq. \eqref{W_tot_fin} into three pieces: the first  piece is defined by taking the conservative part of the
outgoing momenta entering the linear memory, Eq. \eqref{weinb}.
It reads
 \bea
\label{lin_cons}
 \Delta U_2^{{\rm lin, cons} G^3} &=&
\frac{ M^4}{b^3} \bar  m^1\bar m^2\nu \left[
\frac{4}{p_\infty^4}
+\frac{\frac{44 \nu}{7}-\frac{10}{7}}{p_\infty^2}\right.\nonumber\\
&-& \frac{379}{6}+\frac{589 \nu }{21}+\frac{32 \nu ^2}{21}\nonumber\\
&+&\left(-\frac{32 \nu ^3}{77}+\frac{1296 \nu ^2}{77}\right.\nonumber\\
&+&\left.  \frac{5359 \nu }{154}-\frac{12847}{308}\right) p_\infty^2\nonumber\\
&+&\left.  O(p_\infty^4)\right]\,. 
 \eea
The second piece uses the radiation-reaction part of the outgoing momenta entering the linear memory and reads (at orders $O(G^3,\eta^5)$ and $O(G^3,\eta^7)$)
 \bea
\label{lin_mem_rr}
 \Delta U_2^{{\rm lin, rr}, G^3 \eta^5} &=&
-\frac{ M^4}{b^3} \nu^2 p_\infty \left[ \frac{32}{5}\bar  m^1\bar m^2+ \frac{74}{15}\pi (\bar m^2)^2
\right]\nonumber\\
 \Delta U_2^{{\rm lin, rr}, G^3 \eta^7} &=& -\frac{ M^4}{b^3} \nu^2 p_\infty^3 \left[ 
\bar  m^1\bar m^2 \left(\frac{144}{7} + \frac{128 \nu}{35} \right)\right.\nonumber\\ 
&+&\left. 
(\bar m^2)^2\pi \left(\frac{913}{420} + \frac{74\nu}{105}\right)
\right]\,.
 \eea
The third piece, namely the quadrupolar nonlinear memory, is conveniently computed by using the relation (first derived in Ref. \cite{Blanchet:1992br}) between the nonlinear memory and the decomposition of the total radiated energy $E_{\rm gw}({\mathbf n})$ (integrated over time but as a function of the emission angles) in {\it scalar} spherical harmonics. Namely, denoting 
\beq
E^{\rm gw}({\mathbf n})\equiv \int_{-\infty}^{+\infty} dt \frac{dE^{\rm gw}(t,{\mathbf n})}{dtd\Omega}\,,
\eeq
and
\beq
E^{\rm gw}({\mathbf n})=\sum_{l\ge 0}\sum_{m=-l}^l E^{\rm gw}_{lm}Y_{lm}(\theta,\phi)\,,
\eeq
we have (now, only for $l\ge 2$)
\beq
\label{8.12}
W^{\rm nonlin}_l=4\pi N_l \sum_{m} E^{\rm gw}_{lm} \,\, {}_{\bar 2}Y_{lm}(\theta,\phi)\,,
\eeq
where
\beq
N_l=\sqrt{\frac{(l-2)!}{(l+2)!}}\,.
\eeq
The latter result is valid in any Lorentz frame. See Appendix B below (see also Appendix  D  of Ref. \cite{Bini:2024ijq}) for a discussion on how to transform the emitted energy spectrum $E^{\rm gw}({\mathbf n})$ in various Lorentz frames. We used the results displayed in Appendix B, together with a direct computation of the PN expansion of  $E^{\rm gw}({\mathbf n})$  derived from the explicit time-domain tree-level  waveform \cite{Kovacs:1977uw,Kovacs:1978eu} to compute $W^{\rm nonlin}_l$ for $2\le l\le 6$ in the cm frame and up to $O(p_\infty^{12})$.

Let us also mention that the multipolar decomposition of $E^{\rm gw}({\mathbf n})$ in the rest frame of one of the particles has been recently computed to  $O(G^3)$ but to all orders in velocities in Ref. \cite{Georgoudis:2025vkk}. We have checked that the $m_2\to 0$ limit of our results agrees with Ref. \cite{Georgoudis:2025vkk}.

Similarly to the radiation-reaction part of the linear memory, the quadrupolar nonlinear memory is the sum of $O(G^3,\eta^5)$ and $O(G^3,\eta^7)$ contributions. The 
$O(G^3,\eta^5)$ (first computed in Ref. \cite{Wiseman:1991ss})  reads (in our notation)
 \beq
\label{nonlinmemeta5}
 \Delta U_2^{{\rm nonlin} G^3 \eta^5}=
 -M\frac{G^3 M^3 \nu^2 \pi p_\infty}{b^3}\left(\frac{23}{28 }\bar m_1^2+\frac{11}{12}\bar m_2^2\right)\,.
 \eeq
The $O(G^3,\eta^7)$ term is given by
 \bea
\label{nonlinmemeta7}
 \Delta U_2^{{\rm nonlin} G^3 \eta^7}&=&
 -M\frac{ G^3 M^3 \nu^2 \pi p_\infty^3}{b^3}\left[\left(  \frac{3397 }{2688 }\right.\right.\nonumber\\
&+&\left. \frac{ 409}{1344 }\nu \right)\bar m_1^2\nonumber\\
&+& \left.\left(  -\frac{527 }{896 }+\frac{1317 }{448 }\nu \right)\bar m_2^2\right]\,.
 \eea

Finally, we checked that the sum of Eqs. \eqref{lin_cons}, \eqref{nonlinmemeta5}, and \eqref{nonlinmemeta7} agrees with the result ${\mathcal A}_{l=2}^{G^3,\le {\rm 3.5PN}}$ derived above from the soft limit of our $\hat U_2(\omega)$.

\section{Comparison between the $G^2$ truncation of the MPM waveform and the one-loop EFT waveform}

QFT-based EFT methods have been used to compute the 1-loop bremsstrahlung
 waveform in the frequency domain \cite{
Jakobsen:2021smu,Mougiakakos:2021ckm,Brandhuber:2023hhy,Herderschee:2023fxh,Georgoudis:2023lgf,Georgoudis:2023eke,
Bini:2024rsy,Georgoudis:2023eke,Heissenberg:2025fcr,Bohnenblust:2025gir,Brunello:2025eso}.
We recall that in the EFT approach the tree level waveform corresponds to the time-dependent part of the $O(G^2)$ terms in the asymptotic metric (first computed in the time domain by Kovacs and Thorne in Refs. \cite{Kovacs:1977uw,Kovacs:1978eu}), while the 1-loop waveform corresponds to $O(G^3)$ terms in $h_{\mu\nu}$. In terms of our rescaled waveform variable $W$, we have $W^{\rm tree}=O(G)$ and $W^{\rm 1-loop}=O(G^2)$. Our current $O(G^3\eta^7)$ result of $W$ reaches the 2-loop level which has not yet been computed in the EFT approach.
Previous comparisons between the EFT and MPM waveforms have been limited to the 1-loop and $O(\eta^5)$ level \cite{Bini:2023fiz,Bini:2024rsy}. As our present result extended the accuracy of the $G^2$ waveform to the $\eta^7$, an important check on our result is to compare our $O(G^2\eta^7)$ MPM waveform to the corresponding PN expansion of the 1-loop EFT waveform.

Recently Ref. \cite{Heissenberg:2025fcr} has used the mass-polynomiality properties of the bremsstrahlung waveform to derive the comparable mass bremsstrahlung waveform to the 5PN ($O(\eta^{10}$) accuracy from the probe-limit results.

Let us recall the structure of the  1-loop-accurate EFT frequency-domain waveform \cite{Brandhuber:2023hhy,Herderschee:2023fxh,Georgoudis:2023lgf}.
Following Ref. \cite{Bini:2024rsy}, 
\bea
W^{\rm EFT}(\omega,{\mathbf n})&=&W^{G^1}+W^{G^2(0)}+ W^{G^2\epsilon/\epsilon}\nonumber\\
&+& W^{G^2 \rm cut}+W^{G^2 \rm disc, reg'}+\delta W^{G^2 \rm rot}\,.
\eea
We refer to Ref. \cite{Bini:2024rsy} for the precise definition of the various contributions to the waveform. For uniformity with previous sections we have replaced the qualification \lq\lq 1-loop" by the superscript $G^2$.

Let us emphasize here the structure of the contribution \lq\lq $W^{G^2 \rm disc, reg'}$"
\bea
\label{Wdisc}
W^{G^2 \rm disc, reg'}&=&2iG [m_1\omega_1 \ln \left(\frac{\omega_1}{\omega}\right)\nonumber\\
&+& m_2\omega_2 \ln \left(\frac{\omega_2}{\omega}\right)]W^{\rm tree}\,.
\eea
Here, the contribution $W^{G^2 \rm disc, reg'}$ comes from the {\it disconnected} components of the $T$ matrix indicated in Fig. 1b of Ref. \cite{Bini:2024rsy}.
This contribution involves a coupling between zero-frequency gravitons and nonzero frequency gravitons.
As discussed in Sec. VC of Ref. \cite{Bini:2024rsy}, the evaluation of this term is quite delicate, and implies an additional regularization of the zero-frequency gravitons across a cut. The effect of this additional regularization involves an arbitrary infinitesimal parameter $\beta^{\rm reg'}$ (denoted as $\beta$ in \cite{Bini:2024rsy}) as shown in Eq. \eqref{Wdisc} above. We have added the superscript   ${\rm reg}'$   as a reminder of the presence of this additional regularization beyond the usual dimensional regularization one.

In current EFT work it was assumed that the $\beta^{\rm reg'}$ ambiguity can be absorbed in a simple time shift of the waveform and can therefore be ignored.
The rest of the disconnected contribution  to $W$ then  corresponds (when neglecting the angle $\chi^{\rm cons}/2$ between the incoming frame and the barred frame) to the  Veneziano-Vilkovisky supertranslation \cite{Veneziano:2022zwh}, i.e., to the following  {\it angle dependent} time shift 
\bea
\label{beta_VV_def}
\delta t^{VV}(\theta,\phi)&\equiv&\beta^{\rm VV}(\theta,\phi)\nonumber\\
&=&
2 G \left[m_1\frac{\omega_1}{\omega} \ln \left(\frac{\omega_1}{\omega}\right)+m_2\frac{\omega_2}{\omega} \ln \left(\frac{\omega_2}{\omega}\right)\right] \,.\nonumber\\
\eea
Here
\bea
\omega_1&=&-l \cdot u_1=-l \cdot \frac{p_1}{m_1}\,,\qquad\omega_2=-l \cdot u_2=-l \cdot \frac{p_2}{m_2}\,,\nonumber\\
\omega &=&-l \cdot \frac{p_1+p_2}{|p_1+p_2|} =-l \cdot \frac{p_1+p_2}{E}= -l \cdot U\,,
\eea
are covariantly defined in terms of the incoming momenta, four-velocities and the null direction of the emitted radiation, $l^\mu=(1,{\mathbf n})$.

In other words, when ignoring $\beta^{\rm reg'}$ and neglecting $\chi^{\rm cons}/2$,  we can write
\bea
\label{Wdisc}
W^{G^2 \rm disc, reg'} 
&=& i\omega \beta^{\rm VV}(\theta,\phi)W^{\rm tree}\,.
\eea

We recall that $\beta^{\rm VV}(\theta,\phi)$   was obtained in Ref. \cite{Veneziano:2022zwh} as a solution of the problem of adding to the EFT waveform, coming from disconnected diagrams, the $O(G)$ contribution parametrizing the incoming waveform in a scattering situation, i.e., the incoming Weinberg waveform \cite{Weinberg:1965nx}
\beq
\lim_{t_r\to -\infty} W(t_r,{\mathbf n})= \left[\sum_{a=1}^2  \frac{(p_a\cdot \bar m)^2}{E_a -{\mathbf p}_a \cdot  {\mathbf n}}\right]^{\rm incoming}\,.
\eeq
In other words, Ref. \cite{Veneziano:2022zwh}    obtained $\beta^{\rm VV}$ as a solution of the (elliptic) partial differential equation
\beq
\label{eq_sphere}
(2D_aD_b -\Omega_{ab}D^2)\beta^{\rm VV}= 4 W_{ab}^{\rm incoming} \,. 
\eeq
Here the indices $a,b$ are tangent to the 2-sphere at infinity, $\Omega_{ab}$ being the round metric on the 2-sphere, $D_a$ is a covariant derivative on the sphere and $W_{ab}$ is the 2-dimensional symmetric and tracefree tensor representation of the spin-weight-2  object  $W^{\rm incoming}$.

We computed, at our 3.5PN ($O(\eta^7)$) accuracy, the difference 
\beq
\delta U_2=U_{2}^{\rm EFT}-U_{2}^{\rm MPM}\,,
\eeq
when absorbing $\beta^{\rm reg'}$ in a time shift, and  fixing the relation between the EFT infrared energy scale $\mu_{\rm IR}$ and the MPM infrared length scale $b_0$ (entering the MPM tail effects) as in Eq. (11.3) of Ref. \cite{Bini:2024rsy}:
\beq
\ln (2b_0\mu_{\rm IR} )+\gamma_E=0\,.
\eeq
We find the following mismatches at the 3.5PN level after projection of $\delta U_2$ on ${}_{\bar 2}Y_{2m}(\theta,\phi)$
\bea
\label{mis_match}
\delta U_{22}&=& i \nu^2 G^2\eta^7 \sqrt{\frac{\pi}{5}}p_\infty^4 \frac{4}{105 b}(1-4\nu) u [2(3+2u\nonumber\\
&+& 5u^2)K_0(u)+(16+9u+10u^2)K_1(u)]\,,\nonumber\\
\delta U_{20}&=& i \nu^2 G^2 \eta^7\sqrt{\frac{2\pi}{15}}p_\infty^4 \frac{4}{35 b}(1-4\nu) u [2K_0(u)\nonumber\\
&+& 11 u K_1(u)]\nonumber\\
\delta U_{2\bar 2}&=& i \nu^2 G^2\eta^7 \sqrt{\frac{\pi}{5}}p_\infty^4 \frac{4}{105 b}(1-4\nu) u  [2 (3 - 2 u\nonumber\\
& +&  5 u^2) K_0(u)+(-16 + 9 u - 10 u^2) K_1(u) ]\,.
\eea
The latter mismatches can be represented as the effect of a specific supertranslation, namely 
\beq
\delta U_2=i\omega (c_y n^y )W^{\rm tree}\,,
\eeq
where $n^y=\sin\theta \sin\phi$ is the $y$ component of the emission direction ${\mathbf n}$ and
\beq
c_y=+\frac45 \frac{G \eta^6 m_1m_2 (m_1 - m_2) }{(m_1 + m_2)^2}  p_\infty^3\,.
\eeq
Here, the $O(\eta^7)$ mismatches \eqref{mis_match} come from combining the $O(\eta^6)$ $c_y$ with the $O(\eta^1)$ part of $W_{\rm tree}$, and more precisely the $l=3$ 
contribution to $W$, i.e. the term $\eta U_3$ in Eq. \eqref{W_deco}.

The dipolar ($l=1$) supertranslation $c_yn^y$ simply encodes a shift of the spatial origin of the coordinate system with respect to which one is computing the waveform.
In other words, the mismatches  \eqref{mis_match} only correspond to a difference in the choice of the spatial cm origin in the two considered waveforms. 
For a discussion of the effect of a shift of the  spatial origin  on the multipole moments see, e.g.   Appendix B of  Ref. \cite{Damour:1990ji}.
More about this below. 

We recall that the  $y$ axis denotes the spatial direction of the bisector between the incoming and the outgoing spatial momentum of the first particle in the cm frame 
\beq
\bar p_1=\frac12 (p_1^{\rm incoming}+p_1^{\rm outgoing})\,,
\eeq
see Eq. \eqref{bar_pa} above and the associated  discussion as well as Sec. II of Ref. \cite{Bini:2024rsy}.

At our present  PM accuracy, we can identify the $e_y$ axis with the spatial cm direction $e_Y$ of the incoming particles, see Eqs.   \eqref{frame1} and \eqref{frame1inv} above. 
We then notice that the $O(p_\infty^3)$ supertranslation 
\beq
c_y n^y=c_y {\mathbf n}\cdot e_y=c_y{\mathbf n}\cdot e_Y +O(G^2)=c_y n^Y,
\eeq 
coincides with the leading order $l=1$ dipolar projection of the Veneziano-Vilkovisky supertranslation $\beta^{\rm VV}({\mathbf n})$.

We see, from Eq. \eqref{beta_VV_def} in which $\omega_1$ and $\omega_2$ explicitly read
\bea
\frac{\omega_1}{\omega}&=&\frac{E_1-n^Y P_{\rm cm}}{m_1}\,,  
\nonumber\\
\frac{\omega_2}{\omega}&=&\frac{E_2+n^Y P_{\rm cm}}{m_2}\,, 
\eea
with (setting $\eta=1$)
\bea
E_1&=&\frac{m_1}{E}(m_1 + m_2\gamma)\,,\quad E_2=\frac{m_2}{E}(m_2 + m_1\gamma)\,,\nonumber\\
P_{\rm cm}&=& \frac{m_1 m_2 p_\infty}{E}\,,
\eea
that $\beta^{\rm VV}$ is a function of $n^Y={\mathbf n}\cdot e_Y$. Denoting $n^Y=\cos\Theta$ the multipolar decomposition of $\beta^{\rm VV}$ reads
\bea
\beta^{\rm VV}&=& \beta_0^{\rm VV}+\beta_1^{\rm VV}P_1(\cos(\Theta) + \beta_2^{\rm VV}P_2(\cos(\Theta)\nonumber\\
&+&  \beta_3^{\rm VV}P_3(\cos(\Theta)+\ldots
\eea
where $P_l(z)$ are Legendre polynomials.
We recall that such a multipolar expansion is also equivalent to an expansion in STF tensors of the type
\bea
\beta^{\rm VV}({\mathbf n})&=& \sum_{L}\frac{1}{l!}\beta_{\langle L\rangle}^{\rm VV}\hat n^{\langle L\rangle}\,.
\eea

Using the orthogonality property of the Legendre polynomials
\beq
\int_{-1}^1 dz P_l(z)P_{l'}(z)=\frac{2}{2l+1}\delta_{ll'}\,,
\eeq
the coefficients of $l-$th multipolar contribution is given by
\bea
\label{beta_l_VV}
\beta_l^{\rm VV}&=&\frac{2l+1}{2} \int_{-1}^1 dn^Y P_l(n^Y)\beta^{\rm VV}(n^Y)\,.
\eea
In particular, the monopolar and dipolar coefficients are given by
\bea
\beta_0^{\rm VV}&=&\frac12 \int_{-1}^1 dn^Y \beta^{\rm VV}(n^Y)\,,\nonumber\\
\beta_1^{\rm VV}&=&\frac32 \int_{-1}^1 dn^Y n^Y \beta^{\rm VV}(n^Y)\,.
\eea
It easy to see that PN expansion of the $\beta_l^{\rm VV}${}'s has the following structure
\bea
\beta_0^{\rm VV}&\sim & p_\infty^2+p_\infty^4+\ldots\nonumber\\
\beta_1^{\rm VV}&\sim & p_\infty^3+p_\infty^5+\ldots\nonumber\\
\beta_2^{\rm VV}&\sim & p_\infty^2+p_\infty^4+\ldots\nonumber\\
\beta_3^{\rm VV}&\sim & p_\infty^3+p_\infty^5+\ldots\nonumber\\
\beta_4^{\rm VV}&\sim & p_\infty^4+p_\infty^6+\ldots\,.
\eea
Exact expressions for $\beta_0^{\rm VV}$ and $\beta_1^{\rm VV}$ can easily be obtained (see below for $\beta_2^{\rm VV}$).
Defining the following functions
\bea
F(z,a)&=&z^2\ln \left( \frac{z}{a}\right)\,, \nonumber\\
G(z,a,b)&=& z(z-a)\ln \left( \frac{z}{b}\right)=\left(1-\frac{a}{z}\right)F(z,b)\,, \nonumber\\
H(z,a,b)&=& \int dz G(z,a,b)=\frac{z^3}{3}\left(\ln \left( \frac{z}{b}\right)-\frac13 \right)\nonumber\\
&-&\frac{az^2}{2}\left(\ln \left( \frac{z}{b}\right)-\frac12 \right)\,,
\eea
we find
\bea
\beta_0^{\rm VV}&=&-GE\nonumber\\
&-&\frac{G}{2P_{\rm cm}}\left[F(E_1-P_{\rm cm},m_1)-F(E_1+P_{\rm cm},m_1)\right.\nonumber\\ 
&+&\left. F(E_2-P_{\rm cm},m_2)-F(E_2+P_{\rm cm},m_2) \right]\,, \nonumber\\
\beta_1^{\rm VV}&=& \frac{3G}{P_{\rm cm}^2}\left[H(E_1-P_{\rm cm},E_1,m_1)\right.\nonumber\\
&-&H(E_1+P_{\rm cm},E_1,m_1)\nonumber\\
&+&H(E_2+P_{\rm cm},E_2,m_2)\nonumber\\
&-& \left.  H(E_2-P_{\rm cm},E_2,m_2)\right]\,.\qquad
\eea
The PN-expanded version of these results read
\begin{widetext}
\bea
\label{beta0_exp}
\beta_0^{\rm VV}&=&\frac{ G m_1 m_2}{ (m_1 + m_2)}\left[\frac{4}{3} p_\infty^2  
-  
 \frac{2  (m_1^2 + 9 m_1 m_2 + m_2^2)}{15 (m_1 + m_2)^2}p_\infty^4\right. \nonumber\\
&+& \frac{
   (9 m_1^4 + 117 m_1^3 m_2 + 373 m_1^2 m_2^2 + 117 m_1 m_2^3 + 
    9 m_2^4}{210 (m_1 + m_2)^4} p_\infty^6  \nonumber\\
&-& \frac{
  (25 m_1^6 + 431 m_1^5 m_2 + 2131 m_1^4 m_2^2 + 4253 m_1^3 m_2^3 + 
    2131 m_1^2 m_2^4 + 431 m_1 m_2^5 + 25 m_2^6)}{ 1260 (m_1 + m_2)^6} p_\infty^8 
\nonumber\\
&+& \frac{1}{110880 (m_1 + m_2)^8}
 (1225 m_1^8 + 26385 m_1^7 m_2 + 173265 m_1^6 m_2^2 + 
    540393 m_1^5 m_2^3 + 846993 m_1^4 m_2^4 + 540393 m_1^3 m_2^5 \nonumber\\ 
&+&\left. 
    173265 m_1^2 m_2^6 + 26385 m_1 m_2^7 + 1225 m_2^8) p_\infty^{10}+O(p_\infty^{11})\right]\nonumber\\  
&=& G M \nu \left[\frac{4}{3} p_\infty^2  - \frac{2}{15} (1 + 7 \nu) p_\infty^4 
+ 
   \frac{9 + 81 \nu + 157\nu^2}{210}  p_\infty^6 
-\frac{  25 +281 \nu + 632 \nu^2 + 
      803 \nu^3 }{1260}p_\infty^8\right.\nonumber\\ 
&+&\left. \frac{(1225 + 16585 \nu + 39455 \nu^2 + 65198 \nu^3 + 
      62417 \nu^4)}{110880} p_\infty^{10}+O(p_\infty^{11}) \right]\,,
\eea
and
\bea
\beta_1^{\rm VV}&=&\frac{G m_1m_2 (m_1 - m_2) }{(m_1 + m_2)^2} \left[\frac{4 }{5 } p_\infty^3 
- \frac{6   (2 m_1^2 + 7 m_1 m_2 + 2 m_2^2)}{35 (m_1 + m_2)^2}p_\infty^5\right.\nonumber\\
&+& \frac{  (128 m_1^4 + 729 m_1^3 m_2 + 1451 m_1^2 m_2^2 + 
    729 m_1 m_2^3 + 128 m_2^4)}{630 (m_1 + m_2)^4} 
  p_\infty^7  \nonumber\\
&-&\left. \frac{ (1920 m_1^6 + 14905 m_1^5 m_2 + 47361 m_1^4 m_2^2 + 
    73315 m_1^3 m_2^3 + 47361 m_1^2 m_2^4 + 14905 m_1 m_2^5 + 
    1920 m_2^6)}{13860 (m_1 + m_2)^6} 
  p_\infty^9 +O(p_\infty^{11})\right]\nonumber\\
&=& G M\sqrt{1-4\nu}\nu \left(\frac{4}{5} p_\infty^3  - \frac{6(2 + 3 \nu)}{35}  p_\infty^5 + 
   \frac{(128 + 217 \nu + 249 \nu^2)}{630} p_\infty^7 + \frac{(-1920 - 3385 \nu - 
      5021 \nu^2 - 4563 \nu^3)}{13860} p_\infty^9\right) \nonumber\\ 
&+& O(p_\infty^{11})\,.
\eea
\end{widetext}
In particular, the leading order $O(p_\infty^3)$ contribution in $\beta^{\rm VV}_1$ precisely agrees with the supertranslation $c_yn^Y$ needed to align the EFT the MPM waveforms.
The fact that $c_y$ agrees in {\it sign} with $\beta^{\rm VV}_1$ actually means that in order to have agreement at the 3.5PN accuracy between the $l=2$ electric quadrupolar parts of both waveforms one needs to modify the disconnected contribution $W^{G^2 \rm disc, reg'}$, originally obtained as in Eq. \eqref{Wdisc}, into 
\bea
\label{W_disc_new}
W^{G^2 \rm disc, new}= i\omega (\beta^{\rm VV}(\theta,\phi)-\beta_1^{\rm VV}(\theta,\phi)n^Y)W^{\rm tree}\,,
\eea
in which one has {\it subtracted} from  $\beta^{\rm VV}$ its $l=1$ dipolar  part.
We conjecture that this result (here explicitly obtained at the 3.5PN level and for the $l=2$ projected waveform) is exactly valid.
Regarding this subtraction of the dipolar part of $\beta^{\rm VV}$ some remarks are in order.

First, the fact that Eq. \eqref{Wdisc} was obtained by using  an ad-hoc additional regularization of the disconnected contribution to the waveform suggests that a more complete regularization could lead to the new result $W^{G^2 \rm disc, new}$, Eq. \eqref{W_disc_new}.
Second, we wish to emphasize that the partial differential equation, Eq. \eqref{eq_sphere}, {\it defining} $\beta^{\rm VV}(\theta,\phi)$ admits as homogeneous solutions of the operator on the left-hand-side an arbitrary combination of $l=0$ and $l=1$ functions on the sphere. In other words, the most general solution of Eq. 
\eqref{eq_sphere} reads
\beq
\beta^{\rm VV, gen}(\theta,\phi)=\beta^{\rm VV}(\theta,\phi)+c_0+c_x n^X+c_yn^Y +c_zn^Z\,.
\eeq
One could define a unique, \lq\lq canonical" solution of Eq.  \eqref{eq_sphere}, i.e., a unique canonical solution to the problem of adding the initial shear $W_{ab}^{\rm incoming}$ to the EFT waveform,  by projecting away the $l=0$ and $l=1$ parts of $\beta^{\rm VV}(\theta,\phi)$, i.e.
\bea
\beta^{\rm VV, can}(\theta,\phi)&=&\beta_{l\ge 2}^{\rm VV}(\theta,\phi)\nonumber\\
&\equiv& \beta^{\rm VV}(\theta,\phi) - \beta_0^{\rm VV}- \beta_1^{\rm VV} n^Y\,.
\eea 
From this point of view, our result for  $W^{G^2 \rm disc, new}$, Eq. \eqref{W_disc_new}, is compatible with defining $W^{G^2 \rm disc, new}$ in terms of $\beta^{\rm VV, can}(\theta,\phi)$, modulo an ordinary (angle independent) time shift $\propto \beta_0^{\rm VV}$. 
Finally,  let us mention that the choice of spatial origin in the MPM waveform is encoded in various contributions to the radiative moments. Primarily it is encoded in the computation of the source multipole moments (for instance, the presence of explicit $O(\eta^5)$, \cite{Blanchet:1996wx}, and  $O(\eta^7)$, \cite{Faye:2012we}, contributions to $I_{ij}$ corresponds to the choice of specific $O(\eta^5)$ and $O(\eta^7)$ shifts in the definition of the cm origin).
In addition, we note that the gauge terms $M^{W_1I_3}_{ij}$ and $M^{Y_1I_3}_{ij}$ in Eqs. \eqref{blocks} contain effective shifts of the spatial origin, modulo integration by parts.

It is, however, curious to note that we did not need to subtract the monopolar ($l=0$) part of $\beta^{\rm VV}$ to get coincidence (at our accuracy) between $W^{\rm EFT}$ and $W^{\rm MPM}$. In fact, our present EFT-MPM comparison has checked the presence of the monopolar component  $\beta_0^{\rm VV}$ in the MPM waveform both at the $p_\infty^2$ and $p_\infty^4$ level in the PN expansion of Eq. \eqref{beta0_exp} of $\beta_0^{\rm VV}$.

The hidden presence of a $\beta^{\rm VV}_0$ time shift in MPM seems to be related to several of the nonlinear corrections to $I_{ij}^{(2)}$, notably the MPM gauge contributions displayed in Eqs.
\eqref{blocks_0}. For instance, we note that the $l=0$ gauge multipole $W$ entering 
\beq
M^{W_0I_2}_{ij}(t_r)= W^{(2)}(t_r) I^{}_{ij}(t_r)- W^{(1)}(t_r) I^{(1)}_{ij}(t_r)
\eeq
is such that, when $t_r\to -\infty$,   $W^{(1)}(t_r)$ has a finite limit, namely
\beq
\lim_{t_r\to -\infty}W^{(1)}(t_r)=\frac13 M\nu p_\infty^2\,,
\eeq
while $W^{(2)}(t_r)=O(1/t^2)$. This implies that the second time derivative of $M^{W_0I_2}_{ij}$ (which contributes to $U_{ij}$) contains a contribution proportional to $-\frac13 G M \nu^2 p_\infty^2 \dot U_{ij}$. Such a term is equivalent to monopolar time shift of $U_{ij}$, $\delta t_{l=0}\propto G M \nu^2 p_\infty^2$.
The latter monopolar time shift is of the type of the leading PN order contribution to $\beta^{\rm VV, can}_0$. 

We could define an effective monopolar shift (beyond the one encoded in the fraction $11/12$ in the quadrupolar tail) by writing that the total corrections $\Delta I_{ij}^{(2)}=U_{ij}-\overline{U}_{ij}^{\rm 2.5PN}-I_{ij}^{(2)}$ asymptotically behaves in the far past as
$\delta t\,  I_{ij}^{(3)}$.

We leave to future work an investigation of whether the full coefficient $\beta^{\rm VV, can}_0$ (involving $4/3$) can be so explained.

We end this section by displaying the explicit value of $\beta_2^{\rm VV}$, Eq. \eqref{beta_l_VV}, 
namely
\bea
\beta_2^{\rm VV}&=&\frac{5 G E   (3 E^2 - 9 m_1^2 - 9 m_2^2 - 8 P_{\rm cm}^2) }{24 P_{\rm cm}^2}\nonumber\\
&-& \frac{5G}{8P_{\rm cm}^3}\left[m_1^4 \ln \left(\frac{E_1-P_{\rm cm}}{E_1+P_{\rm cm}}\right)+m_2^4\ln \left(\frac{E_2-P_{\rm cm}}{E_2+P_{\rm cm}}\right)\right]  \,.\nonumber\\
\eea
Its PN-expanded form reads
\bea
\beta_2^{\rm VV}&=&GM\nu \left[\frac{2 p_\infty^2}{3}
+\left(\frac{\nu }{21}-\frac{5}{21}\right) p_\infty^4\right.\nonumber\\
&+&\left(-\frac{17 \nu ^2}{252}-\frac{13 \nu }{252}+\frac{5}{36}\right)p_\infty^6\nonumber\\
&+&\left(\frac{103 \nu^3}{1848}+\frac{31 \nu ^2}{462}+\frac{27 \nu }{616}-\frac{25}{264}\right) p_\infty^8\nonumber\\
&+&\left(-\frac{2839 \nu ^4}{64064}-\frac{2089 \nu ^3}{32032}-\frac{3785 \nu ^2}{64064}\right.\nonumber\\
&-&\left.\left.\frac{2335 \nu }{64064}+\frac{175}{2496}\right) p_\infty^{10}
+O(p_\infty^{10})
\right]\,.
\eea
All the higher multipolar contributions $\beta_l^{\rm VV}$ can similarly be easily computed.

\section{Concluding remarks}

We have extended the accuracy of the computation of the quadrupolar part of the time-domain gravitational waveform emitted during the scattering of two masses to the 3.5PN accuracy. At this accuracy the time domain waveform includes up to 3-loop contributions and must involve the 3.5PN radiation reacted hyperbolic motion.
We presented the explicit expressions of the radiation reacted part of the motion at the 3.5PN accuracy and as explicit functions of time at the 2-loop level, i.e. terms of order $O(G^2)$ and $O(G^3)$ in the orbit.

We explicitly evaluated the {\it frequency domain} value of the quadrupolar time domain waveform $h_{ij}(t_r)$ at the 2-loop level, i.e. $O(G^4)$ contributions to   
$\hat h_{ij}(\omega)$. In particular, the nonlinear memory contribution to the waveform has been explicitly computed in the cm frame for multipole orders $2\le l \le 6$ up to $O(p_\infty^{12})$, see Appendix B.

We presented several tests of our results: logarithmic structure, soft limit, first order self-force limit and, most importantly, comparisons with the EFT 1-loop waveform.
All those tests were successful. 
In particular, the comparison with the EFT waveform showed coincidence only  under the application an additional supertranslation 
of the EFT waveform, beyond the standard Veneziano-Vilkovisky one previously considered in the literature as being the necessary ingredient for shifting between the EFT BMS frame and the MPM BMS frame. Interestingly, the latter additional supertranslation consists in subtracting from the standard Veneziano-Vilkovisky supertranslations its dipolar part. We conjecture that the necessity of this subtraction (which corresponds to a simple change of cm spatial origin) will hold at all PN orders and for all multipolar components of the waveform. 

We leave to future work  an extension of the present work to other multipolar components of the bremsstrahlung waveform.

\section*{Acknowledgments}

The authors acknowledge informative discussions with L. Blanchet, S. de Angelis, G: Faye, J. Parra Martinez, P. Mastrolia and R. Roiban. 
T.D. acknowledges useful discussions with C. Heissenberg and R. Russo.
D.B. acknowledges membership to the Italian Gruppo Nazionale per
la Fisica Matematica (GNFM) of the Istituto Nazionale
di Alta Matematica (INDAM), as well as the hospitality
and the highly stimulating environment of the Institut
des Hautes Etudes Scient\'if\'iques.
A.G.  is grateful to the Istituto
per le Applicazioni del Calcolo \lq\lq M. Picone," CNR, Rome (IT) for
past support and hospitality during the development of
the present project.
The present research was partially supported by the
2021 Balzan Prize for Gravitation: Physical and Astrophysical Aspects, awarded to T. Damour.

\appendix

\section{The gauge moments}
\label{gauge_moments}

As a sample of the technical results entering the MPM formalism we give here details of the computation of the gauge moments needed at 3.5PN.
Let us briefly review the definition of the gauge moments $W_L,X_L,Y_L,Z_L$ given e.g., in Ref. \cite{Blanchet:2013haa}, and their consequent evaluation at the needed PN accuracy, 
\begin{widetext}
\bea\label{gaugemoments}
W_L(u)&=& \FP \int  d^3\mathbf{x}\,\widetilde r^B \int^1_{-1}  d
z\biggl\{ {2l+1\over (l+1)(2l+3)} \delta_{l+1} \hat{x}_{iL}
\overline{\Sigma}_i 
-
      {2l+1\over2c^2(l+1)(l+2)(2l+5)} \delta_{l+2}
      \hat{x}_{ijL} {\overline{\Sigma}}_{ij}^{(1)} \biggr\}\,,\nonumber\\
%%%%%%%%%%%
X_L(u) &=& \FP \int  d^3\mathbf{x}\,\widetilde r^B \int^1_{-1}  d
z\biggl\{ {2l+1\over 2(l+1)(l+2)(2l+5)} \delta_{l+2}
\hat{x}_{ijL} \overline{\Sigma}_{ij} \biggr\}\,,\nonumber\\
%%%%%%%%%%%
Y_L(u)&=&\FP \int  d^3\mathbf{x}\,\widetilde r^B \int^1_{-1} d
z\biggl\{ -\delta_l \hat{x}_L \overline{\Sigma}_{ii} + {3(2l+1)\over
  (l+1)(2l+3)} \delta_{l+1} \hat{x}_{iL} {\overline{\Sigma}}_i^{(1)}
- {2(2l+1)\over
  c^2(l+1)(l+2)(2l+5)} \delta_{l+2} \hat{x}_{ijL}
          {\overline{\Sigma}}_{ij}^{(2)} \biggr\}\,,\nonumber\\
%%%%%%%%%%%%
Z_L(u) &=& \FP \int  d^3\mathbf{x}\,\widetilde r^B \int^1_{-1}  d z \,
\epsilon_{ab \langle i_l}\biggl\{- {2l+1\over (l+2)(2l+3)}
\delta_{l+1} \hat{x}_{L-1 \rangle bc} \overline{\Sigma}_{ac} \biggr\}\,.
\eea
\end{widetext}
Here $\widetilde r^B=(r/r_0)^B$, $u$ denotes the retarded time (not to be confused with the frequency variable $u=\frac{\omega b}{p_\infty}$) and 
\beq
\delta_l(z)=\frac{(2l+1)!!}{2^{l+1}l!}(1-z^2)^l\,.
\eeq
The quantities $\bar \Sigma_L=\bar \Sigma_L ({\mathbf x}, u+\frac{zr}{c})$, etc. need to be PN expanded too (as denoted by an over bar on the symbol $\Sigma_L$), namely
e.g., 
\bea
\bar \Sigma_i ({\mathbf x}, u+\frac{zr}{c})&=& \bar \Sigma_i ({\mathbf x}, u)+\frac{zr}{c}\frac{d}{du}\bar \Sigma_i ({\mathbf x}, u)\nonumber\\
&+& \frac12 
\frac{z^2r^2}{c^2}\frac{d^2}{du^2}\bar \Sigma_i ({\mathbf x}, u)+O(\eta^3)\,.
\eea
Their definitions are listed in Eqs. (4.11) of Ref. \cite{Blanchet:1995fg} and briefly recalled below:
\bea
\bar \Sigma  
&=& \left(1+\frac{4}{c^2}V\right)\sigma+\eta^2{\mathcal G}+O(\eta^4)\,,  \nonumber\\
\bar \Sigma_i
&=& \left(1+\frac{4}{c^2}V\right)\sigma_i+\eta^2{\mathcal G}_i +O(\eta^4)\,,\nonumber\\
\bar \Sigma_{ij}
&=& \sigma_{ij}+{\mathcal G}_{ij}+O(\eta^2)\,,
\eea
where
\bea
{\mathcal G}&=& \frac{1}{\pi G}\partial_iV \partial_i V\,,\nonumber\\
{\mathcal G}_i&=& \frac{1}{\pi G}\left[\partial_k V (\partial_i V_k-\partial_k V_i)+\frac34 \partial_t V \partial_i V  \right]\,,\nonumber\\
{\mathcal G}_{ij}&=&\frac{1}{4\pi G}\left[\partial_i V\partial_jV -\frac12 \delta_{ij}\partial_k V \partial_k V \right] \,,
\eea
with the potentials 
\beq
V=\frac{Gm_1}{r_1}+\frac{Gm_2}{r_2}\,,\qquad V_i=\frac{Gm_1v_1^i}{r_1}+\frac{Gm_2v_2^i}{r_2}\,,
\eeq
needed here only at the Newtonian accuracy.

For completeness, let us recall the energy-momentum tensor $T^{\mu\nu}(x)$ of the two-body system
(see Eq. (2.18) of Ref. \cite{Blanchet:1995fg}), namely
\beq
T^{\mu\nu}(x)=\sum_{A=1}^2 m_A \frac{dy_A^\mu}{dt}\frac{dy_A^\nu}{dt}\frac{1}{\sqrt{-g}}\left(\frac{dt}{d\tau}\right)_A\delta({\mathbf x}-{\mathbf y}_A)\,.
\eeq
The latter (in contravariant components) is used to form the \lq\lq source variables," $\sigma_L=\sigma_L(t,{\rm x})$, 
\bea
\sigma&=&\frac{T^{00}+T^{ii}}{c^2}\,,\quad \sigma_i =\frac{T^{0i}}{c} \,,\quad
\sigma_{ij} =T^{ij}\,. 
\eea
At the 1PN accuracy level,  they are given by  
\bea
\sigma (t,{\rm x})&=& \sum_A \mu_A (t)\left(1+\frac{v_A^2}{c^2} \right)\delta({\bf x}-{\bf y}_A)\,,\nonumber\\
\sigma_i (t,{\rm x})&=& \sum_A \mu_A (t)v_A^i \delta({\bf x}-{\bf y}_A)\,,\nonumber\\
\sigma_{ij}(t,{\rm x}) &=& \sum_A \mu_A (t)v_A^iv_A^j \delta({\bf x}-{\bf y}_A)\,,
\eea
where 
\bea
\mu_A (t)&=& m_A+\frac{m_A}{c^2}\left(\frac12 v_A^2-U_A \right)\,,\nonumber\\
U_A  
&=& \frac{G\mu_B(t)}{r_{AB}}\left(1+\frac{v_B^2}{c^2}\right)\,,
\eea
having denoted $r_{AB}=|{\bf y}_A-{\bf y}_B|$.
Consequently,
\beq
\mu_A (t)=m_A+\frac{m_A}{c^2}\left(\frac12 v_A^2- \frac{m_B}{r_{AB}} \right)\,.
\eeq
For a later use let us also recall the 1PN form of the metric $g_{\alpha\beta}=\eta_{\alpha\beta}+h_{\alpha\beta}$ (see e.g. Eqs. (3.2) of Ref. \cite{Bini:2012gu}),  given by
\bea
g_{00}&=& -1+\frac{2}{c^2}V-\frac{2}{c^4}V^2\,,\nonumber\\
g_{0i}&=& -\frac{4}{c^3}V_i \,,\nonumber\\
g_{ij}&=& \delta_{ij}\left(1+\frac{2}{c^2}V\right)\,,
\eea
implying
\bea
h_{00}&=& \frac{2}{c^2}V-\frac{2}{c^4}V^2\,,\nonumber\\
h_{0i}&=&-\frac{4}{c^3}V_i\,,\quad
h_{ij}=\frac{2}{c^2}V \delta_{ij} \,, 
\eea
with
\bea
 h&=& 4 \frac{V}{c^2}+\frac{2}{c^4}V^2\,,\nonumber\\
{\rm det}(g)&=& {\sf g}=-1-\frac{4}{c^2}V-\frac{2}{c^4}V^2=-1-h\,,\nonumber\\
\sqrt{-{\sf g}}&=& 1+\frac{2}{c^2}V-\frac{1}{c^4}V^2 \,.
\eea
The passage to the  cm system is achieved through  
the relations
\bea
\label{CMdef}
{\bf y}_1&=& [X_2 + \eta^2 \nu  (X_1-X_2){\mathcal P}]{\bf x}\,,\nonumber\\
{\bf y}_2&=& [-X_1 + \eta^2 \nu (X_1-X_2){\mathcal P}]{\bf x}\,,
\eea
with
\beq
{\mathcal P}=\frac12 v^2 -\frac12 \frac{GM}{r}\,.
\eeq

Finally, one needs certain integral relations, e.g. the identity (see Ref. \cite{Damour:1990ji} and Eq. (4.22b) of Ref. \cite{Blanchet:1995fg}; see also Ref. \cite{Blanchet:1995fr} for an independent proof of this formula)
\bea
{\mathcal I}_L&=&{\rm FP}_{B=0}\int d^3x \widetilde r^B  
\frac{\hat x_L}{r_1r_2}\nonumber\\
&=& -2\pi \frac{|{\bf y}_1-{\bf y}_2|}{l+1}\sum_{p=0}^l y_1^{\langle L-P}y_2^{P\rangle}\,,
\eea
where $|{\bf y}_1-{\bf y}_2|=r_{12}$ and
\beq
y_2^{P}=y_2^{i_1}\ldots y_2^{i_p}\,,\qquad y_1^{L-P}=y_1^{i_{p+1}}\ldots y_1^{i_l}\,.
\eeq
For, example, for $l=1$  
\bea
{\mathcal I}_i &=& {\rm FP}_{B=0}\int d^3x \widetilde r^B  
\frac{ x_i}{r_1r_2}\nonumber\\
&=& - \pi \,r_{12} (y_1^i+y_2^i) \,;  
\eea
for $l=2$
\bea
{\mathcal I}_{ij}&=& {\rm FP}_{B=0}\int d^3x \widetilde r^B  
\frac{\hat x_{ij}}{r_1r_2}\nonumber\\
&=&- \frac{2\pi}{3}r_{12} ( \hat y_1^{ij}+y_1^{\langle i}y_2^{j\rangle}+\hat y_2^{ij})\,,%\nonumber\\
\eea
and for $l=3$
\bea
\label{x_3_indici}
{\mathcal I}_{ijk}&=&{\rm FP}_{B=0}\int d^3x \widetilde r^B %%%|x|^B
\frac{\hat x_{ijk}}{r_1r_2}\nonumber\\
&=& - \frac{\pi}{2}r_{12} (\hat y_1^{ijk}+y_1^{\langle ij}y_2^{k\rangle}+y_1^{\langle i}y_2^{jk\rangle}+\hat y_2^{ijk})\,. \qquad
\eea

Several identities are easily derived. For example, 
\beq
\frac{\partial}{\partial y_1^i}\frac{1}{r_1}=\frac{n_1^i}{r_1^2}=-\frac{\partial}{\partial x^i}\frac{1}{r_1}\,.
\eeq
Similarly, for for $r_{12}=|{\mathbf y}_1-{\mathbf y}_2|$ we find
\beq
\frac{\partial}{\partial y_1^i}r_{12}=\frac{y_1^i-y_2^i}{r_{12}}=n_{12}^i=-\frac{\partial}{\partial y_2^i}r_{12}\,,
\eeq
and
\beq
\frac{\partial}{\partial y_2^j}\frac{\partial}{\partial y_1^i}r_{12}=-\frac{1}{r_{12}}(\delta^{ij}-n_{12}^in_{12}^j)=-\frac{1}{r_{12}}P(n_{12})^{ij}\,,
\eeq
where $P(n_{12})^{ij}$ projects orthogonally to $n_{12}^i$.

Proceeding in successive steps, one needs to compute (at the Newtonian order level) the expressions of $\partial_t V$, $\partial_i V$, $\partial_{i}V\partial_j V$, $\partial_i V_k-\partial_kV_i$.

For their non singular parts (denoted below by the symbol $\sim$) we find
\bea
\partial_t V&=& \frac{Gm_1}{r_1^2}(n_1\cdot v_1)+\frac{Gm_2}{r_2^2}(n_2\cdot v_2)\nonumber\\
\partial_i V&=& -\frac{Gm_1}{r_1^2}n_1^i- \frac{Gm_2}{r_2^2}n_2^i\,,\nonumber\\
\partial_i V\partial_j V&=& \left(\frac{Gm_1}{r_1^2}n_1^i+\frac{Gm_2}{r_1^2}n_2^i  \right)\left(\frac{Gm_1}{r_1^2}n_1^j+\frac{Gm_2}{r_1^2}n_2^j  \right)\nonumber\\
&\sim & \frac{G^2 m_1m_2}{r_1^2 r_2^2}(n_1^i n_2^j+n_1^jn_2^i)\nonumber\\
&\sim & G^2 m_1 m_2 \left[\frac{\partial^2}{\partial y_1^i\partial y_2^j} +\frac{\partial^2}{\partial y_1^j\partial y_2^i} \right]\left( \frac{1}{r_1r_2}\right)\,,\nonumber\\
\eea
implying
\bea
\partial_{k}V\partial_k V 
&\sim & 2\frac{G^2 m_1m_2}{r_1^2 r_2^2}(n_1^k n_2^k)\nonumber\\
&\sim & 2 G^2 m_1 m_2  \frac{\partial^2}{\partial y_1^k\partial y_2^k}\left(\frac{1}{r_1r_2}\right)\,.
\eea
Finally
\bea
\partial_i V_k-\partial_k V_i &=& -Gm_1  \frac{(v_1^k n_1^i-v_1^i n_1^k)}{r_1^2}\nonumber\\
&-&  Gm_2  \frac{(v_2^k n_2^i-v_2^i n_2^k)}{r_2^2}\,.
\eea
Introducing the operator  notation
\beq
A_{ij}=\frac{\partial^2}{\partial y_1^i\partial y_2^j}\,,
\eeq
together with 
\bea
\label{calD_notation}
{\mathcal D}_{ij}
&=& A_{ij}+A_{ji}-\delta_{ij}A_{kk}\,,\nonumber\\
{\mathcal D}_i &=& v_2^k A_{ki} + v_1^k A_{ik}- (v_1^i+v_2^i) A_{kk}\nonumber\\
&-& \frac34 v_1^k A_{ki}-\frac34 v_2^k A_{ik}\,,
\eea
implying
\bea
{\mathcal G}_{ij}&=&\frac{Gm_1m_2}{4\pi}{\mathcal D}_{ij}\left(\frac{1}{r_1r_2}\right)\,,\nonumber\\
{\mathcal G}_i&=&  \frac{G m_1m_2}{\pi}{\mathcal D}_{i}\left(\frac{1}{r_1r_2}\right)\,,
\eea
we get
\bea
\label{all_cal_G}
{\mathcal G}&=& \frac{2 G m_1m_2}{\pi}A_{kk}\left(\frac{1}{r_1 r_2}\right)\,,
\eea

For the purposes of the present paper we need to compute (at the Newtonian level) the following gauge moments
\bea
W_{ij}
&=&  \frac{5}{21}\, {\rm FP}_{B=0}\int d^3 x \tilde r^B \hat x^{aij} \sigma_a\,,  \nonumber\\
W_i 
&=& \frac{3}{10} \, {\rm FP}_{B=0}\int d^3 x \tilde r^B \hat x^{ai}\sigma_a\,,
\eea
as well as
\bea
X  
&=& \frac{1}{20}{\rm FP}_{B=0}\int d^3 x \tilde r^B \hat x^{ij}(\sigma_{ij}+{\mathcal G}_{ij})\,,
\eea
\bea
Y_{ij}
&=& -{\rm FP}_{B=0}\int d^3 x \tilde r^B \hat x^{ij}(\sigma_{aa}+{\mathcal G}_{aa})\nonumber\\
&+&\frac57 \frac{d}{du}\, {\rm FP}_{B=0}\int d^3 x \tilde r^B \hat x^{aij}\sigma_a\,, \nonumber\\
Y_i
&=& -{\rm FP}_{B=0}\int d^3 x \tilde r^B \hat x^{i}(\sigma_{aa}+{\mathcal G}_{aa})\nonumber\\
&+&\frac{9}{10} \frac{d}{du}\,  {\rm FP}_{B=0}\int d^3 x \tilde r^B  \hat x^{ai}\sigma_a \,,
\eea
and 
\bea
Z_i
&=&
-\frac15 \epsilon_{abi}\, {\rm FP}_{B=0}\int d^3 x \tilde r^B \hat x^{bc}(\sigma_{ac}+{\mathcal G}_{ac})\,.
\eea
For each integral (say $Q$) in the computation we will distinguish the $\sigma$-part ($Q^{(1)}$) and the ${\mathcal G}$-part ($Q^{(2)}$). 

Using the notation \eqref{calD_notation} for the gauge moments $X,Y_i,Y_{ij},Z_i$ we can write
\bea
X&=&   \frac{1}{20}{\rm FP}_{B=0}\int d^3 x \tilde r^B \hat x^{ij} \sigma_{ij}\nonumber\\ 
&+& \frac{1}{20}\frac{Gm_1m_2}{4\pi}{\mathcal D}_{ij}({\mathcal I}_{ij})\,,
\eea
\bea
Y_{ij} 
&=& -{\rm FP}_{B=0}\int d^3 x \tilde r^B \hat x^{ij} \sigma_{aa} \nonumber\\
 &+&\frac57 \frac{d}{du}\, {\rm FP}_{B=0}\int d^3 x \tilde r^B \hat x^{aij}\sigma_a -\frac{Gm_1m_2}{4\pi}{\mathcal D}_{aa}({\mathcal I}_{ij})\nonumber\\
Y_i 
&=& -{\rm FP}_{B=0}\int d^3 x \tilde r^B \hat x^{i} \sigma_{aa}  \nonumber\\
&+&\frac{9}{10} \frac{d}{du}\,  {\rm FP}_{B=0}\int d^3 x \tilde r^B  \hat x^{ai}\sigma_a 
-\frac{G m_1 m_2}{4\pi}{\mathcal D}_{aa}({\mathcal I}_{i}) \,,\nonumber\\
\eea
\bea
Z_i&=& 
-\frac15 \epsilon_{abi}\, {\rm FP}_{B=0}\int d^3 x \tilde r^B \hat x^{bc} \sigma_{ac} \nonumber\\
&-& \frac15 \frac{Gm_1m_2}{4\pi}\epsilon_{abi}\,{\mathcal D}_{ac}({\mathcal I}_{bc})\,.
\eea
This leads to the following explicit results for $W_{ij}$ and $W_i$
\bea
W_{ij}&=& \frac{5}{21}M\nu (1-3\nu) r^3 [\dot r \hat n^{ij}-\frac25 n^{\langle i}v^{j\rangle}]  \nonumber\\
W_i &=&  -\frac{3}{10}M\nu X_{12}r^2 [\dot r n^i-\frac13 v^i]\,.
\eea
In addition,
\bea
X
&=& \frac{1}{20}M\nu (1-3\nu)r^2 \left(\dot r^2-\frac13 v^2\right)\nonumber\\
&+&\frac{1}{20}\frac{Gm_1m_2}{4\pi}{\mathcal D}_{ij}({\mathcal I}_{ij}) \nonumber\\
Y_{ij} 
&=& -M\nu (1-3\nu)\left[\hat x^{ij}v^2-\frac57 \frac{d}{du}(\hat x^{aij}v_a)\right]\nonumber\\
&-&\frac{Gm_1m_2}{4\pi}{\mathcal D}_{aa}({\mathcal I}_{ij})\nonumber\\
Y_i 
&=& -M\nu X_{12}v^2 x^i -\frac{9}{10}M\nu X_{12}\frac{d}{du}\left(r^2(\dot r n^i-\frac13 v^i) \right)\nonumber\\
&-& \frac{G m_1 m_2}{4\pi}{\mathcal D}_{aa}({\mathcal I}_{i})\nonumber\\
Z_i
&=& -\frac15 M\nu (1-3\nu)r^2 \dot r \epsilon_{abi}v^an^b\nonumber\\
&-&\frac15 \frac{Gm_1m_2}{4\pi}\epsilon_{abi}\,{\mathcal D}_{ac}({\mathcal I}_{bc})\,,
\eea
where
\bea
&&\frac{d}{du}\hat x^{aij}v_a  
=r^2 \left[\left(v^2-\frac35 \frac{GM}{r} \right) \hat n^{ij}+\frac65  \dot r v^{\langle i}n^{j\rangle}-\frac25  \hat v^{ij}\right]\,,\nonumber\\
&&\frac{d}{du}\left(r^2(\dot r n^i-\frac13 v^i) \right) = r \left[\left(v^2 -\frac23 \frac{GM}{r}\right)n^i +\frac13  \dot r v^i\right]\,.\nonumber\\
\eea
Further using the following cm, Newtonian level relations
\bea
{\mathcal D}_{ij}({\mathcal I}_{ij})&=& \frac83 (\nu-2)\pi r\,,\nonumber\\
{\mathcal D}_{aa}({\mathcal I}_{ij})&=& -2\pi r (1-2\nu) \hat n^{ij}\,,\nonumber\\
{\mathcal D}_{aa}({\mathcal I}_{i})&=& 2\pi X_{12}n^i\,,\nonumber\\
\epsilon_{abi}\,{\mathcal D}_{ac}({\mathcal I}_{bc})&=& 0\,,
\eea
we finally get the following explicit results for most of the  remaining needed gauge moments.
Finally,
\bea
X
&=& \frac{M\nu r^2}{20}\left[ (1-3\nu)\dot r^2 -\frac{1-3\nu}{3}v^2 -\frac{2(2-\nu)}{3}\frac{GM}{r}\right]\,,\nonumber\\
Y_{ij} 
&=& \frac{M\nu r^2}{7} \left[-2(1-3\nu)\hat v^{ij}    +6(1-3\nu)\dot r v^{\langle i}n^{j\rangle}\right.\nonumber\\
&+&\left. \left(\frac{1+4\nu}{2}\frac{GM}{r}-2v^2(1-3\nu) \right)\hat n^{ij}\right]\,,\nonumber\\
Y_i 
&=& \frac{1}{10}M\nu X_{12} r \left[ \left( v^2+ \frac{GM}{r} \right)n^i-3\dot r v^i\right] \,,\nonumber\\
Z_i
&=& -\frac15 M\nu (1-3\nu)r^2 \dot r \epsilon_{abi}v^an^b \,,  
\eea
where  $Z_i$ does not get contributions from the ${\mathcal G}_{ab}$.

The last gauge moment  to determine is $W$, which  should be obtained at 1PN level of accuracy.
Let us start from its definition
\bea
W&=&{\rm FP}_{B=0}\int d^3x \tilde r^B \int_{-1}^1 dz \left(\frac13 \delta_1 x^i \Sigma_i({\mathbf x}, u+\frac{zr}{c})\right.\nonumber\\
&-&\left. \frac{1}{20}\eta^2 \delta_2 \hat x^{ij}\Sigma_{ij}^{(1)}({\mathbf x}, u+\frac{zr}{c}) \right)\,,
\eea
where $\Sigma_i, \Sigma_{ij}$ are functions of $({\mathbf x}, u+\frac{zr}{c})$.
Since we need to be at 1PN,   $\Sigma_i$ should be PN expanded:
\bea
\Sigma_i({\mathbf x}, u+\frac{zr}{c})&=&\Sigma_i({\mathbf x},u)+\frac{zr}{c}\Sigma_i^{(1)}({\mathbf x},u)\nonumber\\
&+& \frac12 \frac{z^2r^2}{c^2}\Sigma_i^{(2)}({\mathbf x},u)+O(\eta^3)\,,
\eea
and to simplify notation we do not display the arguments of $\Sigma_i, \Sigma_{ij}$ when they are $({\mathbf x},u)$.
The integration in $z$ kills $\Sigma_i^{(1)}$ since
\beq
\int_{-1}^1 z \delta_1 dz=0\,,\qquad \int_{-1}^1 z^2 \delta_1 dz=\frac15\,.
\eeq
Therefore
\bea
W&=& {\rm FP}_{B=0}\int d^3x \tilde r^B  \left[\frac13   x^i \left(\bar \Sigma_i+\frac{1}{10}r^2 \eta^2 \bar \Sigma_i^{(2)}\right)\right.\nonumber\\
&-&\left. \frac{1}{20}\eta^2 \hat x^{ij}\bar \Sigma_{ij}^{(1)} \right]\,.
\eea
In turn, $\bar \Sigma_i,\bar \Sigma_{ij}$ are related to the sources $\sigma_i, \sigma_{ij}$ via the 1PN relations
\bea
\Sigma_i&=& \sigma_i (1+4 V\eta^2)+\eta^2 {\mathcal G}_i\nonumber\\
\Sigma_{ij}&=& \sigma_{ij}+{\mathcal G}_{ij}\,,
\eea
namely
\begin{widetext}
\bea
W&=& {\rm FP}_{B=0}\int d^3x \tilde r^B \left[\frac13   x^i \left(\sigma_i (1+4 V\eta^2)+\eta^2 {\mathcal G}_i+\frac{1}{10}r^2 \eta^2 \sigma_i^{(2)}\right)-\frac{1}{20}\eta^2 \hat x^{ij}(\sigma_{ij}^{(1)}+{\mathcal G}_{ij}^{(1)}) \right]\nonumber\\
&=& W^{(\sigma)}+W^{(g)} \,,
\eea
with
\bea
W^{(\sigma)}&=&  {\rm FP}_{B=0}\int d^3x \tilde r^B \left[\frac13   x^i \left(\sigma_i (1+4 V\eta^2)+\frac{1}{10}r^2 \eta^2 \sigma_i^{(2)}\right)-\frac{1}{20}\eta^2 \hat x^{ij} \sigma_{ij}^{(1)} \right]\nonumber\\
&=& \frac13    {\rm FP}_{B=0}\int d^3x \tilde r^B x^i  \sigma_i (1+4 V\eta^2)
+\frac{1}{30} \eta^2 \frac{d^2}{du^2} {\rm FP}_{B=0}\int d^3 \tilde r^B  r^2 x^i\sigma_i
-\frac{1}{20} \eta^2\frac{d}{du} {\rm FP}_{B=0}\int d^3x \tilde r^B   \hat x^{ij} \sigma_{ij} \nonumber\\
&\equiv & W^{(\sigma a)}+W^{(\sigma b)}+W^{(\sigma c)}\nonumber\\
W^{(g)}&=&  \eta^2 {\rm FP}_{B=0}\int d^3x \tilde r^B \left[\frac13   x^i   {\mathcal G}_i  -\frac{1}{20} \hat x^{ij} {\mathcal G}_{ij}^{(1)}  \right]\nonumber\\
&=&   \frac13   \eta^2 {\rm FP}_{B=0}\int d^3x \tilde r^B x^i   {\mathcal G}_i  
 -\frac{1}{20} \eta^2\frac{d}{du} {\rm FP}_{B=0}\int d^3x \tilde r^B  \hat x^{ij} {\mathcal G}_{ij} \nonumber\\
&\equiv &  W^{(g a)}+W^{(g b)}\,.
\eea
In the term $W^{(\sigma a)}$  one has to use 1PN expression for $\sigma_i$ so that
\bea
\sigma_i (1+4 V\eta^2)&=& m_1v_1^i \left[1+\eta^2 \left(\frac12 v_1^2-V_1\right)\right](1+4 V_1\eta^2)\delta({\mathbf x}-{\mathbf y}_1)+1\to 2\nonumber\\
&=& m_1v_1^i \left[1+\eta^2 \left(\frac12 v_1^2+3 V_1\right)\right] \delta({\mathbf x}-{\mathbf y}_1)+1\to 2\,,
\eea
\end{widetext}
with 
\bea
V_1&=&\frac{Gm_2}{r_{12}}+O(\eta^2)\,,\quad
V_2=\frac{Gm_1}{r_{12}}+O(\eta^2).\qquad
\eea
Therefore, 
\bea
W^{(\sigma a)}&=&\frac13 y_1^i m_1 v_1^i\left[1+\eta^2 \left(\frac12 v_1^2+3 V_1\right)\right]+1\to 2 \nonumber\\
&=&\frac13 m_1 y_1^i v_1^i + \frac{\eta^2}{3} m_1 y_1^i v_1^i\left(\frac12 v_1^2+3 V_1\right)\nonumber\\
&+&
\frac13 m_2 y_2^i v_2^i  +\frac{\eta^2}{3} m_2 y_2^i v_2^i\left(\frac12 v_2^2+3 V_2\right)\,.
\eea
Passing to the cm in the $O(\eta^0)$ terms should be done at 1PN   by using Eqs. \eqref{CMdef}.
Consequently,
\bea
W^{(\sigma a)} &=& \frac13 M\nu r\dot r \nonumber\\
&+& M\nu r \dot r \left(\frac16  v^2(1-3\nu)  +   \frac{ G M}{r}(1 -2 \nu ) \right)\eta^2\,.\nonumber\\
\eea
Let us compute now at the Newtonian level
\bea
W^{(\sigma b)}&=& \frac{1}{30} \eta^2 \frac{d^2}{du^2} {\rm FP}_{B=0}\int d^3 \tilde r^B  r^2 x^i \sigma_i\,,\nonumber\\
W^{(\sigma c)}&=& -\frac{1}{20} \eta^2\frac{d}{du} {\rm FP}_{B=0}\int d^3x \tilde r^B   \hat x^{ij} \sigma_{ij} \,.
\eea
We find
\bea
{\rm FP}_{B=0}\int d^3 x \tilde r^B  r^2 x^i \sigma_i
&=& M\nu (1-3\nu)r^3 \dot r
+ O(\eta^2)\,,\nonumber\\
{\rm FP}_{B=0}\int d^3x \tilde r^B   \hat x^{ij} \sigma_{ij}
&=&  M\nu (1-3\nu) r^2 \left(\dot r^2-\frac13 v^2 \right)\nonumber\\
&+& O(\eta^2)\,.
\eea
Recalling that at the Newtonian level $r\ddot r=v^2-\dot r^2-\frac{GM}{r}$, we find
\beq
W^{(\sigma b)}+W^{(\sigma c)}=-\frac16 M\nu r\dot r (1-3\nu)\eta^2 \left(\frac{GM}{r}-\frac45 v^2\right)\,.
\eeq
Therefore
\bea
W^{(\sigma)}
&=& \frac13 M\nu r \dot r \nonumber\\
&+& \eta^2 M\nu r \dot r \left[\frac{3}{10}(1-3\nu)v^2 +\frac12 \left(\frac53-3\nu \right)\frac{GM}{r} \right]\,.\nonumber\\
\eea
Let us consider now $W^{(g a)}$, recalling Eqs. \eqref{all_cal_G} as well as \eqref{calD_notation}
\bea
W^{(g a)}&=& \frac13   \eta^2 {\rm FP}_{B=0}\int d^3x \tilde r^B x^i   {\mathcal G}_i  \nonumber\\
&=& \frac13  \frac{1}{\pi G  } \eta^2 {\rm FP}_{B=0}\int d^3x \tilde r^B x^i  \left[\partial_k V (\partial_i V_k-\partial_k V_i)\right.\nonumber\\
&+&\left.\frac34 \partial_t V \partial_i V  \right]\nonumber\\
&=& \frac13  \frac{1}{\pi G  } \eta^2 G^2 m_1m_2 {\mathcal D}_{i}\,  {\rm FP}_{B=0}\int d^3x \tilde r^B \frac{x^i}{r_1r_2}\nonumber\\
&=& M\nu \frac{GM}{r} r\dot r \eta^2 \left(\frac12 +\frac83 \nu \right)\,.
\eea

Let us consider now $W^{(g b)}$
\bea
W^{(g b)}&=&  
 -\frac{1}{20} \eta^2\frac{d}{du} {\rm FP}_{B=0}\int d^3x \tilde r^B  \hat x^{ij} {\mathcal G}_{ij} \,,
\eea
with
\bea
{\rm FP}_{B=0}\int d^3x \tilde r^B  \hat x^{ij} {\mathcal G}_{ij}
&=&\frac{Gm_1m_2}{4\pi}{\mathcal D}_{ij} {\rm FP}_{B=0}\int d^3x \tilde r^B  \frac{\hat x^{ij}}{r_1r_2}\nonumber\\
&=& \frac{Gm_1m_2}{4\pi}{\mathcal D}_{ij}{\mathcal I}^{ij}\,,
\eea
where ${\mathcal D}_{ij}$ has been defined in Eq. \eqref{calD_notation}.
We find
\beq
W^{(g b)}=-\frac{1}{30}M\nu\eta^2  r\dot r \frac{ G M}{r}  (-2 + \nu) \,,
\eeq
and then
\bea
W^{(g)}&=&W^{(g a)}+W^{(g b)}\nonumber\\
&=& \nu M \eta^2 \frac{GM}{r}   r \dot r  \frac{17+79\nu}{30}\,.
\eea
Finally
\bea
W&=& \frac13 M\nu r \dot r \nonumber\\
&+&\frac15 \eta^2 M\nu r \dot r \left( \frac{21+17\nu}{3} \frac{GM}{r}+\frac{3}{2}(1-3\nu)v^2\right)\,.\nonumber\\
\eea 

The 2PN-accurate values of these gauge moments have  been recently presented in Appendix D of  Ref. \cite{Blanchet:2026suq}.

\section{Nonlinear memory in the frame of body A vs cm frames}

As a starting point we computed (from the explicit tree-level time-domain waveform \cite{Kovacs:1977uw,Kovacs:1978eu})
the angular distribution of the time-integrated gravitational-wave radiation in the frame of body A, up to the fourth multipolar level, i.e.:
\bea
\frac{dE^{\rm gw}(\Omega)}{d\Omega}\bigg|_{A}= \sum_{l\ge 0} E_l^{\rm gw}|_{\rm A}\,,
\eea
where
\bea
E_{l}^{\rm gw}|_{\rm A}=\sum_{m=-l}^l E_{lm}^{\rm A} Y_{lm}(\theta_A,\phi_A)\,.
\eea
In this $A$-frame multipolar expansion, the first two terms (monopolar and dipolar) respectively encode the total
$A$-frame radiated energy, 
\beq
E^{\rm gw}_{l=0}\bigg|_{A}=\frac{E_0^{\rm gw, A}}{4 \pi},
\eeq
and the total radiated linear momentum,
\beq
E^{\rm gw}_{l=1}\bigg|_{A}= \frac{3}{4\pi} P^{\rm gw, A}_i  n^i\,,
\eeq
where the unit vector $n^i$ is the direction of gravitational wave emission.
We recall the result
\beq
P^\mu_{\rm gw\,3PM}=\frac{G^3 m_1^2 m_2^2}{b^3}\frac{{\mathcal E}(\gamma)}{\gamma+1}(u_1^\mu+u_2^\mu)\,,
\eeq
where the PN-exact value of ${\mathcal E}(\gamma)$ has been first computed in \cite{Herrmann:2021tct} (see also 
Eq. (5.19) of Ref. \cite{Bini:2021gat} and Eq. (D15) of Ref. \cite{Bini:2024ijq} for it high PN expansion).
Explicitly, the monopolar and dipolar parts of A-frame emitted energy reads
\beq
\frac{dE^{l\le 1}_{\rm gw}(\Omega)}{d\Omega}\bigg|_{A}= \frac{1}{4\pi}  \frac{G^3 m_1^2 m_2^2}{b^3} {\mathcal E}(\gamma) \left(1+3\sqrt{\frac{\gamma-1}{\gamma+1}}{\mathbf n}\cdot \hat {\mathbf V} \right)\,,
\eeq
where $\hat V^\mu$ is the following unit 4-vector
\beq
\hat V^\mu = \frac{u_2^\mu-\gamma u_1^\mu}{\sqrt{\gamma^2-1}}\,.
\eeq

The next step is  the transformation  from the frame of body A to the cm frame. We have the following relations~\footnote{These relations are PM-exact and do not contain $G$.}, see Ref. \cite{Bini:2024tft}, between the  retarded times $t_r^A$, $t_r^{\rm cm}$ 
and the quantities 
\bea
\alpha_A&=&\cos\theta_A\,,\nonumber\\
\beta_A &=& \sin\theta_A \cos\phi_A\,,\nonumber\\
\delta_A &=&  \sin\theta_A \sin\phi_A\,,
\eea
and their  cm-frame analogs:
\bea
t_r^A&=&f_{\rm cm}  t_r^{\rm cm}-\beta_A b_A\,,\nonumber\\
\alpha_A&=&\frac{\alpha_{\rm cm}+v_{\rm cm}}{1+\alpha_{\rm cm}v_{\rm cm}}\,,\nonumber\\
\phi_A&=& \phi_{\rm cm}\,,
\eea
where
\bea
f_{\rm cm}&=&\gamma_{\rm cm}(1-\alpha_A v_{\rm cm})=\frac{1}{\gamma_{\rm cm}(1+\alpha_{\rm cm} v_{\rm cm})}\,.
\eea
Note that the polar coordinates used in the A frame uses as $z$-axis the A-frame relative velocity vector $\hat V$.
However, after transforming to the cm frame, we will use as $z$-axis our usual one, i.e. orthogonal to the plane of motion.
 
Moreover,
\beq
b_A=\frac{E_2}{E}b\,,
\eeq
and
\bea
v_{\rm cm}&=& \frac{m_2 \gamma v}{m_1+m_2\gamma}\,,\quad
\gamma_{\rm cm}=\frac{E_1}{m_1}=\frac{m_1+\gamma m_2}{E}\,.\qquad
\eea
Starting from  Eq. 3.20 of Ref. \cite{Bini:2024tft}, namely
\beq
\label{B13}
\frac{dE^{\rm gw}_{\rm cm}(t_r^{\rm cm},\Omega_{\rm cm})}{dt_r^{\rm cm}d\Omega_{\rm cm}}  dt_r^{\rm cm} d\Omega_{\rm cm}
=f_{\rm cm} \frac{dE^{\rm gw}_{A}(t_r^{A},\Omega_{A})}{dt_r^{A}d\Omega_{A}} dt_r^{A} d\Omega_{A}\,,
\eeq
and integrating over the retarded times gives 
\bea
\label{B14}
\frac{dE^{\rm gw}_{\rm cm}(\Omega_{\rm cm})}{d\Omega_{\rm cm}} d\Omega_{\rm cm}
&=& f_{\rm cm} \frac{dE^{\rm gw}_{A}(\Omega_{A})}{d\Omega_{A}} d\Omega_{A}\nonumber\\
&=& f_{\rm cm}^3 \frac{dE^{\rm gw}_{A}(\Omega_{A})}{d\Omega_{A}}  d\Omega_{\rm cm}\,,
\eea
where we used 
\beq
d\Omega_A=f_{\rm cm}^2 d\Omega_{\rm cm}\,.
\eeq
Note that the last Eq. \eqref{B14} gives a simple, explicit relation between the angular spectrum in the A-frame and the angular spectrum in the cm-frame.

Using these relations we computed the angular spectrum $\frac{dE^{\rm gw}_{\rm cm}(\Omega_{\rm cm})}{d\Omega_{\rm cm}}$ as function of our usual cm-frame polar coordinates, i.e.,
\beq
\label{cm_new}
\alpha_{\rm cm}=\sin\theta_{\rm cm}\sin\phi_{\rm cm},\quad \beta_{\rm cm}=\sin\theta_{\rm cm}\cos\phi_{\rm cm}\,.
\eeq
We computed the cm-frame multipolar decomposition
\bea
\frac{dE^{\rm gw}(\Omega)}{d\Omega}\bigg|_{\rm cm}= \sum_{l\ge 0} \sum_{m=-l}^l E_{lm}^{\rm cm} Y_{lm}(\theta_{\rm cm},\phi_{\rm cm})\,,
\eea
as functions of $\theta_{\rm cm}$ and $\phi_{\rm cm}$ defined in Eq. \eqref{cm_new}. 

We computed $E_{lm}^{\rm cm}$ for $2\le l\le 6$ as a power series expansion in the relative velocity $v=\frac{p_\infty}{\sqrt{1+p_\infty^2}}$,
\beq
E_{l}^{\rm cm}=\frac{G^3m_1^2m_2^2}{b^3}\sum_{k= 1}^{11} \sum_{m=-l}^2E ^{{\rm cm} v^k}_{lm} v^k Y_{lm}(\theta_{\rm cm},\phi_{\rm cm})\,.
\eeq
In Tables \ref{tab1:Ecm}-\ref{tab1:Ecml6} we display the nonzero expansion coefficients $E_{lm}^{{\rm cm} v^k}$.  
$E_{lm}^{{\rm cm} v^k}$ vanishes when  $l+m$ is odd and $k+l$ is even.  
Moreover, $E_{l\bar m}^{{\rm cm} v^k}$ is not shown because it is the complex conjugate of   $E_{lm}^{{\rm cm} v^k}$ and  $E_{lm}^{{\rm cm} v^k}$ is either real or purely imaginary.
These results  allow one to immediately compute the corresponding contributions to the nonlinear memory by using Eq. \eqref{8.12} (see Ref. \cite{Blanchet:1992br}). 

\begin{table*}
\caption{\label{tab1:Ecm} The various nonzero multipolar coefficients of $E^{{\rm cm}v^k}_{lm}$  in an expansion in powers of $v=\frac{p_\infty}{\sqrt{1+p_\infty^2}}$, evaluated for $l=2$ and all $m$  corresponding to nonzero values. The highest power of $v$ included here is $v^{11}$. 
}
\begin{ruledtabular}
\begin{tabular}{ll}
$E^{{\rm cm}v^1}_{22}$ &  $\frac17 \sqrt{\frac{2\pi}{15}}  $\\
$E^{{\rm cm}v^1}_{20}$ &  $\frac{73}{42} \sqrt{\frac{\pi}{5}} $\\
\hline
$E^{{\rm cm}v^3}_{22}$ &  $\sqrt{\frac{\pi }{30}} \left(\frac{253 \nu }{32}-\frac{2425}{448}\right)$\\
$E^{{\rm cm}v^3}_{20}$ &  $\sqrt{\frac{\pi }{5}} \left(\frac{545 \nu }{168}+\frac{173}{112}\right)$\\
\hline
$E^{{\rm cm}v^5}_{22}$ &  $\sqrt{\frac{\pi }{30}} \left(-\frac{12007 \nu ^2}{2464}+\frac{15055 \nu }{1232}-\frac{37189}{7392}\right)$\\
$E^{{\rm cm}v^5}_{20}$ &  $\sqrt{\frac{\pi }{5}} \left(-\frac{359 \nu ^2}{192}+\frac{48883 \nu }{8448}+\frac{18035}{12672}\right)$\\
\hline
$E^{{\rm cm}v^7}_{22}$ &  $\sqrt{\frac{\pi }{30}} \left(\frac{943505 \nu ^3}{256256}-\frac{6161559 \nu ^2}{512512}+\frac{5106931 \nu }{192192}-\frac{49820795}{6150144}\right)$\\
$E^{{\rm cm}v^7}_{20}$ &  $\sqrt{\frac{\pi }{5}} \left(\frac{523165 \nu ^3}{384384}-\frac{312451 \nu ^2}{64064}+\frac{52371623 \nu }{4612608}+\frac{3008189}{9225216}\right)$\\
\hline
$E^{{\rm cm}v^9}_{22}$ &  $\sqrt{\frac{\pi }{30}} \left(-\frac{389995 \nu ^4}{128128}+\frac{1773625 \nu ^3}{146432}-\frac{3466781 \nu ^2}{128128}+\frac{558360659 \nu
   }{12300288}-\frac{151626931}{12300288}\right)$\\
$E^{{\rm cm}v^9}_{20}$ &  $\sqrt{\frac{\pi }{5}} \left(-\frac{563677 \nu ^4}{512512}+\frac{4050245 \nu ^3}{878592}-\frac{32849965 \nu ^2}{3075072}+\frac{122634311 \nu }{6709248}-\frac{79970861}{73801728}\right)$\\
\hline
$E^{{\rm cm}v^{11}}_{22}$ &  $\sqrt{\frac{\pi }{30}} \left(\frac{184121491 \nu ^5}{69701632}-\frac{1729366757 \nu ^4}{139403264}+\frac{6359959343 \nu ^3}{209104896}-\frac{7131379795 \nu
   ^2}{139403264}+\frac{1426604723 \nu }{21446656}-\frac{7065028081}{418209792}\right)$\\
$E^{{\rm cm}v^{11}}_{20}$ &  $ \sqrt{\frac{\pi }{5}} \left(\frac{12288335 \nu ^5}{13069056}-\frac{477098315 \nu ^4}{104552448}+\frac{7199811893 \nu ^3}{627314688}-\frac{8203732039 \nu
   ^2}{418209792}+\frac{21664352711 \nu }{836419584}-\frac{248946721}{96509952}\right)$\\
\end{tabular}
\end{ruledtabular}
\end{table*}

Let us explicitly display the quadrupolar nonlinear memory ($l=2$). 
We use   the relations
\bea
{}_{\bar 2}Y_{22}(\theta_{\rm cm},\phi_{\rm cm})&=&\frac14 \sqrt{\frac{5}{\pi}}[{\bar m}_1^2+2i {\bar m}_1{\bar m}_2-{\bar m}_2^2]\,, \nonumber\\
{}_{\bar 2}Y_{20}(\theta_{\rm cm},\phi_{\rm cm})&=& -\frac12 \sqrt{\frac{15}{2\pi}}[{\bar m}_1^2+{\bar m}_2^2]\,,\nonumber\\
{}_{\bar 2}Y_{2\bar 2}(\theta_{\rm cm},\phi_{\rm cm})&=& \frac14 \sqrt{\frac{5}{\pi}}[{\bar m}_1^2-2i {\bar m}_1{\bar m}_2-{\bar m}_2^2]\,,
\eea
where
\bea
\label{choicebarms}
\bar m_1&=& \frac{1}{\sqrt{2}} (\cos \theta_{\rm cm} \cos\phi_{\rm cm} +i \sin \phi_{\rm cm})), \nonumber\\
\bar m_2&=&  \frac{1}{\sqrt{2}}  (\cos \theta_{\rm cm} \sin \phi_{\rm cm}  -i\cos\phi_{\rm cm}), \nonumber\\
\bar m_3&=&  - \frac{1}{\sqrt{2}} \sin \theta_{\rm cm}\,,
\eea
should not be confused with the masses $m_1$ and $m_2$.
Re-expressing the  PN expansion in terms of $p_\infty$ (up to $O(p_\infty^{13})$) the quadrupolar nonlinear memory in the cm frame can be presented in the form  
\bea
U_2^{\rm nonlin G^3} &=& \nu^2 \left(\frac{G M  }{b}\right)^3 \pi \sum_{k=0}^5 p_\infty^{2k+1}{\mathcal M}^{\rm nl}_{2k+1}({\bar m}_1,{\bar m}_2, \nu)\nonumber\\
&+& O(p_\infty^{13})\,, 
\eea
where the coefficients ${\mathcal M}^{\rm nl}_{k}({\bar m}_1,{\bar m}_2,\nu)$ are listed below:
\begin{widetext}
\bea
{\mathcal M}^{\rm nl}_{1}&=& -\frac{23}{28}  \bar m_1^2 - \frac{11}{12}  {\bar m}_2^2 \nonumber\\
{\mathcal M}^{\rm nl}_{3}&=& \left(-\frac{3397}{2688}  - \frac{409}{1344} \nu  \right)\bar m_1^2  
 + \left(\frac{527}{896}  - \frac{1317}{448} \nu  \right){\bar m}_2^2\nonumber\\
{\mathcal M}^{\rm nl}_{5}&=&\bar m_1^2 \left(\frac{19325}{29568} - \frac{2253}{5632} \nu + \frac{3629}{29568} \nu^2     \right) + 
{\bar m}_2^2 \left(-\frac{163}{396} - \frac{61529}{118272} \nu + \frac{ 17219}{9856} \nu^2  \right)\nonumber\\
{\mathcal M}^{\rm nl}_{7}&=&  
\bar m_1^2 \left(-\frac{6406691}{12300288}+ \frac{165205}{512512} \nu   + \frac{
    393725}{3075072} \nu^2  - \frac{34275}{512512} \nu^3 \right)\nonumber\\
&+&{\bar m}_2^2 \left(\frac{7378931}{5271552} - \frac{15189773}{4612608} \nu  
+ \frac{76521}{1025024} \nu^2  - \frac{1989835}{1537536} \nu^3  \right) \nonumber\\
{\mathcal M}^{\rm nl}_{9}&=& \bar m_1^2 \left(\frac{21646927}{49201152} - \frac{14065201}{49201152} \nu   
- \frac{235687}{1537536} \nu^2   - \frac{30565}{585728} \nu^3  + \frac{131051}{3075072} \nu^4   \right)\nonumber\\ 
&+& 
{\bar m}_2^2 \left(-\frac{32430845}{21086208} + \frac{8264959}{2342912} \nu   +  
    \frac{5984701}{3075072} \nu^2  + \frac{10965}{53248} \nu^3 +  \frac{3251011}{3075072} \nu^4  \right)\nonumber\\
{\mathcal M}^{\rm nl}_{11}&=&\bar m_1^2 \left(-\frac{6636573}{17425408}+\frac{143813107}{557613056} \nu
+\frac{22044585}{139403264} \nu^2+\frac{12835215}{139403264} \nu^3+\frac{6206693}{278806528} \nu^4
-\frac{960913 }{32169984}\nu^5
\right)\nonumber\\
&+& {\bar m}_2^2 \left(\frac{3720247871}{2509258752}-\frac{5570263849}{1672839168} \nu
-\frac{3807889}{1537536} \nu^2-\frac{1935699511}{1254629376} \nu^3
-\frac{341477447}{836419584} \nu^4
-\frac{18130231}{19914752} \nu^5
\right)
\,.
\eea
When $\nu \to 0$ we have checked that our result for the quadrupolar nonlinear memory  agrees with the PN expansion of the corresponding A-frame result of Ref.
\cite{Georgoudis:2025vkk}.

\begin{table*}
\caption{\label{tab1:Ecml3} The various nonzero multipolar coefficients of $E^{{\rm cm}v^k}_{lm}$  in an expansion in powers of $v=\frac{p_\infty}{\sqrt{1+p_\infty^2}}$, evaluated for $l=3$ and all $m$  corresponding to nonzero values.  
}
\begin{ruledtabular}
\begin{tabular}{ll}
$E^{{\rm cm}v^2}_{33}$ &  $ \frac{79}{288} i \sqrt{\frac{\pi }{35}} \sqrt{1-4 \nu } $\\
$E^{{\rm cm}v^2}_{31}$ &  $ \frac{653}{240} i \sqrt{\frac{\pi }{21}} \sqrt{1-4 \nu } $\\
\hline
$E^{{\rm cm}v^4}_{33}$ &  $ i \sqrt{\frac{\pi }{35}} \sqrt{1-4 \nu } \left(\frac{22049 \nu }{6336}-\frac{293197}{50688}\right)$\\
$E^{{\rm cm}v^4}_{31}$ &  $ i \sqrt{\frac{\pi }{21}} \sqrt{1-4 \nu } \left(\frac{2479 \nu }{960}+\frac{206633}{84480}\right)$\\
\hline
$E^{{\rm cm}v^6}_{33}$ &  $ i \sqrt{\frac{\pi }{35}} \sqrt{1-4 \nu } \left(-\frac{455329 \nu ^2}{164736}+\frac{3556165 \nu }{658944}-\frac{115727}{20592}\right)$\\
$E^{{\rm cm}v^6}_{31}$ &  $ i \sqrt{\frac{\pi }{21}} \sqrt{1-4 \nu } \left(-\frac{1117481 \nu ^2}{549120}+\frac{1053629 \nu }{219648}+\frac{351125}{109824}\right)$\\
\hline
$E^{{\rm cm}v^8}_{33}$ &  $i \sqrt{\frac{\pi }{35}} \sqrt{1-4 \nu } \left(\frac{734681 \nu ^3}{329472}-\frac{36090845 \nu ^2}{5271552}+\frac{10767013 \nu}{878592}-\frac{148376231}{21086208}\right)$\\
$E^{{\rm cm}v^8}_{31}$ &  $i \sqrt{\frac{\pi }{21}} \sqrt{1-4 \nu } \left(\frac{21845 \nu ^3}{13728}-\frac{48834587 \nu ^2}{8785920}+\frac{14487243 \nu}{1464320}+\frac{94656511}{35143680}\right)$\\
\hline
$E^{{\rm cm}v^{10}}_{33}$ &  $i \sqrt{\frac{\pi }{35}} \sqrt{1-4 \nu } \left(-\frac{168230203 \nu ^4}{89616384}+\frac{51609589 \nu ^3}{6893568}-\frac{1416696299 \nu^2}{89616384}+\frac{742451165 \nu }{32587776}-\frac{1153428515}{119488512}\right)$\\
$E^{{\rm cm}v^{10}}_{31}$ &  $i \sqrt{\frac{\pi }{21}} \sqrt{1-4 \nu } \left(-\frac{4868185 \nu ^4}{3734016}+\frac{168809489 \nu ^3}{29872128}-\frac{466455443 \nu^2}{37340160}+\frac{10215279229 \nu }{597442560}+\frac{69618979}{49786880}\right)$\\
\hline
\end{tabular}
\end{ruledtabular}
\end{table*}

\begin{table*}
\caption{\label{tab1:Ecml4} The various nonzero multipolar coefficients of $E^{{\rm cm}v^k}_{lm}$  in an expansion in powers of $v=\frac{p_\infty}{\sqrt{1+p_\infty^2}}$, evaluated for $l=4$ and all $m$  corresponding to nonzero values.  
}
\begin{ruledtabular}
\begin{tabular}{ll}
$E^{{\rm cm}v^1}_{44}$ &  $ -\frac{5}{96} \sqrt{\frac{5\pi}{14}}$\\
$E^{{\rm cm}v^1}_{42}$ &  $ \frac{1}{42} \sqrt{\frac{5\pi}{10}}$\\
$E^{{\rm cm}v^1}_{40}$ &  $ \frac{17}{560}\sqrt{\pi}$\\
\hline
$E^{{\rm cm}v^3}_{44}$ &  $\sqrt{\frac{\pi }{70}} \left(\frac{1013 \nu }{2112}-\frac{16321}{8448}\right)$\\
$E^{{\rm cm}v^3}_{42}$ &  $\sqrt{\frac{\pi }{10}} \left(\frac{8509 \nu }{2112}-\frac{89171}{59136}\right) $\\
$E^{{\rm cm}v^3}_{40}$ &  $\sqrt{\pi } \left(\frac{19843 \nu }{12320}-\frac{1861}{4480}\right)$\\
\hline
$E^{{\rm cm}v^5}_{44}$ &  $\sqrt{\frac{\pi }{70}} \left(\frac{30131 \nu ^2}{4224}-\frac{1022961 \nu }{36608}+\frac{57085}{9984}\right)$\\
$E^{{\rm cm}v^5}_{42}$ &  $\sqrt{\frac{\pi }{10}} \left(\frac{384413 \nu ^2}{192192}+\frac{122307 \nu }{64064}-\frac{59867}{48048}\right)$\\
$E^{{\rm cm}v^5}_{40}$ &  $ \sqrt{\pi } \left(\frac{761 \nu ^2}{1280}+\frac{56517 \nu }{33280}-\frac{136503}{366080}\right)$\\
\hline
$E^{{\rm cm}v^7}_{44}$ &  $\sqrt{\frac{\pi }{70}} \left(-\frac{792361 \nu ^3}{109824}+\frac{2273097 \nu ^2}{146432}-\frac{4837891 \nu }{146432}+\frac{373509}{53248}\right)$\\
$E^{{\rm cm}v^7}_{42}$ &  $\sqrt{\frac{\pi }{10}} \left(-\frac{3766261 \nu ^3}{1537536}+\frac{9239289 \nu ^2}{2050048}+\frac{436553 \nu }{157696}-\frac{1111637}{630784}\right)$\\
$E^{{\rm cm}v^7}_{40}$ &  $\sqrt{\pi } \left(-\frac{396457 \nu ^3}{512512}+\frac{13731639 \nu ^2}{10250240}+\frac{4975185 \nu }{2050048}-\frac{4509055}{8200192}\right)$\\
\hline
$E^{{\rm cm}v^9}_{44}$ &  $\sqrt{\frac{\pi }{70}} \left(\frac{17503399 \nu ^4}{2715648}-\frac{196759965 \nu ^3}{9957376}+\frac{178315755 \nu ^2}{4978688}-\frac{4956673363 \nu
   }{119488512}+\frac{22556821}{2715648}\right)$\\
$E^{{\rm cm}v^9}_{42}$ &  $\sqrt{\frac{\pi }{10}} \left(\frac{29041657 \nu ^4}{13069056}-\frac{18353655 \nu ^3}{2489344}+\frac{372385071 \nu ^2}{34850816}+\frac{12191183 \nu
   }{8042496}-\frac{12667267}{6534528}\right)$\\
$E^{{\rm cm}v^9}_{40}$ &  $\sqrt{\pi } \left(\frac{1116599 \nu ^4}{1584128}-\frac{2966181 \nu ^3}{1244672}+\frac{138681993 \nu ^2}{43563520}+\frac{419869753 \nu }{174254080}-\frac{54545377}{87127040}\right)$\\
\hline
$E^{{\rm cm}v^{11}}_{44}$ &  $\sqrt{\frac{\pi }{70}} \left(-\frac{6462415477 \nu ^5}{1135140864}+\frac{1843606357 \nu ^4}{79659008}-\frac{36616623269 \nu ^3}{756760576}+\frac{5062489397 \nu
   ^2}{70946304}-\frac{256615122113 \nu }{4540563456}+\frac{11540048199}{1100742656}\right)$\\
$E^{{\rm cm}v^{11}}_{42}$ &  $\sqrt{\frac{\pi }{10}} \left(-\frac{15433710293 \nu ^5}{7945986048}+\frac{90593025779 \nu ^4}{10594648064}-\frac{95018918347 \nu ^3}{5297324032}+\frac{340360736759 \nu
   ^2}{15891972096}-\frac{33139413665 \nu }{15891972096}-\frac{145175465583}{84757184512}\right)$\\
$E^{{\rm cm}v^{11}}_{40}$ &  $ \sqrt{\pi } \left(-\frac{156487549 \nu ^5}{254679040}+\frac{562547577 \nu ^4}{203743232}-\frac{7662159255 \nu ^3}{1324331008}+\frac{84547069619 \nu ^2}{13243310080}+\frac{42900059333
   \nu }{26486620160}-\frac{12333832361}{21189296128}\right)$\\
\hline
\end{tabular}
\end{ruledtabular}
\end{table*}

\begin{table*}
\caption{\label{tab1:Ecml5} The various nonzero multipolar coefficients of $E^{{\rm cm}v^k}_{lm}$  in an expansion in powers of $v=\frac{p_\infty}{\sqrt{1+p_\infty^2}}$, evaluated for $l=5$ and all $m$  corresponding to nonzero values.  }
\begin{ruledtabular}
\begin{tabular}{ll}
$E^{{\rm cm}v^2}_{55}$ &  $-\frac{25}{384} i \sqrt{\frac{7 \pi }{11}} \sqrt{1-4 \nu }$\\
$E^{{\rm cm}v^2}_{53}$ &  $\frac{7 i \sqrt{\frac{7 \pi }{55}} \sqrt{1-4 \nu }}{1152}$\\
$E^{{\rm cm}v^2}_{51}$ &  $\frac{59}{96} i \sqrt{\frac{\pi }{330}} \sqrt{1-4 \nu }$\\
\hline
$E^{{\rm cm}v^4}_{55}$ &  $i \sqrt{\frac{7 \pi }{11}} \sqrt{1-4 \nu } \left(-\frac{193 \nu }{9984}-\frac{18095}{39936}\right)$\\
$E^{{\rm cm}v^4}_{53}$ &  $i \sqrt{\frac{\pi }{385}} \sqrt{1-4 \nu } \left(\frac{333067 \nu }{29952}-\frac{950239}{119808}\right)$\\
$E^{{\rm cm}v^4}_{51}$ &  $i \sqrt{\frac{\pi }{330}} \sqrt{1-4 \nu } \left(\frac{5593 \nu }{384}-\frac{776101}{139776}\right)$\\
\hline
$E^{{\rm cm}v^6}_{55}$ &  $i \sqrt{\frac{7 \pi }{11}} \sqrt{1-4 \nu } \left(\frac{36811 \nu ^2}{99840}-\frac{467539 \nu }{174720}+\frac{165421}{399360}\right)$\\
$E^{{\rm cm}v^6}_{53}$ &  $i \sqrt{\frac{\pi }{385}} \sqrt{1-4 \nu } \left(\frac{104545 \nu ^2}{59904}+\frac{29897 \nu }{59904}-\frac{1582769}{239616}\right)$\\
$E^{{\rm cm}v^6}_{51}$ &  $i \sqrt{\frac{\pi }{330}} \sqrt{1-4 \nu } \left(\frac{6869 \nu ^2}{19968}+\frac{4442099 \nu }{279552}-\frac{2770819}{559104}\right)$\\
\hline
$E^{{\rm cm}v^8}_{55}$ &  $i \sqrt{\frac{7 \pi }{11}} \sqrt{1-4 \nu } \left(-\frac{1530067 \nu ^3}{3394560}+\frac{7549403 \nu ^2}{5591040}-\frac{115368121 \nu}{31682560}+\frac{295392673}{380190720}\right)$\\
$E^{{\rm cm}v^8}_{53}$ &  $i \sqrt{\frac{\pi }{385}} \sqrt{1-4 \nu } \left(-\frac{335993 \nu ^3}{63648}+\frac{29174993 \nu ^2}{4073472}+\frac{229397 \nu}{208896}-\frac{138016747}{16293888}\right)$\\
$E^{{\rm cm}v^8}_{51}$ &  $i \sqrt{\frac{\pi }{330}} \sqrt{1-4 \nu } \left(-\frac{1532821 \nu ^3}{339456}+\frac{8559353 \nu ^2}{4752384}+\frac{11341131 \nu}{452608}-\frac{137160775}{19009536}\right)$\\
\hline
$E^{{\rm cm}v^{10}}_{55}$ &  $i \sqrt{\frac{7 \pi }{11}} \sqrt{1-4 \nu } \left(\frac{463068001 \nu ^4}{1031946240}-\frac{10149536767 \nu ^3}{7223623680}+\frac{1283793899 \nu^2}{412778496}-\frac{1226078291 \nu }{277831680}+\frac{142735991}{141639680}\right)$\\
$E^{{\rm cm}v^{10}}_{53}$ &  $i \sqrt{\frac{\pi }{385}} \sqrt{1-4 \nu } \left(\frac{3690332107 \nu ^4}{619167744}-\frac{10504683127 \nu ^3}{619167744}+\frac{26026310365 \nu^2}{1238335488}-\frac{549521725 \nu }{619167744}-\frac{2098909867}{206389248}\right)$\\
$E^{{\rm cm}v^{10}}_{51}$ &  $i \sqrt{\frac{\pi }{330}} \sqrt{1-4 \nu } \left(\frac{139682059 \nu ^4}{25798656}-\frac{1315205249 \nu ^3}{90295296}+\frac{194290415 \nu^2}{21245952}+\frac{85651103 \nu }{2821728}-\frac{141792619}{15049216}\right)$\\
\hline
\end{tabular}
\end{ruledtabular}
\end{table*}
 
\begin{table*}
\caption{\label{tab1:Ecml6} The various nonzero multipolar coefficients of $E^{{\rm cm}v^k}_{lm}$  in an expansion in powers of $v=\frac{p_\infty}{\sqrt{1+p_\infty^2}}$, evaluated for $l=6$ and all $m$  corresponding to nonzero values.  }
\begin{ruledtabular}
\begin{tabular}{ll}
$E^{{\rm cm}v^3}_{66}$ &  $\sqrt{\frac{7 \pi }{429}} \left(\frac{475}{1024}-\frac{25 \nu }{16}\right)$\\
$E^{{\rm cm}v^3}_{64}$ &  $\sqrt{\frac{7 \pi }{26}} \left(\frac{27}{2816}-\frac{3 \nu }{88}\right)$\\
$E^{{\rm cm}v^3}_{62}$ &  $\sqrt{\frac{3 \pi }{455}} \left(\frac{189 \nu }{176}-\frac{3627}{11264}\right)$\\
$E^{{\rm cm}v^3}_{60}$ &  $\sqrt{\frac{\pi }{13}} \left(\frac{59 \nu }{132}-\frac{347}{2688}\right)$\\
\hline
$E^{{\rm cm}v^5}_{66}$ &  $\sqrt{\frac{7 \pi }{429}} \left(-\frac{27 \nu ^2}{32}-\frac{5411 \nu }{512}+\frac{16683}{5120}\right)$\\
$E^{{\rm cm}v^5}_{64}$ &  $\sqrt{\frac{7 \pi }{26}} \left(\frac{151 \nu ^2}{88}-\frac{110395 \nu }{39424}+\frac{23917}{35840}\right) $\\
$E^{{\rm cm}v^5}_{62}$ &  $\sqrt{\frac{3 \pi }{455}} \left(\frac{17689 \nu ^2}{1056}-\frac{259279 \nu }{16896}+\frac{293377}{101376}\right)$\\
$E^{{\rm cm}v^5}_{60}$ &  $\sqrt{\frac{\pi }{13}} \left(\frac{87 \nu ^2}{16}-\frac{87953 \nu }{21504}+\frac{1017721}{1419264}\right)$\\
\hline
$E^{{\rm cm}v^7}_{66}$ &  $\sqrt{\frac{7 \pi }{429}} \left(\frac{18053 \nu ^3}{5440}-\frac{90639513 \nu ^2}{2437120}+\frac{1827739 \nu }{143360}-\frac{96571}{696320}\right)$\\
$E^{{\rm cm}v^7}_{64}$ &  $\sqrt{\frac{7 \pi }{26}} \left(-\frac{7207 \nu ^3}{29920}-\frac{400611 \nu ^2}{478720}-\frac{11981961 \nu }{6702080}+\frac{386227}{609280}\right)$\\
$E^{{\rm cm}v^7}_{62}$ &  $\sqrt{\frac{3 \pi }{455}} \left(-\frac{80423 \nu ^3}{17952}+\frac{6933805 \nu ^2}{382976}-\frac{52514477 \nu }{3446784}+\frac{19000141}{6893568}\right)$\\
$E^{{\rm cm}v^7}_{60}$ &  $\sqrt{\frac{\pi }{13}} \left(-\frac{56675 \nu ^3}{35904}+\frac{10012941 \nu ^2}{1340416}-\frac{103322815 \nu }{24127488}+\frac{15881779}{24127488}\right)$\\
\hline
$E^{{\rm cm}v^9}_{66}$ &  $\sqrt{\frac{7 \pi }{429}} \left(-\frac{923387 \nu ^4}{206720}+\frac{1081324949 \nu ^3}{46305280}-\frac{5710176791 \nu ^2}{92610560}+\frac{10408823317 \nu
   }{370442240}-\frac{69701849}{26460160}\right)$\\
$E^{{\rm cm}v^9}_{64}$ &  $\sqrt{\frac{7 \pi }{26}} \left(-\frac{5761 \nu ^4}{10880}+\frac{1066702421 \nu ^3}{1018716160}-\frac{510201797 \nu ^2}{291061760}-\frac{14394064737 \nu
   }{8149729280}+\frac{38091899}{52920320}\right)$\\
$E^{{\rm cm}v^9}_{62}$ &  $\sqrt{\frac{3 \pi }{455}} \left(-\frac{1392167 \nu ^4}{682176}-\frac{142485923 \nu ^3}{21829632}+\frac{435864713 \nu ^2}{14553088}-\frac{11501325185 \nu
   }{523911168}+\frac{938996735}{261955584}\right)$\\
$E^{{\rm cm}v^9}_{60}$ &  $\sqrt{\frac{\pi }{13}} \left(-\frac{29797 \nu ^4}{62016}-\frac{479434981 \nu ^3}{152807424}+\frac{648440301 \nu ^2}{50935808}-\frac{1267023463 \nu
   }{193019904}+\frac{5155151}{5677056}\right)$\\
\hline
$E^{{\rm cm}v^{11}}_{66}$ &  $\sqrt{\frac{7 \pi }{429}} \left(\frac{16053429 \nu ^5}{3307520}-\frac{29718746549 \nu ^4}{1481768960}+\frac{18210805387 \nu ^3}{317521920}-\frac{732636233177 \nu
   ^2}{8890613760}+\frac{129346607 \nu }{3342336}-\frac{31344450671}{7112491008}\right)$\\
$E^{{\rm cm}v^{11}}_{64}$ &  $\sqrt{\frac{7 \pi }{26}} \left(\frac{15469159 \nu ^5}{18191360}-\frac{9950936827 \nu ^4}{4074864640}+\frac{12309646381 \nu ^3}{3056148480}-\frac{16524977899 \nu
   ^2}{6112296960}-\frac{1029659509 \nu }{514719744}+\frac{1507438225}{1778122752}\right)$\\
$E^{{\rm cm}v^{11}}_{62}$ &  $\sqrt{\frac{3 \pi }{455}} \left(\frac{105225589 \nu ^5}{21829632}-\frac{12966599807 \nu ^4}{1397096448}-\frac{9012762697 \nu ^3}{2095644672}+\frac{37761109781 \nu
   ^2}{931397632}-\frac{69264935 \nu }{2297856}+\frac{161397384799}{33530314752}\right)$\\
$E^{{\rm cm}v^{11}}_{60}$ &  $\sqrt{\frac{\pi }{13}} \left(\frac{1239349 \nu ^5}{909568}-\frac{187124233 \nu ^4}{87318528}-\frac{1898578879 \nu ^3}{458422272}+\frac{3589598939 \nu ^2}{203743232}-\frac{3255068345
   \nu }{349274112}+\frac{18976660691}{14669512704}\right)$\\
\end{tabular}
\end{ruledtabular}
\end{table*}

\end{widetext}


\begin{thebibliography}{99}
\bibitem{ancillary}
Ancillary file. It contains the complete results for $U_2$ up to $O(G^3\eta^7)$ as well as all the  \lq\lq Iterated Bessel Function" master integrals needed for their explicit evaluation.
\end{thebibliography}
\end{document}